\documentstyle[aps,epsfig]{revtex}
\def\Journal#1#2#3#4{{#1} {\bf#2}, #3 (#4)}

\def\NPA{{\rm Nucl. Phys.} A}
\def\NPB{{\rm Nucl. Phys.} B}
\def\PLB{{\rm Phys. Lett.}  B}

\def\PRD{{\rm Phys. Rev.} D}
\def\PRC{{\rm Phys. Rev.} C}

\def\epp{\epsilon^{\prime}}
\def\ep{\epsilon}
\def\vep{\varepsilon}
\def\la{\langle}
\def\ra{\rangle}

\def\be{\begin{equation}}
\def\ee{\end{equation}}
\def\bea{\begin{eqnarray}}
\def\eea{\end{eqnarray}}
\begin{document}
\draft
\title{The Vector Meson Form Factor Analysis in Light-Front Dynamics}
\author{ Bernard L. G. Bakker$^{a}$, Ho-Meoyng Choi$^{b}$
and Chueng-Ryong Ji$^{c}$\\
$^a$ Department of Physics and Astrophysics, Vrije Universiteit,
     De Boelelaan 1081, NL-1081 HV Amsterdam, \\
     The Netherlands\\
$^b$ Department of Physics, Carnegie-Mellon University,
     Pittsburgh, PA 15213\\
$^c$ Department of Physics, North Carolina State University,
Raleigh, NC 27695-8202}
\maketitle

\begin{abstract}
We study the form factors of vector mesons using
a covariant fermion field theory model in $(3+1)$ dimensions.
Performing a light-front calculation in the $q^+ =0$
frame in parallel with a manifestly covariant calculation, 
we note the existence of a nonvanishing
zero-mode contribution to the light-front current $J^+$ and find
a way of avoiding the zero-mode
in the form factor calculations.
Upon choosing the light-front gauge ($\ep^+_{h=\pm}=0$) with circular
polarization and with spin projection $h=\uparrow\downarrow=\pm$,
only the helicity zero to zero matrix element of the plus current
receives zero-mode contributions. Therefore, one can obtain the
exact light-front solution of the form factors using only the valence
contribution if only the helicity components,
$(h'h)=(++),(+-)$, and $(+0)$, are used.
We also compare our results obtained
from the light-front gauge in the light-front helicity basis (i.e.
$h=\pm,0$) with those obtained from the non-LF gauge in the instant
form linear polarization basis (i.e. $h=x,y,z$) where the zero-mode
contributions to the form factors are unavoidable.
\end{abstract}
\date{}

\section{Introduction}
\label{sect.I}
One of the great challenges in hadronic physics is to calculate the
structure of hadrons starting from QCD alone. Presently this task is
very difficult and one relies on specific models to gain some
understanding of hadronic structure at low energies and momentum
transfer values. A popular model is the constituent quark model (CQM)
which in its relativized form has met with quite some success.  A first
test of this model is the comparison of the mass spectra it predicts to
the experimental data. Such a test provides some constraints on the
wave functions. A more stringent test for the wave functions is found
when one also calculates the form factors of a hadron. It lies in the
nature of the CQM that only valence wave functions are determined
easily. However, in a fully covariant calculation of the form factors
one needs the full structure of the hadron-quark vertex.

It has been known for some time that there are situations where the
form factors can be expressed correctly as convolutions of the wave
functions.  Such is the case for certain components of the currents. In
particular one finds within the formalism of light-front dynamics (LFD)
\cite{BCJ1} that the so-called plus component of
the currents for a scalar or pseudoscalar meson can be expressed
in terms of the wave functions alone for spacelike momentum transfer.
The matrix elements obtained this way we call the {\em valence parts}.
The parts arising from vertices that can not be expressed in the wave
functions we call the {\em nonvalence parts}.
(The plus component of a four-vector being a particular combination of
its usual components: $p^+ = (p^0 + p^3)/\sqrt 2$ where the factor $\sqrt 
2$ is conventional.)

In the case of vector mesons the situation is more complicated. Till
now there have been several
recipes~\cite{GK,CCKP,BH}
for the extraction of the invariant
form factors from the matrix elements of the currents. It turns out that
even when one limits oneself to the plus component, these different
ways of extracting the form factors do not produce the same
results~\cite{Card}.
One realizes that
since the nine complex matrix elements of the current($J^+$),
corresponding to the possible combinations of the polarizations of the
initial and final spin-1 particles, can be expressed in terms of three
real invariants only, it becomes clear that there must be relations
between these matrix elements. This was of course known for a long time
and many authors have used this knowledge to sort out the invariants from
the calculated matrix elements. In reference frames where the plus-component
of the momentum transfer, $q^+$, vanishes these relations 
can be reduced to just one
besides the relations provided by Hermiticity, parity and rotation
about the $z$-axis. The latter relation is known as the {\em angular
condition}~\cite{GK}.  In general reference frames the situation was not so
clear. In a previous paper \cite{BJ2} we completely analyzed these
conditions for the spin-1 case and found besides the angular condition
given before another one. There we gave only the formal expressions for
these consistency conditions.
In the $q^+=0$ frame, however, the
additional condition is very simple involving only two helicity
amplitudes and doesn't seem to provide as strong a constraint as the
usual condition since most constituent quark models are expected to satisfy it
rather easily.  Nevertheless, the $q^+=0$ frame is in principle
restricted to the spacelike region of the form factors and it may be
useful to impose this additional condition in processes involving the
timelike region which must be analyzed in the $q^+\neq 0$ frame. Thus,
it is important to analyze both angular conditions in different frames
calculating actually the form factors with existing theoretical
models. In the present paper, we demonstrate their usefulness for
theoretical/phenomenological models for spin-1 objects.  In order that
the matrix elements satisfy these constraints, the current operator
must transform properly and the state vectors must be eigenstates of
total spin.  If the models do not have these properties, the angular
conditions will not be met.

In this work, we use a simple but exactly solvable model for the spin-1
(e.g. $\rho$) meson and separate the valence and nonvalence
contributions to the three physical form factors to investigate the
degree of violations in the two angular conditions for each
contribution in different frames.  Although the quantitative results
that we find in this model may differ in other models depending on the
details of the dynamics in each model, the basic structure of model
calculations is common and we expect the essential findings from this model
calculation may apply to realistic models.

In particular, we compared two different types of polarization vectors,
the one obtained from the LF gauge ($\ep^+_{h=\pm}=0)$, which is
usually used in the LF CQM analysis, and the other obtained from the
instant form (IF) 
polarization, which is not associated with the LF gauge,
i.e. $\ep^+_{x}\neq 0$, 
but used in some recent
papers\cite{Melo1,MT,MS}.
In both cases, there is a zero-mode contribution, i.e.  a contribution
from the nonvalence part that remains finite for $q^+ \to 0$, even if
the plus-component of the currents is used. Specifically, there is a
zero mode contribution in the LF helicity case ($h=+,-,0$) to the
$(h',h)=(0,0)$ amplitude, where $h$ and $h'$ are the initial and final
helicities, respectively, but there is no zero-mode for other helicity
combinations such as $(+,+),(+,-)$ and $(+,0)$. On the other hand, in
the instant form case, only $(yy)$ is immune to the zero-mode but
others such as $(xx),(zz)$ and $(zx)$ do receive zero-mode
contributions. Of course, the two results are exactly the same if one
properly includes the zero-mode contribution.

Now turning to the angular conditions, there are several different
prescriptions~\cite{GK,CCKP,BH} in choosing the matrix elements to
extract the three physical form factors.  We compare three different
types of helicity combinations, GK\cite{GK}, CCKP\cite{CCKP}, and
BH~\cite{BH}, using both LF and instant form helicity bases in a
reference frame where $q^+ = 0$.
One of our very interesting findings of
the analysis in the LF helicity basis is that the prescription using
the plus-component of the current but not involving the $(h',h)=(0,0)$
helicity amplitude in the LF gauge is preferred for model calculation.
Especially, the GK prescription uses only
$(h',h)=(+,+),(+,-)$ and $(+,0)$ but not the pure $(0,0)$ component and
thus achieves the goal of not involving the zero-modes.
On the other hand, the
longitudinal (0,0) component is the most dominant contribution in the
high momentum transfer region and thus it may be better to use the BH
prescription, involving the $(0,0),(+,0)$ and $(+,-)$ amplitudes only,
in the high momentum perturbative QCD analysis. The CCKP prescription,
however, involves all helicity states, i.e. $(+,+),(0,0), (+,0)$ and
$(+,-)$ and one needs a quantitative analysis of the angular conditions
to pin down the momentum transfer region for the validity of this
prescription.  Our quantitative analyses indeed verify that the GK
prescription is remarkably free from the zero-mode contribution but
others are not.  
In the recent work by Melikhov and Simula\cite{MS}, we see that the 
result using the GK prescription is not in complete agreement with their 
covariant model calculation, which is due to the dependence on a light-like
four-vector called $\omega (\omega^2=0)$ in their formulation that 
necessarily involves unphysical form factors. The covariant formulation
presented in our work should be intrinsically distinguished from 
theirs\cite{MS}, since our formulation involves neither $\omega$ nor any
unphysical form factor. 

If we use the instant form basis, however, all three
prescriptions receive the zero-mode contribution.   
Since the instant form helicity is not obtained from the LF
gauge, 
i.e.  $\ep^+_{x}\neq 0$, 
even the GK prescription gets the
zero-mode contribution especially for the magnetic form
factor as we shall show in this work.
Thus, the instant form basis used in the LF formulation seems quite
dangerous, because it can lead to a wrong interpretation of the physics
involved in LF dynamical models. Our solvable model calculation clearly
indicates that one can avoid the zero-mode contribution if the LF basis
when the LF gauge is used without using the longitudinal to
longitudinal helicity amplitude.

This paper is organized as follows.  In Section \ref{sect.II}, we
summarize the angular conditions for spin-1 systems using the LF
helicity basis and the kinematics for the reference frames
Drell-Yan-West (DWY), Breit (BRT), and target-rest frame (TRF) used in
this work.  The three prescriptions (GK, CCKP, BH) used in extracting
the physical form factors are also briefly discussed in that Section.
In Section \ref{sect.III}, we present our covariant model calculations
of physical quantities such as the three electromagnetic form factors
and the decay constant of the spin-1 meson system using both the manifestly
covariant Feynman method and the LF technique. In the $q^+=0$
frame, we separate the full amplitudes into the valence contribution
and the zero-mode contribution to show explicitly that only the
helicity zero to zero amplitude is contaminated by the zero-mode.  In
Section \ref{sect.IV}, we present the numerical results for the form
factors and the angular conditions and analyze the dependences on the
prescriptions, reference frames, and helicity bases.  The taxonomical
decompositions of the full results into valence and  non-valence
contributions are used wherever possible to make a quantitative
comparison of these dependences.  Conclusions follow in Section
\ref{sect.V}. The details of the instant form analysis
and a derivation of the zero-mode are summarized in the
Appendices \ref{sect.A.1} and \ref{sect.A.2}, respectively.

\section{Spin-1 form factors in light-front helicity basis}
\label{sect.II}

The Lorentz-invariant electromagnetic form factors $F_1$, $F_2$, and $F_3$
for a spin-1 particle of mass $m$ are defined~\cite{ACG}
by the matrix elements of the currents $J^\mu$ between the initial
$|p,h\ra$ and the final
$|p',h'\ra$ eigenstates of the momentum $p$ and the helicity $h$
as follows:
\begin{equation}
 G^\mu_{h' h} = \la p',h'|J^\mu|p,h\ra
 = -\ep^{*}_{h'}\cdot\ep_h(p+p')^\mu F_1(Q^2) + (\ep^\mu_h\;q\cdot\ep^{*}_{h'}
 - \ep^{*\mu}_{h'}\;q\cdot\ep_h)F_2(Q^2)
 + \frac{(\ep^{*}_{h'}\cdot q)(\ep_h\cdot q)}{2m^2}
 (p+p')^\mu F_3(Q^2),
 \label{eq.1}
\end{equation}
where $Q^2 = -q^2$, $q=p'-p$ and $\ep_h(\ep_{h'})$ is
the polarization vector of the initial(final) meson.

The physical form factors, charge, magnetic, and quadrupole, are related
in a well-known way to the form factors $F_i$, viz,
\begin{eqnarray}
G_C &=& (1 + \frac{2}{3}\eta) F_1 +\frac{2}{3}\eta F_2
         +\frac{2}{3}\eta (1 + \eta) F_3 \nonumber \\
G_M &=& -F_2 \nonumber \\
G_Q &=& F_1 + F_2 + (1 + \eta) F_3,
 \label{eq.2}
\end{eqnarray}
where $\eta = Q^2/(4m^2)$. 

Using the convention $\varepsilon^\mu = (\varepsilon^+, \varepsilon^-,
\varepsilon^1, \varepsilon^2)$,
the general form of the LF polarization vectors is given by
\begin{equation}
 \left.
 \begin{array}{c}
 \varepsilon_{\rm LF} (p^+,p^1,p^2 ;+) \\
 \varepsilon_{\rm LF} (p^+,p^1,p^2 ;0) \\
 \varepsilon_{\rm LF} (p^+,p^1,p^2 ;-)
 \end{array}
 \right\} = \left\{
 \begin{array}{c}
 \left( 0, \frac{p^r}{p^+}, \frac{-1}{\surd 2}, \frac{-i}{\surd 2},
 \right) \\
 \left( \frac{p^+}{m}, \frac{\vec{p}^{\perp \, 2} - m^2}{2 m p^+},
 \frac{p^1}{m}, \frac{p^2}{m}
 \right) . \\
 \left( 0, \frac{p^l}{p^+}, \frac{ 1}{\surd 2}, \frac{-i}{\surd 2} \right)
 \end{array} \right.
 \label{eq.3}
\end{equation}
Here $p^r (p^l) = \mp (p_x \pm ip_y)/\surd 2$.
Using Eqs.(\ref{eq.1}) and (\ref{eq.3}), we obtain the matrix elements:
\begin{eqnarray}
 G^+_{h' h} & = & \left(
 \begin{array}{ccc}
 G^+_{++} & G^+_{+0} & G^+_{+-} \\
 G^+_{0+} & G^+_{00} & G^+_{0-} \\
 G^+_{-+} & G^+_{-0} & G^+_{--}
 \end{array}
 \right),
 \nonumber \\
  & = & \left(
 \begin{array}{ccc}
 a_1 F_1 + a_3 F_3 & c_1 F_1 + c_2 F_2 + c_3 F_3 & e^*_3 F_3 \\
 b_1 F_1 + b_2 F_2 + b_3 F_3 & d_1 F_1 + d_2 F_2 + d_3 F_3 &
 -(b_1 F_1 + b_2 F_2 + b_3 F_3)^* \\
 e_3 F_3 & -(c_1 F_1 + c_2 F_2 + c_3 F_3)^* & a_1 F_1 + a_3 F_3 \end{array}
 \right).
 \label{eq.4}
\end{eqnarray}
Since we are working only with the plus component of the current,
we shall use the following short-hand notations
\[
 G_a = G^+_{++} = G^{+*}_{--}, \quad G_b = G^+_{0+} = -G^{+*}_{0-},
\]
\begin{equation}
 G_c = G^+_{+0} = -G^{+*}_{-0}, \; G_d = G^+_{00}, \;
 G_e = G^+_{-+} = G^{+*}_{+-}.
 \label{eq.5}
\end{equation}

The invariant form factors can be extracted in a straightforward
way.  The simplest procedure is to solve first for $F_3$ from $G_e$.
Next $F_1$ is obtained from $G_a$ and $F_3$. Then there are three
options for obtaining $F_2$ from $G_b$, $G_c$, and $G_d$. These
solutions are denoted by $F^b_2$, $F^c_2$, and $F^d_2$, respectively.
The full solutions are then
\begin{eqnarray}\
 F_1 & = & \frac{1}{a_1} G_a - \frac{a_3}{a_1 e_3} G_e , \nonumber \\
 F_3 & = & \frac{1}{e_3} G_e, \nonumber \\
 F^b_2 & = & \frac{1}{b_2}
 \left[ -\frac{b_1}{a_1} \, G_a + G_b +
 \frac{a_3 b_1 - a_1 b_3}{a_1 e_3} \, G_e \right], \nonumber \\
 F^c_2 & = & \frac{1}{c_2}
 \left[ -\frac{c_1}{a_1} \, G_a + G_c +
 \frac{a_3 c_1 - a_1 c_3}{a_1 e_3} \, G_e \right], \nonumber \\
 F^d_2 & = & \frac{1}{d_2}
 \left[ -\frac{d_1}{a_1} \, G_a + G_d +
 \frac{a_3 d_1 - a_1 d_3}{a_1 e_3} \, G_e \right].
 \label{eq.6}
\end{eqnarray}
This procedure makes it clear that the covariant form factors of a
spin-1 hadron in Eq.~(\ref{eq.1}) can be determined using only the plus
component of the currents, $G^+_{h'h}(0)\equiv\la P',h'|J^\mu|P,h\ra$,
in any chosen Lorentz frame.  The nine elements of the current operator
$G^+_{h'h}(0)$ must be constrained by the invariance under (i)
time-reversal, (ii) rotation about $\hat{z}$ and (iii) reflection in
the plane perpendicular to $\hat{z}$, and rotational covariance, i.e.
invariance under the rotations about an axis perpendicular to
$\hat{z}$.  So two additional constraints on the current operator are
required. These consistency conditions are the angular conditions,
which we define as \cite{BJ2}
\begin{equation}
 \Delta_{\rm bc} = F^b_2 - F^c_2, \quad \Delta_{\rm bd} = F^b_2 - F^d_2.
 \label{eq.7}
\end{equation}
We know that the form of these conditions depends on the reference frame
\cite{BJ2}. In this work we consider three different frames, which
we define in the following subsection \ref{sect.II.1}. Especially, in the 
frames where $q^+=0$, our angular condition $\Delta_{\rm bd}$ is 
equivalent to the usual angular condition relating the four helicity 
amplitudes discussed in the literature\cite{GK} 
modulo an overall factor as we discuss in the subsection \ref{sect.II.2}.

\subsection{Kinematics}
\label{sect.II.1}

Our conventions for the momenta of the initial and final
state mesons in the three different reference frames, (DYW), (BRT), and (TRF)
are given below. We use the notation
$p^\mu = (p^+, p^-, p_x, p_y) = (p^+, p^-, \vec{p}_\perp)$
and the metric convention $p\cdot q= p^+q^- + p^-q^+
-\vec{p}_\perp\cdot\vec{q}_\perp$. \\

\noindent{\em DYW}
\begin{eqnarray}
 p & = & (p^+, m^2/(2 p^+), 0,0) \nonumber \\
 p^\prime
 & = & (p^+, (Q^2 + m^2)/(2 p^+), Q \cos \phi, Q \sin \phi)
 \label{eq.8}
\end{eqnarray}

\noindent{\em BRT}
\begin{equation}
 \beta = \sqrt{1+\left(\frac{Q}{2m}\right)^2}.
 \label{eq.9}
\end{equation}
\begin{eqnarray}
 p & = &
 \left(\frac{2 m \beta - Q \cos \theta}{2 \surd 2},
       \frac{2 m \beta + Q\cos \theta}{2\surd 2},
         -\frac{Q \sin \theta \cos \phi}{2},
         -\frac{Q \sin \theta \sin \phi}{2}
 \right),
 \nonumber \\
 p^\prime & = &
 \left(\frac{2 m \beta + Q \cos \theta}{2 \surd 2},
       \frac{2 m \beta - Q\cos \theta}{2\surd 2},
         \frac{Q \sin \theta \cos \phi}{2},
         \frac{Q \sin \theta \sin \phi}{2}
 \right).
 \label{eq.10}
\end{eqnarray}

\noindent{\em TRF}
\begin{equation}
 \kappa = \frac{Q^2}{2 m},
 \label{eq.11}
\end{equation}
\begin{eqnarray}
 p & = & \left(\frac{m}{\surd 2}, \frac{m}{\surd 2}, 0,0 \right). \nonumber \\
 p^\prime & = &
 \left(\frac{m + \kappa + \beta Q \cos \theta}{\surd 2},
 \frac{m + \kappa - \beta Q \cos \theta}{\surd 2},
 \beta Q \sin \theta \cos \phi,
 \beta Q \sin \theta \sin \phi
 \right) .
 \label{eq.12}
\end{eqnarray}

In the literature usually the reference frames used are limited to ones
where $q^+ = 0\; (q^2=2q^+q^-- \vec{q}^2_\perp<0)$.
One of such reference frames is the special Breit frame used in
Refs.~\cite{GK,CCKP,BH,Card,Kei,CJNPA},
where $q^+=0, q_y=0, q_x=Q$, and $\vec{p}_\perp=-\vec{p'}_\perp$
{\it i.e.} $\theta = \frac{\pi}{2}, \phi =0$ in Eq.(\ref{eq.10});\\

\noindent{\em Special Breit}
\begin{equation}
 q^\mu = (0,0,Q, 0),\;
 p^\mu=(m\sqrt{1+\eta}/\surd 2, m\sqrt{1+\eta}/\surd 2, -Q/2,0),\;
 p'^\mu = (m\sqrt{1+\eta}/\surd 2, m\sqrt{1+\eta}/\surd 2, Q/2,0),
 \label{eq.13}
\end{equation}
where $\eta=Q^2/4m^2$ is a kinematic factor.
The corresponding polarization vectors are obtained by substituting these
four vectors in Eq.~(\ref{eq.3}) and the transverse($h=\pm$) and
longitudinal($h=0$) polarization vectors in this special Breit frame
are given by
\begin{eqnarray}
 &&\ep^\mu(p,\pm)=
 \frac{\mp 1}{\sqrt{2}}\biggl(0, \frac{-Q}{2p^+}, 1, \pm i\biggr),\;
 \ep^\mu(p,0)=
 \frac{1}{m}\biggl(p^+, \frac{-m^2+ Q^2/4}{2p^+}, \frac{-Q}{2}, 0\biggr),
 \nonumber\\
 &&\ep^\mu(p',\pm)=
 \frac{\mp 1}{\sqrt{2}}\biggl(0, \frac{Q}{2p^+}, 1, \pm i\biggr),\;
 \ep^\mu(p',0)=
 \frac{1}{m}\biggl(p^+, \frac{-m^2+ Q^2/4}{2p^+},\frac{Q}{2}, 0\biggr).
 \label{eq.14}
\end{eqnarray}

\subsection{Angular condition in $q^+ = 0$ frame and prescriptions of 
choosing helicity amplitudes} 
\label{sect.II.2}

In the $q^+=0$ frame, one can reduce the independent matrix elements of
the current down to four, e.g. $G^+_{++},G^+_{+-},G^+_{+0}$ and
$G^+_{00}$ using the front-form helicity
basis~\cite{GK,CCKP,BH,Card,CJNPA} and the rotational covariance requires
one additional constraint on the
current operator. This is what these authors call the angular condition
$\Delta(Q^2)$ and can be obtained from the explicit representations of
the helicity amplitudes in terms of the physical form factors.
Using the relation between the covariant form factors $F_i$ and
the current matrix elements given by Eq.~(\ref{eq.1}),
one can obtain the following helicity amplitudes in the $q^+=0$ frame
\begin{eqnarray}
 G^+_{++}&=& 2p^+(F_1 + \eta F_3),\;
 G^+_{+0}= p^+\sqrt{2\eta} (2F_1 + F_2 + 2\eta F_3),\nonumber\\
 G^+_{+-}&=&-2p^+\eta F_3,\;\;
 G^+_{00}= 2p^+ \{ (1-2\eta)F_1 - 2\eta F_2 - 2\eta^2 F_3 \}.
 \label{eq.15}
\end{eqnarray}
Thus, the usual angular condition relating the four helicity amplitudes
is given by~\cite{GK}
\begin{equation}
 \Delta(Q^2)=(1+2\eta) G^+_{++} + G^+_{+-}
 -\sqrt{8\eta}G^+_{+0} - G^+_{00}=0,
 \label{eq.16}
\end{equation}
where we note an overall factor difference between $\Delta(Q^2)$ and
$\Delta_{bd}(Q^2)$, i.e. $\Delta = d_2 \Delta_{bd}$ 
(See Section \ref{sect.IV.2.5} for the discussion of the factor $d_2$.).

In a practical computation, instead of calculating the Lorentz-invariant
form factors $F_i(Q^2)$, the physical charge ($G_C$), magnetic ($G_M$),
and quadrupole ($G_Q$) form factors are often
used\footnote{In Refs.~\cite{Card,MT}, the form factors $G_0$, $G_1$, and
$G_2$ are used and the two definitions are related by
$G_C=G_0/2p^+$, $G_M=G_1/2p^+$, and $(\eta\sqrt{8}/3)G_Q=G_2/2p^+$}.
However, the relations between the physical invariant form factors and
the matrix elements $G^+_{h'h}$~\cite{Card} are not unique.
Only if the matrix elements fulfill the angular condition Eq.~(\ref{eq.16})
the extracted form factors would not depend on the choice made.
So one may choose which matrix elements to use to extract the form factors.
Perhaps the most popular choices are \cite{GK,CCKP,BH}:
\begin{eqnarray}
 \label{G_GK}
G_C^{\rm GK} &=& \frac{1}{2p^+}
\biggl[ \frac{(3-2\eta)}{3}G^+_{++}
+ \frac{4\eta}{3}\frac{G^+_{+0}}{\sqrt{2\eta}}
+ \frac{1}{3}G^+_{+-} \biggr],
\nonumber\\
G_M^{\rm GK} &=& \frac{2}{2p^+}
\biggl[ G^+_{++} - \frac{1}{\sqrt{2\eta}}G^+_{+0} \biggr],
\nonumber\\
G_Q^{\rm GK} &=& \frac{1}{2p^+}
\biggl[ - G^+_{++}
+ 2\frac{G^+_{+0}}{\sqrt{2\eta}} - \frac{G^+_{+-}}{\eta} \biggr],
\end{eqnarray}
\begin{eqnarray}
 \label{G_CCKP}
G_C^{\rm CCKP} &=& \frac{1}{2p^+(1+\eta)}
\biggl[\frac{3-2\eta}{6}(G^+_{++} + G^+_{00})
+ \frac{10\eta}{3} \frac{G^+_{+0}}{\sqrt{2\eta}}
+ \frac{4\eta-1}{6}G^+_{+-} \biggr],
\nonumber\\
G_M^{\rm CCKP}&=& \frac{1}{2p^+(1+\eta)}
\biggl[G^+_{++} + G^+_{00} - G^+_{+-}
- \frac{2(1-\eta)}{\sqrt{2\eta}}G^+_{+0} \biggr],
\nonumber\\
G_Q^{\rm CCKP} &=& \frac{1}{2p^+(1+\eta)}
\biggl[-\frac{1}{2} (G^+_{++} + G^+_{00})
+ 2\frac{G^+_{+0}}{\sqrt{2\eta}}
- \frac{(\eta + 2)}{2\eta} G^+_{+-} \biggr],
\end{eqnarray}
and
\begin{eqnarray}
 \label{G_BH}
G_C^{\rm BH} &=& \frac{1}{2p^+(1+2\eta)}
\biggl[ \frac{(3-2\eta)}{3}G^+_{00}
+ \frac{16\eta}{3}\frac{G^+_{+0}}{\sqrt{2\eta}}
+ \frac{2(2\eta-1)}{3}G^+_{+-} \biggr],
\nonumber\\
G_M^{\rm BH} &=& \frac{2}{2p^+(1+2\eta)}
\biggl[ G^+_{00} - G^+_{+-}
+ \frac{(2\eta -1)}{\sqrt{2\eta}}G^+_{+0} \biggr],
\nonumber\\
G_Q^{\rm BH} &=& \frac{1}{2p^+(1+2\eta)}
\biggl[2\frac{G^+_{+0}}{\sqrt{2\eta}} - G^+_{00}
- \frac{(1+\eta)}{\eta}G^+_{+-} \biggr].
\end{eqnarray}
The relation between $F's$ and $G's$ given by Eq.~(\ref{eq.2})
holds for any prescription given above.

It is interesting to note that while Grach and Kondratyuk in~\cite{GK}
regarded $G^+_{00}$ as the ``worst" element and took care not to use it
writing the relations Eq.~(\ref{G_GK}), Brodsky and Hiller\cite{BH}
included $G^+_{00}$ instead of $G^+_{++}$ expecting the helicity zero
to zero component of the current matrix element to be the dominant one
in the perturbative QCD regime.  Chung et al.~\cite{CCKP} used all four
independent helicity components of the current matrix elements.  On the
other hand, in the instant form basis  used by some
authors~\cite{Melo1,MT,FFS}, the independent matrix elements of the
current operator are $G^+_{xx},G^+_{yy},G^+_{zz}$ and $G^+_{zx}$ and
the angular condition becomes $\Delta(Q^2)=G^+_{yy}-G^+_{zz}$.  In
appendix A, we show the relevant expressions for the form factors in
the instant form spin basis~\cite{Melo1,MT,FFS}.  Although the authors
in Refs.~\cite{Melo1,MT,FFS} argued that this basis is
completely equivalent to the LF helicity basis, the relation between
them, which will be discussed later, is not trivial.

\section{Calculation in a solvable covariant model}
\label{sect.III}
The solvable model, based on the covariant Bethe-Salpeter(BS)
model of ($3+1$)-dimensional fermion field theory, enables us to derive
the form factors of a spin-1 particle exactly.

\begin{figure}
\centerline{\psfig{figure=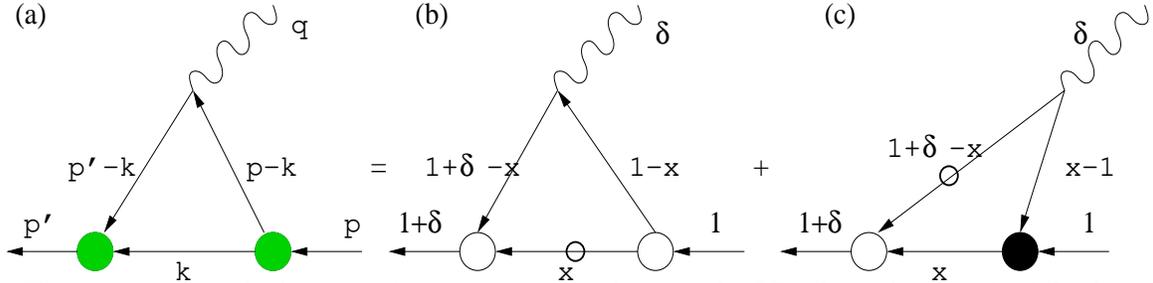,height=1.5in,width=6in}}
\caption{The covariant triangle diagram (a) is represented as the sum
of a LF valence diagram (b) defined in the region $0<k^+<p^+$ and the
nonvalence diagram (c) defined in $p^+<k^+<p'^+$.
$\delta=q^+/p^+=p'^+/p^+-1$. The white and black blobs at
the meson-quark vertices in (b) and (c) represent the
LF wave-function and non-wave-function vertices, respectively.
The small circles in (b) and (c) represent the (on-shell)
mass pole of the quark propagator determined from the $k^-$-integration.
\label{Fig1}}
\end{figure}

The covariant diagram shown in Fig.~\ref{Fig1}(a) is in general
equivalent to the sum of the LF valence diagram (b) and the nonvalence
diagram (c), where $\delta=q^+/p^+=p'^+/p^+-1$.  The matrix element
$G^\mu_{h'h}(0)$ of the electromagnetic (EM) current
of a spin-1 particle
with equal mass constituents ($m_q=m_{\bar q}$) obtained from the
covariant diagram of Fig.~\ref{Fig1}(a) is given by
\begin{eqnarray}
 \label{jji}
G^\mu_{h'h}(0)&=&
iN_cg^2\int\frac{d^4k}{(2\pi)^4}
\frac{S_\Lambda(k-p)S^\mu_{h'h}S_\Lambda(k-p')}
{[(k-p)^2-m^2_q+i\vep][k^2-m^2_q+i\vep][(k-p')^2-m^2_q+i\vep]},
\end{eqnarray}
where $N_c$ is the number of colors and $g$, modulo the charge factor
$e_q$, is the normalization constant, which can be fixed by requiring
the charge form factor to be unity at zero momentum transfer.
$S^\mu_{h'h}$ is the trace term of the quark propagators.  To
regularize the covariant fermion triangle-loop in ($3+1$) dimension,
we replace the point photon-vertex $\gamma^\mu$ by a non-local
(smeared) photon-vertex $S_\Lambda(p)\gamma^\mu S_\Lambda(p')$, where
$S_\Lambda(p)=\Lambda^2/(p^2-\Lambda^2+i\vep)$ and $\Lambda$ plays the
role of a momentum cut-off similar to the Pauli-Villars
regularization\cite{BCJ1}.

When we do the Cauchy integration over $k^-$ to obtain the LF
time-ordered diagrams, we want to avoid the complexity of treating
double $k^-$-poles, so we decompose the product of five energy denominators
in Eq.~(\ref{jji}) into a sum of terms with three energy denominators
only:
\begin{eqnarray}
\label{eq.17}
\frac{1}{D_\Lambda D_0 D_k D'_0 D'_\Lambda}&=&
\frac{1}{(\Lambda^2-m^2_q)^2}\frac{1}{D_k}
\biggl( \frac{1}{D_\Lambda} - \frac{1}{D_0}\biggr)
\biggl(\frac{1}{D'_\Lambda} - \frac{1}{D'_0} \biggr),
\end{eqnarray}
where
\begin{equation}\label{DK}
D_\Lambda = (k-p)^2-\Lambda^2 + i\ep, \; D_0=(k-p)^2-m^2_q+i\ep,\;
D_k= k^2-m^2_q + i\ep,
\end{equation}
and $D'_{0[\Lambda]}=D_{0[\Lambda]}(p\to p')$.

Our treatment of $S_\Lambda$ as the non-local smearing photon-vertex
remedies~\cite{BCJ1} the conceptual difficulty associated with the
asymmetry appearing if the fermion-loop were regulated by smearing the
$q\bar{q}$ bound-state vertex. As will be discussed later, the two
methods lead to different results for the calculation of the decay
constant even though they give the same result for the form
factors.

The vector meson decay constant $f_V$ in this covariant model
with the nonlocal gauge boson vertex $S_\Lambda(k)\gamma^\mu S_\Lambda(k-p)$
is defined by
\begin{eqnarray}\label{fV}
A^\mu&=&\la 0|\bar{q}\gamma^\mu q|p;1J_3\ra=i\sqrt{2}f_V m\ep^\mu(J_3),
\end{eqnarray}
where
\begin{eqnarray}\label{fV_A}
A^\mu&=& N_c g\Lambda^4\int\frac{d^4k}{(2\pi)^4}
\frac{ {\rm Tr}[\not\!\ep(\not\!k-\not\!p+m_q)\gamma^\mu(\not\!k+m_q)] }
{[k^2-m_q^2+i\vep][(k-p)^2-m_q^2+i\vep][k^2-\Lambda^2+i\vep]
[(k-p)^2-\Lambda^2+i\vep]}.
\end{eqnarray}

\subsection{Manifestly covariant calculation}
\label{sect.III.1}
In the manifestly covariant calculation, we obtain the form factors
$F_i(i=1,2,3)$ using dimensional regularization.
Although the splitting procedure Eq.~(\ref{eq.17}) may not be neccessary
in the covariant calculation, it seems more effective in practical
computation. Here we describe some essential steps for the
derivation of the covariant form factors:
We $(i)$ reduce the five propagators into the sum of three propagators
using Eq.~(\ref{eq.17}), $(ii)$ use the Feynman parametrization for
the three propagators, e.g.,
\begin{eqnarray}\label{Feyn}
\frac{1}{D_k D_0 D'_0}&=&
2\int^1_0 dx\int^{1-x}_0 dy
\frac{1}{[D_k + (D_0 - D_k) x + (D'_0-D_k) y]^3},
\end{eqnarray}
and $(iii)$ make a Wick rotation of Eq.~(\ref{jji}) in $D$-dimension to
regularize the integral, since otherwise one encounters missing the
logarithmic divergent terms in Eq.~(\ref{jji}).  Following the above
procedures $(i)$ -$(iii)$ we
finally obtain the  covariant form factors as follows:

\begin{eqnarray}\label{Form_Cov}
F_1(Q^2)&=&\frac{N_cg^2\Lambda^4}{8\pi^2(\Lambda^2-m^2_q)^2}
\int^1_0dx\int^{1-x}_0dy
\biggl\{ (2-x-y)
{\rm ln}\biggl(\frac{C^2_{k\Lambda 0}C^2_{k0\Lambda}}
{C^2_{k\Lambda\Lambda}C^2_{k00}} \biggr)
\nonumber\\
&&\;+\biggl[ -(x+y)(x+y-1)^2 m^2 + (2-x-y)xy\; Q^2
-(2-x-y)m^2_q\biggr]C^2 \biggr\},
\nonumber\\
F_2(Q^2)&=&-\frac{N_cg^2\Lambda^4}{8\pi^2(\Lambda^2-m^2_q)^2}
\int^1_0dx\int^{1-x}_0dy
\biggl\{ (2+x+y)
{\rm ln}\biggl(\frac{C^2_{k\Lambda 0}C^2_{k0\Lambda}}
{C^2_{k\Lambda\Lambda}C^2_{k00}}\biggr)
\nonumber\\
&&\; +\biggl[ (x+y)[(x+y)^2 -1]m^2 + (x+y)xy\; Q^2
-(2+x+y)m^2_q\biggr]C^2 \biggr\},
\nonumber\\
F_3(Q^2)&=&\frac{N_cg^2\Lambda^4}{8\pi^2(\Lambda^2-m^2_q)^2}
\int^1_0dx\int^{1-x}_0dy\;
8xy(x+y-1)m^2 C^2,
\end{eqnarray}
where
\begin{eqnarray}\label{CKK}
C^2_{k\Lambda\Lambda}&=&(x+y)(1-x-y)m^2 - xy\;Q^2
-(x+y)\Lambda^2 - (1-x-y)m^2_q,
\nonumber\\
C^2_{k\Lambda0}&=& (x+y)(1-x-y)m^2 - xy\;Q^2
-(x\Lambda^2 + ym^2_q) - (1-x-y)m^2_q,\nonumber\\
C^2_{k0\Lambda}&=&C^2_{k\Lambda0}(x\leftrightarrow y),
\nonumber\\
C^2_{k00}&=&(x+y)(1-x-y)m^2 - xy\;Q^2 - m^2_q,
\end{eqnarray}
and  $C^2=(1/C^2_{k\Lambda\Lambda} -
1/C^2_{k\Lambda0} - 1/C^2_{k0\Lambda} + 1/C^2_{k00})$.
Note that the logarithmic terms in $F_1$ and $F_2$ are obtained from the
dimensional regularization.

Following a similar procedure for the form factor calculation,
the covariant result for the decay constant is obtained as
\begin{eqnarray}\label{fv_Cov}
f^{\rm COV}_V
=\frac{N_c g\Lambda^4}{4\sqrt{2}\pi^2m(\Lambda^2-m^2_q)^2}
\int^1_0 dx &&
\biggl\{
[m^2_q+{\tilde M}^2]\;
{\rm ln}\frac{[{\tilde M}^2-xm^2_q-(1-x)\Lambda^2]
                [{\tilde M}^2-x\Lambda^2-(1-x)m^2_q]}
             {[{\tilde M}^2-\Lambda^2][{\tilde M}^2-m^2_q]}
\nonumber\\
&&\;\;- [{\tilde M}^2-\Lambda^2]\; \ln[-{\tilde M}^2+\Lambda^2]
- [{\tilde M}^2-m^2_q]\; \ln[-{\tilde M}^2+m^2_q]
\nonumber\\
&&\;\; + [{\tilde M}^2-xm^2_q-(1-x)\Lambda^2]\;
    \ln[-{\tilde M}^2 + xm^2_q+(1-x)\Lambda^2]
\nonumber\\
&&\;\; + [{\tilde M}^2-x\Lambda^2-(1-x)m^2_q]\;
    \ln[-{\tilde M}^2 + x\Lambda^2+(1-x)m^2_q]
\biggr\},
\end{eqnarray}
where ${\tilde M}^2=x(1-x)m^2$.

\subsection{Light-front calculation}
\label{sect.III.2}

We shall use only the plus-component of the current matrix element
$G^+_{h'h}$ in the calculation of the form factors.  In principle,
one can directly calculate the trace term $S^+_{h'h}$ with
$k^-=k^-_{\rm pole}$, which depends on the integration region of $k^+$.
However, for the purpose of a clear understanding of the physics
implied in LF dynamics, we instead separate $S^+_{h'h}$ into the
on-mass shell propagating part and the (off-mass shell) instantaneous
one using the following identity
\begin{equation}
 \label{iden}
\not\!p + m_q = (\not\!{p_{\rm on}} + m_q)
+ \gamma^+(p^- - p^-_{\rm on}),
\end{equation}
where the subscript (on) denotes the on-mass shell ($p^2=m^2_q$) quark
propagator, i.e.  $p^-=p^-_{\rm on}=(m^2_q+\vec{p}^2_\perp)/2p^+$.
Then the trace term $S^+_{h'h}$ of the quark propagators
in Eq.~(\ref{jji}) is given by
\begin{eqnarray}
 \label{Sji}
S^+_{h'h}(P=k-p,P'=k-p',k)= (S^+_{h'h})_{\rm on} +
(S^+_{h'h})_{\rm inst.},
\end{eqnarray}
where
\begin{eqnarray}
 \label{Sji_on}
(S^+_{h'h})_{\rm on}&=&{\rm Tr}
[\not\!\ep^{*}_{h'}(\not\!P'_{\rm on}+m_q)\gamma^+
(\not\!P_{\rm on}+m_q)\not\!\ep_h(\not\!k_{\rm on} + m_q)],
\nonumber\\
&=&\; 4\; P^+ \biggl[
(k_{\rm on}\cdot\ep^{*}_{h'})(P'_{\rm on}\cdot\ep_h)
+ (P'_{\rm on}\cdot\ep^{*}_{h'})(k_{\rm on}\cdot\ep_h)
+ (\ep^{*}_{h'}\cdot\ep_h)(m_q^2 - k_{\rm on}\cdot P'_{\rm on})\biggr]
\nonumber\\
&&+\; 4\; P'^+ \biggl[
(k_{\rm on}\cdot\ep^{*}_{h'})(P_{\rm on}\cdot\ep_h)
+ (P_{\rm on}\cdot\ep^{*}_{h'})(k_{\rm on}\cdot\ep_h)
+ (\ep^{*}_{h'}\cdot\ep_h)(m_q^2 - k_{\rm on}\cdot P_{\rm on})\biggr]
\nonumber\\
&&+\; 4\; k^+ \biggl[
(P'_{\rm on}\cdot\ep^{*}_{h'})(P_{\rm on}\cdot\ep_h)
- (P_{\rm on}\cdot\ep^{*}_{h'})(P'_{\rm on}\cdot\ep_h)
- (\ep^{*}_{h'}\cdot\ep_h)(m_q^2 - P_{\rm on}\cdot P'_{\rm on})\biggr]
\nonumber\\
&&-\; 4\;\ep^{*+}_{h'}\biggl[
(P_{\rm on}\cdot\ep_h)(k_{\rm on}\cdot P'_{\rm on} - m^2_q)
-(P'_{\rm on}\cdot\ep_h)(k_{\rm on}\cdot P_{\rm on} -m^2_q)
+(k_{\rm on}\cdot\ep_h)(p_{\rm on}\cdot P'_{\rm on} - m^2_q) \biggr]
\nonumber\\
&&-\; 4\;\ep^{+}_h\biggl[
(P'_{\rm on}\cdot\ep^{*}_{h'})(k_{\rm on}\cdot P_{\rm on} - m^2_q)
-(P_{\rm on} \cdot\ep^{*}_{h'})(k_{\rm on}\cdot P'+_{\rm on} -m^2_q)
+(k_{\rm on}\cdot\ep^{*}_{h'})(P_{\rm on}\cdot P'_{\rm on} - m^2_q) \biggr],
\end{eqnarray}
and
\begin{eqnarray}\label{Sji_inst}
(S^+_{h'h})_{\rm inst.} &=&
(k^- - k^-_{\rm on})
{\rm Tr}[\not\!\ep^{*}_{h'}(\not\!{P'_{\rm on}} + m_q)\gamma^+
(\not\!{P_{\rm on}}+m_q)\not\!\ep_h\gamma^+],\nonumber\\
&=& 8(k^- - k^-_{\rm on}) \biggl[
\ep^{*+}_{h'}P^+_{\rm on}(\ep_h\cdot P'_{\rm on})
+ \ep^+_h P'^+_{\rm on}(\ep^{*}_{h'}\cdot P_{\rm on})
- P'^+_{\rm on}P^+_{\rm on}(\ep^{*}_{h'}\cdot\ep_h)
+ \ep^{*+}_{h'} \ep^+_h (m^2_q - P'_{\rm on}\cdot P_{\rm on})
\biggr].
\end{eqnarray}
As we shall show below, the LF {\em valence} contribution comes
exclusively from the {\em on-mass} shell propagating part,
Eq.~(\ref{Sji_on}), and the {\it zero-mode} (if it exists) from
the {\it instantaneous} part, Eq.~(\ref{Sji_inst}).

Using the special Breit frame (See Eq.(\ref{eq.13}).) with the LF gauge, we
obtain for the trace terms
$(S^+_{h'h})_{\rm on}$ and $(S^+_{h'h})_{\rm inst.}$ given by
Eqs.~(\ref{Sji_on}) and (\ref{Sji_inst}) the expressions
\begin{eqnarray}\label{spp_on}
(S^+_{++})_{\rm on}&=& \frac{4 p^+}{x}
\biggl[ m^2_q + (2x^2-2x + 1)
\biggl(\vec{k}^2_\perp - \frac{x^2}{4}Q^2
+ i x (\vec{k}_\perp\times \vec{q}_{\perp})\cdot\hat{z}\biggr)
\biggr],\nonumber\\
(S^+_{+-})_{\rm on}&=&
8(1-x)p^+\biggl[ (k_x -ik_y)^2 - \frac{x^2}{4}Q^2\bigg],
\nonumber\\
(S^+_{+0})_{\rm on}&=&
\frac{\sqrt{8\eta}}{Q}p^+(2(k_x-ik_y) - xQ)(2x-1)
\biggl[(1-x)(m^2 + M^2_0) + \frac{x}{4}Q^2
+ \vec{k}_\perp\cdot \vec{q}_\perp
\biggr],\nonumber\\
(S^+_{00})_{\rm on}&=& \frac{4p^+}{m^2}\biggl[
x\biggl( (1-x)(m^2 + M^2_0) + \frac{x}{4}Q^2\biggr)^2
- x (\vec{k}_\perp\cdot \vec{q}_\perp)^2
\biggr],
\end{eqnarray}
and
\begin{eqnarray}\label{spp_inst}
(S^+_{++})_{\rm inst.}&=&(S^+_{+-})_{\rm inst.}=0,\nonumber\\
(S^+_{+0})_{\rm inst.}&=& \frac{8(p^+)^2}{m\sqrt{2}}(k^- - k^-_{\rm on})
(x-1)\biggl[k_x-ik_y + \biggl(1-\frac{x}{2}\biggr) Q\biggr],\nonumber\\
(S^+_{00})_{\rm inst.}&=&
\frac{8(p^+)^2}{m^2}(k^- - k^-_{\rm on})
\biggl[-\frac{x^2}{4}Q^2 + m^2_q + \vec{k}^2_\perp \biggr],
\end{eqnarray}
where $x=k^+/p^+$, and
$M^2_0=(m^2_q+ \vec{k}^2_\perp)/[x(1-x)]$.
We note that the terms proportional to an odd power of $\vec{k}_\perp$
do not contribute to the integral.

By doing the integration over $k^-$ in Eq.~(\ref{jji}), one finds the
two LF time-ordered contributions to the residue calculations
corresponding to two poles in $k^-$, the one coming from the interval
(I) $0<k^+<p^+$ (See Fig.~\ref{Fig1}(b).), the ``valence diagram", and
the other from (II) $p^+<k^+<p'^+$ (See Fig.~\ref{Fig1}(c).), the
``nonvalence diagram" or ``Z" graph.  These diagrams are expressed in
terms of energy denominators.

\subsubsection{Valence contribution}
\label{sect.III.2.1}
In the region 
$0<k^+<p^+$, the pole $k^-=k^-_{\rm on}=(m^2_q +
\vec{k}^2_\perp -i\vep)/2k^+$ (i.e the spectator quark), is located in
the lower half of the complex $k^-$-plane.  Thus, the Cauchy integration
formula for the $k^-$-integral in Eq.~(\ref{jji}) gives in this region
for the plus current, $G^+_{h'h}(0)$,
\begin{eqnarray}\label{jjiv}
G^{+val}_{h'h}&=& \frac{N_c}{2(2\pi)^3}
\int^{1}_{0}\frac{dx}{x(1-x)^4}\int d^2 \vec{k}_\perp
\frac{ g\Lambda^2}{(m^2-{\cal M}^2_{0})
(m^2-{\cal M}^2_{\Lambda})}
S^{+val}_{h'h}
\frac{g\Lambda^2}{(m^2-{\cal M}'^{2}_{0})
(m^2-{\cal M}'^{2}_{\Lambda})},
\end{eqnarray}
where
\begin{eqnarray}\label{inmass}
{\cal M}^2_{0}&=&
\frac{m^2_q + (\vec{k}_\perp - x\vec{p}_\perp)^2}{x}
+ \frac{m^2_q + (\vec{k}_\perp - x\vec{p}_\perp)^2}{1-x},\nonumber\\
{\cal M}^2_{\Lambda}&=&
\frac{m^2_q + (\vec{k}_\perp - x\vec{p}_\perp)^2}{x}
+ \frac{\Lambda^2+ (\vec{k}_\perp - x\vec{p}_\perp)^2}{1-x},
\end{eqnarray}
are the invariant masses of the initial meson state. The invariant
masses of the final state, i.e.  ${\cal M}'^2_0$ and
${\cal M}'^2_\Lambda$ in Eq.~(\ref{jjiv}), can be obtained by replacing
$\vec{p}_\perp\to -\vec{p}_\perp$ in Eq.~(\ref{inmass}).  As one can
easily see from Eqs.~(\ref{spp_on}) and (\ref{spp_inst}), only the
on-mass shell quark propagator part contributes to the valence diagram,
i.e. $S^{+val}_{h'h}=(S^{+}_{h'h})_{\rm on}$.  Note, however, that this
relation does not hold in general for other components of the currents,
e.g. $S^{-val}_{h'h}\neq(S^{-}_{h'h})_{\rm on}$.  One of the
distinguished features of the LF plus current matrix element given by
Eq.~(\ref{jjiv}) is that the physical interpretation is manifest in
terms of the LF wave function, i.e. a convolution of the initial and the
final state LF wave functions, which is not possible for the covariant
calculation.

\subsubsection{Zero-mode contribution}
\label{sect.III.2.2}
In the region  
$p^+<k^+<p'^+$, the poles are at
$k^-=p'^- + [m^2_q +(\vec{k}_\perp -\vec{p'}_\perp)^2 -i\vep]/2(k^+-p'^+)$
(from the struck quark propagator) and
$k^-=p'^- + [\Lambda^2+(\vec{k}_\perp -\vec{p'}_\perp)^2 -i\vep]/2(k^+-p'^+)$
(from the smeared quark-photon vertex $S_\Lambda(k-p')$),
and are located in the upper half of the complex $k^-$-plane.

Since the integration range of the nonvalence region,
$p^+<k^+<p'^+(=p^++q^+)$, shrinks to zero in the $q^+\to 0$ limit, the
nonvalence contribution is sometimes mistakenly thought to be always
vanishing for $q^+ \to 0$.  However, in reality it may not vanish but give a
finite contribution,
\begin{eqnarray}\label{limit}
\lim_{q^+\to 0}\int^{p^++q^+}_{p^+}dk^+(\cdots)
\equiv\lim_{\delta\to 0}\int^{1+\delta}_1 dx(\cdots)\neq 0.
\end{eqnarray}
Then it is called the ``zero-mode"~\cite{CM,Bur,BHw,SB,CJZ,BJ1} in the $q^+=0$
frame. The nonvanishing zero-mode contribution occurs only if the integrand
$(\cdots)$ in Eq.~(\ref{limit}) behaves $\sim k^-({\rm i.e.}(1-x)^{-1})$.
Note that there is no zero-mode contribution either in the case the
integrand behaves like $k^-(k^+-p^+)^n(n\geq 1)$ or is $k^-$-independent.

For the plus current, the zero-mode contribution
comes from the spin structure of the fermion propagator,
specifically only from the instantaneous part given by Eq.~(\ref{spp_inst})
and neither from the on-mass shell propagating part nor the energy denomimator.
Thus, without detailed knowledge of the energy denominator, it is easy
to find from Eqs.~(\ref{spp_inst}) and~(\ref{limit}) that
only the helicity zero to zero component gives a nonvanishing
zero-mode contribution:
\begin{equation}\label{S00_zm}
\lim_{x\to 1}(S^+_{00})_{\rm inst.}
= \frac{8(p^+)^2}{m^2}k^-(m^2_q + \vec{k}^2_\perp - Q^2/4),
\end{equation}
where $k^-\sim 1/(1-x)\to\infty$ as $x\to 1$. In other words, while
the integration region shrinks to zero, the integrand for the helicity
zero to zero component goes to infinity leading to a finite
zero mode contribution.

As we said before, we avoided the complexity of the Cauchy integration
over double $k^-$-poles by decomposing the product of five energy
denominators in Eq.~(\ref{jji}) into a sum of terms with three energy
denominators.  In this way, we perform the Cauchy integration of
$G^+_{00}$ over the single $k^-$-pole, either
$D'_\Lambda$ or $D'_0$, instead of double $k^-$-poles.

For example, the $1/ (D_k D_\Lambda D'_\Lambda)$ term in
Eq.~(\ref{eq.17}) combined with the pole position
$k^- = p'^- + [\Lambda^2+(\vec{k}_\perp -\vec{p'}_\perp)^2]/2(k^+-p'^+)$
appearing in $(S^+_{00})_{\rm inst.}$ gives
(See Appendix~\ref{sect.A.2} for the detailed derivation.)
\begin{equation}\label{jjinv}
i\int\frac{d^4k}{(2\pi)^4} \frac{k^-}{D_k D_\Lambda D'_\Lambda}
= \frac{1}{2(2\pi)^3p^+} \int d^2\vec{k}_\perp
\frac{{\rm ln}\biggl[ \frac{(\vec{k}_\perp -\vec{p'}_\perp)^2
+ \Lambda^2}
{(\vec{k}_\perp -\vec{p}_\perp)^2 + \Lambda^2}\biggr]}
{[(\vec{k}_\perp -\vec{p}_\perp)^2 + \Lambda^2]
- [(\vec{k}_\perp -\vec{p'}_\perp)^2 + \Lambda^2] }.
\end{equation}
Similarly, one can obtain a
nonvanishing zero-mode contributions for the other energy denominator
terms given by Eq.~(\ref{eq.17}).
Explicitly, the zero-mode contribution from $S^+_{00}$ in this special
Breit frame is given by
\begin{eqnarray}\label{S00}
G^{+ \,{\rm z.m.}}_{00}&=&
\frac{N_cg^2\Lambda^4}{2p^+(2\pi)^3(\Lambda^2-m^2_q)^2}
\int d^2\vec{k}_\perp
\frac{8(p^+)^2}{m^2}(m^2_q+\vec{k}^2_\perp-Q^2/4)
\nonumber\\
&\times&\biggl\{
\frac{\ln \biggl[ \frac{(\vec{k}_\perp -\vec{p'}_\perp)^2 + \Lambda^2}
{(\vec{k}_\perp -\vec{p}_\perp)^2 + \Lambda^2}\biggr] }
{[(\vec{k}_\perp -\vec{p}_\perp)^2 + \Lambda^2]
- [(\vec{k}_\perp -\vec{p'}_\perp)^2 + \Lambda^2]}
-
\frac{\ln \biggl[ \frac{(\vec{k}_\perp -\vec{p'}_\perp)^2 + m^2_q}
{(\vec{k}_\perp -\vec{p}_\perp)^2 + \Lambda^2}\biggr] }
{[(\vec{k}_\perp -\vec{p}_\perp)^2 + \Lambda^2]
- [(\vec{k}_\perp -\vec{p'}_\perp)^2 + m^2_q]}
+ ( m_q\leftrightarrow \Lambda)
\biggr\}.
\end{eqnarray}
The angular condition $\Delta(Q^2)$ given by Eq.~(\ref{eq.16}) is
satisfied only if the zero-mode contribution for
$G^{+ \,{\rm z.m.}}_{00}$ in Eq.~(\ref{S00}) is included,
i.e. $G^+_{00}=G^{+ \,{\rm val}}_{00}+G^{+ \,{\rm z.m.}}_{00}$.

A similar analysis has been made by de Melo et al.~\cite{Melo1}, where
the authors found the zero-mode contribution using the instant form
basis~\cite{Melo1,MT,FFS} instead of the LF helicity
basis~\cite{GK,CCKP,BH,Card,ACG,CJNPA} for the polarization vectors of
a spin-1 particle.  In principle, the LF helicity basis can be related
to the instant form spin basis by some transformation.
Interestingly, however, we find that since the authors in
Ref.~\cite{Melo1} used the non-LF gauge(i.e. $\ep^+_x\neq 0$)
polarization vectors, the three polarization components, i.e.
$G^+_{xx},G^+_{zz}$ and $G^+_{zx}$, receive zero-mode contributions as
we explicitly show in appendix A.  In other words, using the instant
form basis with a non-LF gauge used in~\cite{Melo1}, one cannot avoid
the zero-mode contribution to the form factors of a spin-1 particle no
matter what prescription is used.

We use the results of our numerical calculations to compare the form
factors obtained in the LF helicity basis (in LF gauge) with those
obtained in the instant form linear polarization basis (in non-LF gauge)
as well as the covariant ones.

In the LF calculation of the vector meson decay constant,
the plus current with the longitudinal($h=0$) polarization vector is
usually used. In the special Breit frame (See Eqs.(\ref{eq.13}) and 
(\ref{eq.14}).), we thus obtain 
\begin{eqnarray}\label{fv_LF}
f^{\rm LF}_V&=& \frac{N_c g\Lambda^4}{4\sqrt{2}\pi^3m}
\int^1_0\frac{dx}{x^3(1-x)^3}\int d^2\vec{k}_\perp
\biggl[x(1-x)(p^+)^2 + m^2_q + \vec{k}^2_\perp -
\vec{k}_\perp\cdot \vec{p}_\perp\biggr]
\nonumber\\
&&\times
\frac{2x(1-x)m^2 - m^2_q - \Lambda^2
- 2( \vec{k}_\perp-x \vec{p}_\perp)^2}
{[m^2-M^2_{0m}][m^2-M^2_{0\Lambda}]
[m^2-M^2_{0m}(m_q\leftrightarrow\Lambda)]
[m^2-M^2_{0\Lambda}(m_q\leftrightarrow\Lambda)]}.
\end{eqnarray}
Our LF calculation of the decay constant in Eq.~(\ref{fv_LF}) is
exactly the same as the covariant result in Eq.~(\ref{fv_Cov}).  We
also note that there is no zero mode contribution to $f^{\rm LF}_V$ in
our model calculation. This can be easily seen from the trace
calculation, because ${\rm
Tr}[\not\!\ep(\not\!k-\not\!p+m_q)\gamma^+(\not\!k+m_q)]$
=$4\{(\ep\cdot k)(2k^+-p^+) +\ep^+(m^2_q-k^2 + k\cdot p)\}$ and the
$k^-$-terms cancel each other, so only the good component is left in the
numerator.  It is interesting to note that
while our calculation of the decay constant with a non-local (but
symmetric) gauge boson vertex is immune to the zero-mode, the same
calculation by Jaus~\cite{Ja} is not, where the author used a local
gauge boson vertex and an asymmetric smearing meson vertex.

\section{Numerical Results}
\label{sect.IV}

In this section, we present the numerical results for the form factors and
angular conditions and analyze the dependences on prescriptions,
helicity bases and reference frames. However, we do not aim at finding the
best-fit parameters to describe the experimental data of the $\rho$ meson
properties. Rather, we simply take the parameters used by others\cite{MT}
with which we were able to reproduce the results in
that particular work. Nevertheless, as we mentioned earlier,
our model calculations have a generic structure and the essential
findings from our calculations may apply to the more realistic
models, although the quantitative results would differ in other models
depending on the details of the dynamics in each model.

In our numerical calculations, we thus use $m=0.77$ GeV, $m_q=0.43$
GeV, and $\Lambda=1.8$ GeV~\cite{MT} and make the taxonomical
decompositions of the full results into the valence and nonvalence
contributions to facilitate a quantitative comparison of the various
dependences such as the prescriptions (GK,CCKP,BH), the helicity bases
(LF,IF) and the reference frames (DYW,BRT,TRF).  We first present the
dependences on the prescriptions and the helicity bases in the $q^+=0$
frame (See subsection \ref{sect.IV.1}.). Then, in subsection
\ref{sect.IV.2}, we present the frame dependences using exclusively the
LF helicity basis.

\subsection{Dependences on the helicity bases and the prescriptions}
\label{sect.IV.1}

In Fig.~\ref{Fig_GC}, we show the charge form factor $|G_C(Q^2)|$
obtained from the light-front (left) and the instant-form (right) spin
bases.  The full solutions(thick solid line) are obtained from three
different prescriptions~\cite{GK,CCKP,BH} given by
Eqs.~(\ref{G_GK})-(\ref{G_BH}) for the light-front basis and
Eqs.~(\ref{G_GK_I})-(\ref{G_BH_I}) for the instant-form basis,
respectively, and they all turn out to give exactly the same result as
the covariant one as they should be.  The slope of the full solution
gives the charge radius of the bound-state as defined in Eq.(\ref{eq.x.1}) 
and we obtained $\langle r_C^2 \rangle = 7.63$ GeV$^{-2}$ with the 
parameter set we used. More detailed discussions on the charge, magnetic 
and quadrupole radii can be found in subsection \ref{sect.IV.2.6}. In the
$q^+=0$ frame, the full solutions can be decomposed into the valence
contribution and the zero-mode contribution since the nonvalence
diagram reduces in the limit $q^+ \to 0$ to the zero mode.  To estimate
it, we plot the valence contribution for each prescription, i.e. the dotted
line for GK~\cite{GK}, long-dashed line for CCKP~\cite{CCKP}, and
dot-dashed line for BH~\cite{BH}, respectively. The normalization
constant $g$ is fixed by
requiring the full solution to be
normalized to $G_C(0)=1$. As one can see in Fig.\ref{Fig_GC}, the two
results for the valence contributions obtained from the light-front and
the instant-form bases exactly coincide with each other. However, only
the GK prescription is immune to the zero-mode contribution for both
helicity bases. The dotted curve cannot be seen because it is on top of
the solid curve.  Other prescriptions, CCKP and BH, receive large
amounts of zero-mode contributions (i.e. the difference between the
full solution and the valence one).  As we discussed earlier, the GK
prescription does not involve the $G^+_{00}$ component which is the
only source of the zero-mode for the light-front helicity basis and the
zero-modes from the $G^+_{xx}$ and $\eta G^+_{zz}$ terms in
Eq.~(\ref{G_GK_I}) for the instant-form basis cancel each other(See
Eq.~(\ref{Sxx}).).  We also show the angular condition(small squares)
given by Eq.~(\ref{eq.16}) without including the zero-mode contributions.
If we include the zero-mode contributions, then it is of course exactly
zero.

The situation is rather different for the calculation of the magnetic
form factor $G_M$ as shown in Fig.~\ref{Fig_GM}.  For the full
solution, the two (LF and IF helicity bases) results are again exactly
the same as they should be.  The magnetic moment (in units of $e/2m$)
and its radius given by Eq.(\ref{eq.x.1}) are obtained as $\mu_1= 2.1$ and 
$\langle r_M^2 \rangle = 9.73$ GeV$^{-2}$ (See also subsection 
\ref{sect.IV.2.6}.), respectively.  However, the valence (or for that matter
the zero-mode) contributions to the full solution are quite different
depending on the helicity bases.  For the light-front helicity basis,
the GK prescription is again immune to the zero-mode and the dotted
curve is exactly on top of the solid curve.  Also, the other
prescriptions, CCKP and BH, receive large amounts of the zero-mode
contributions as in the case of the $G_C$ calculation.  However, for
the instant form spin basis used in~\cite{MT}, not only the CCKP and BH
prescriptions but also the GK prescription are affected by the
zero-mode, because the zero-mode terms $-G^+_{zz}$ and
$G^+_{zx}/\sqrt{\eta}$ in Eq.~(\ref{G_GK_I}) do not cancel each other
(See Eq.~(\ref{Sxx}).) but rather add up.

We show in Fig.~\ref{Fig_GQ} the quadrupole form factor $G_Q(Q^2)$
obtained from the light-front (left) and the instant-form (right)
helicity bases. The quadrupole moments (in units of $e/m^2$) and the
corresponding radius given by Eq.(\ref{eq.x.1}) are obtained as $Q_1 = 
0.91$ and $\langle r^2_Q \rangle = 12.6$ GeV$^{-2}$ (See also subsection 
\ref{sect.IV.2.6}.), respectively. As in the 
case of $G_C(Q^2)$, the two (LF and IF helicity bases) results coincide 
and the dotted curves are exactly on top of the solid curves because of 
the absence of the zero-mode in the GK prescription.

The decay constant (See Eqs.(\ref{fv_Cov}) and (\ref{fv_LF}).)
using the same parameters yields the result $f_v = 133.7$ MeV,
while the experimental data $f_{\rho^0} = 152.8 \pm 3.6$ MeV
and $f_{\rho^\pm} = 147.3 \pm 0.7$ MeV are
obtained from the width $\Gamma (\rho \rightarrow e^+ e^-)$
and the branching ratio $Br(\tau \to \rho \nu_\tau)
= (25.02 \pm 0.16)\%$ \cite{pdg}, respectively.

\subsection{Light-front valence parts}
\label{sect.IV.2}

We checked that in all reference frames the sum of the valence and
nonvalence contributions to the form factors is equal to the covariant
result. Therefore henceforth we plot the valence contributions only.
The valence parts will in general depend on the polar angle $\theta$
in BRT and TRF,
but are independent of the azimuthal angle $\phi$
in all three reference frames(DYW,BRT,TRF). We used the latter
property as a check of the accuracy of our codes.

Using Eq.~(\ref{eq.1}) for the matrix elements and the kinematics specified in
Eqs.~(\ref{eq.10}) and (\ref{eq.12}) for BRT and TRF, respectively, one 
finds that the coefficients $b_i$ and $c_i$, $i=1,2,3$, vanish for 
$\theta = 0$.  Therefore we illustrate the angular dependence of the
valence parts in BRT and TRF by giving them for the small but nonvanishing 
value $\theta = \pi/20$.  On the other hand $\theta = \pi/2$ is singled out
for the BRT frame, so we chose for a larger value of $\theta$ the
value $9\pi/20$.  There is a symmetry about $\theta = \pi/2$, so the
amplitudes for $\pi/2 \leq \theta \leq \pi$ don't contain any additional
information.

Eventually we plot the momentum dependence of the valence parts of the
form factors and of the violation of the angular conditions for two
values of the polar angle $\theta$. 
We see that in all reference frames and for all angles the angular
condition $\Delta_{\rm bd}$ diverges for $Q^2 \to 0$. This is due to
the fact that the coefficient $d_2$ of $F_2$ in the matrix element
$G_d$ (See Eq.~(\ref{eq.4}).) vanishes for $Q^2 = 0$. For that reason
there is a finite contribution of the nonvalence part or zero mode to
the charge form factor even in the limit $Q^2 \to 0$, which shows up in
all $d$ variants of the physical form factors. For finite values of
$Q^2$ an accidental singularity in $\Delta_{\rm bd}$ may occur.  To
follow that up we plot the angular variation of the angular conditions
for two values of the momentum transfer $Q^2$. We can also explain the
occurrence of these singularities as due to the vanishing of
the coefficient $d_2$ of $F_2$ in the matrix element $G_d$.

\subsubsection{Drell-Yan-West kinematics}
\label{sect.IV.2.1}

In the DYW reference frame there is no dependence on $\theta$. The
dependence on $\phi$ amounts to simple phase factors, $e^{\pm i\phi}$ for
$G_b$ and $G_c$, and $e^{2i\phi}$ for $G_e$.
In Figs. \ref{fig.05} and \ref{fig.06},
the results for the valence parts of the
invariant form factors $F_1$, $F_2$, and $F_3$ and the physical ones
$G_C$, $G_M$ and $G_Q$ are shown. $F_1(0)$ is normalized to 1, which is
not affected by the zero mode.  For the same reason 
$G^b_C(0) = G^c_C(0) = 1$.
However, $G^d_C(0) \neq 1$ and for positive $Q^2$ $G^d_C$ deviates from the 
correct one by the zero-mode contribution to $F^d_2$.
It is clear that the zero mode is very important if one does choose
the $F^d_2$-prescription. Neither of the $F^b_2$- or $F^c_2$-prescription
contains the zero mode.  As mentioned before, they correspond to the
GK-prescription in the DYW-frame for $\phi = 0$.

\subsubsection{Breit frame kinematics}
\label{sect.IV.2.2}

Our convention for the BRT frame entails both $\theta$- and
$\phi$-dependences of the matrix elements. The latter being trivial, we
fixed $\phi = 0$ in all our calculations, after checking that indeed
the form factors are independent of this angle. For $\theta = \pi/2$
the BRT frame and the DYW frame can be connected by a kinematical
transformation, so the results for the form factors become identical.
(See the discussion in \cite{BJ2}.) We chose two values for the angle
$\theta$, slightly different from $0$ and $\pi/2$ to illustrate the
angular dependence of the valence parts of the form factors.  The
results shown 
in Figs.\ref{fig.07}-\ref{fig.10}
are for $\theta = \pi/20$ and $\theta = 9\pi/20$. 
As shown in Fig.\ref{fig.07}, it
is immediately clear that only $F_1$ and $F_3$ are rather insensitive
to the choice of the polar angle, but the three prescriptions for $F_2$
are dramatically changing with $\theta$ going from a small value to one
near $\pi/2$. This strong angle-dependence is found also in the
physical form factors shown in Figs.\ref{fig.08}-\ref{fig.10}, although the 
$b$- and $c$-variants are much less
changed than the $d$-variant. In all cases the charge form factor shows
the least angular variation.

\subsubsection{Target-rest-frame kinematics}
\label{sect.IV.2.3}

The results shown 
in Figs.\ref{fig.11}-\ref{fig.14} are again for $\theta = \pi/20$ and 
$\theta = 9\pi/20$.
Everything we said for the results obtained in the Breit frame can be 
repeated for the target-rest frame. 
In Figs.\ref{fig.12}-\ref{fig.14},
we see for the $b$- and $c$-variants similar angular
dependences, but for the $d$-variant the variation with $\theta$ is even more
dramatic than in the Breit frame. 
The results in Fig.\ref{fig.14} for $\theta = 9\pi/20$ hint at
a singular behaviour of the $d$-variant that is explained by the fact that
for some combinations of $Q^2$ and $\theta$ the coefficient $d_2$ vanishes.
We discuss more details of the situation below in Section~\ref{sect.IV.2.5}.

\subsubsection{Angular condition}
\label{sect.IV.2.4}

In the next plots of Figs.\ref{fig.15}-\ref{fig.17},
we show the two angular conditions for a somewhat
longer $Q^2$ interval, up till 10 GeV${}^2$. In the case of the Breit
(Fig.\ref{fig.16}) and target-rest(Fig.\ref{fig.17}) frames we plot the 
differences $\Delta_{bc} = F^b_2 - F^c_2$ and $\Delta_
{bd} = F^b_2 - F^d_2$ for four angles $\theta =
\pi/8, \, \pi/4, \, 3\pi/8, \pi/2$. For the DYW frame(Fig.\ref{fig.15}) 
$\Delta_{bc}= 0$, so in the plot $\Delta_{bc}$ coincides with the $Q^2$-axis.

In the Breit frame the angular conditions do depend on $\theta$ and we
show their behaviour for the same angles as in
Sec.~\ref{sect.IV.2.2}.  The angular condition that was trivially
fulfilled in the DYW frame turns out to be only weakly violated in the
Breit frame. The other one however, $\Delta_{bd}$ is strongly violated
for small values of $Q^2$. It demonstrates clearly the importance of
including the nonvalence parts in a calculation of the matrix elements
of the current. For large values of $Q^2$ it tends very quickly to
zero, corroborating the expectation that in perturbative QCD one may
ignore largely the nonvalence parts for sufficiently high momentum
tranfers.

Again, the discussion of the behaviour of the angular conditions in the
target-rest frame can be very similar to the one for the Breit frame,
so we shall not repeat it. We only mention that the overall behaviour
is similar in these two cases, but the details differ. In particular,
in the following subsection \ref{sect.IV.2.5} we consider two $Q^2$ values
($Q^2 = 1.0$ GeV$^2$ and $10.0$ GeV$^2$) and find that 
the singularity in $\Delta_{bd}$
occurs in the Breit frame for $\theta$ close to $\pi/4$ while in the
target-rest frame it shows up for $\theta$ close to $3\pi/8$.

\subsubsection{Singular behavior}
\label{sect.IV.2.5}
Figure \ref{fig.14} shows that $G^d_C$ starts to drop significantly
when $Q^2$ is near to 4 GeV${}^2$. Such a behavior can be
understood from the dependence of the coefficient $d_2$ occurring in
Eqs.~(\ref{eq.4}) and (\ref{eq.5}), on the momentum transfer $Q$ and the
angle $\theta$. It appears that both in the BRT frame and the TRF this
coefficient may vanish for a particular combination of $Q^2$ and
$\theta$.

The singularity of $\Delta_{bd}$ in the Breit frame is illustrated in
Fig.~\ref{fig.18}. We see that it occurs for $Q^2 = 1.0$ GeV${}^2$ close
to $\theta = \pi/4$.  The other angular condition remains flat in
$\theta$ and the same is true for both conditions for $Q^2 = 10.0$
GeV${}^2$. A similar picture is found in Fig. \ref{fig.19} for the 
target-rest frame, only the position of the singularity being different.

We can understand this behaviour very easily if we consider the expressions for
$d_2$, that can be derived in a straightforward way from the kinematics and the
expressions for the polarization vectors inserted in Eq.~(\ref{eq.1}). We
find
\begin{equation}
 d^{\rm BRT}_2 =
 \frac{\surd 2 \beta m Q^2 [1-\beta^2 + (1+\beta^2)\cos 2\theta]}
 {4\beta^2 m^2 - Q^2 \cos^2\theta}
 \label{eq.IV.01}
\end{equation}
for the Breit frame and
\begin{equation}
 d^{TRF}_2 =
 \frac{(2m + \kappa + \beta Q \cos\theta)
       (\kappa^2 + 2\beta\kappa Q \cos\theta+ \beta^2 Q^2 \cos 2\theta)}
 {2\surd 2m(m+\kappa + \beta Q \cos\theta)}
 \label{eq.IV.02}
\end{equation}
for the target-rest frame.

Solving the equation $d_2 = 0$ for $Q^2 = 1.0$ GeV$^2$, we find for BRT 
$\theta = 0.698 = 0.222 \pi$  and for TRF $\theta = 1.181 = 0.376 \pi$.
These angles coincide with the positions of singularities shown in 
Figs.\ref{fig.18} and \ref{fig.19}. Also, we note that for $\theta = 
\frac{\pi}{8}$ in the BRT frame there exists a singularity at 
$Q^2 = 8 m^2/(\sqrt{2}-1) \approx 11.45$ GeV$^2$ which is not shown in 
Fig.\ref{fig.18} due to the restricted interval of $Q^2$ only up to $10$ 
GeV$^2$. However, except the tiny region near this singularity position, 
the angular condition is very well satisfied at the larger $Q^2$ region.

\subsubsection{Charge, Magnetic and Quadrupole Radii}
\label{sect.IV.2.6}

From the slope of physical form factors 
for $Q^2 \to 0$, the corresponding radii ($\langle r^2_C \rangle,
\langle r^2_M \rangle,\langle r^2_Q \rangle$) 
can be defined\footnote{
These relations are associated with the interpretation of the Fourier
transform of the form factors for spacelike momentum transfer as densities
and the behaviour of the spherical Bessel functions for small argument
$j_l(x) \sim \frac{x^l}{(2l+1)!!}\left[ 1 - \frac{x^2}{2 (2l+3)} + \cdots 
\right]$.} 
as 
\begin{eqnarray}
 G_C(Q^2) & \sim & G_C(0)\left[1 - \frac{1}{6} \langle r^2_C \rangle Q^2 
\right] , \nonumber \\
 G_M(Q^2) & \sim & G_M(0)\left[1 - \frac{1}{10} \langle r^2_M \rangle Q^2 
\right] , \nonumber \\
 G_Q(Q^2) & \sim & G_Q(0)\left[1 - \frac{1}{14} \langle r^2_Q \rangle Q^2 
\right] . \label{eq.x.1}
\end{eqnarray}

When one considers only the valence parts of charge, magnetic and quadrupole 
form factors, one should be careful in obtaining the corresponding radii 
given by Eq.(\ref{eq.x.1}) because some of them exhibit singular 
behaviors as $Q^2 \to 0$. In order to determine the radii, one may in 
general try to take the limit $(G(Q^2) - G(0))/Q^2$ as $Q^2 \to 0$. In 
practice, however, this gives a rather unreliable value as for very 
small values of $Q^2$ the calculations may have numerical noise that is 
amplified by taking the difference of two almost equal numbers and 
dividing the result by the small number $Q^2$. A more stable procedure 
is to make a linear fit to the form factors in a domain close to $Q^2 = 
0$. We have chosen the interval $0.01 \leq Q^2 \leq 0.1$ and took ten 
equidistant values for $Q^2$. In order to check whether it makes sense 
to fit the form factors to a linear function of $Q^2$ in this domain, we 
plotted the form factors as shown in Figs.\ref{fig.1}-\ref{fig.7} and
also checked the quality of the fit. 

It turned out that only $G_M$ and $G_Q$ in the $d$-variants could not be 
fitted with a straight line. The reason is that in these variants the 
influence of the zero mode is very big and only including the zero mode 
or the nonvalence part can correct the nonlinear behaviour.
Because of this reason, we do not show the $d$-variant valence results for 
$G_M$ and $G_Q$ which are anyway out of scale in 
Figs.\ref{fig.1},\ref{fig.4} 
and \ref{fig.7}. However, in the case that includes the nonvalence part 
as shown in Figs.\ref{fig.2}-\ref{fig.4} (total in BRT) and 
Figs.\ref{fig.5}-\ref{fig.7} (total in TRF),  
all values obtained for the radii do agree. 
The same is true for the case where the zero mode does not occur 
as shown in Fig.\ref{fig.1} ($b$- and
$c$-variants in DYW). An indication of
the accuracy of the results is obtained if one includes the values at
$Q^2 = 0$ in the fit. Then the radii do not change by more than 1.5\%.
If one would try do determine these quantities by fitting the form
factors in a much smaller interval, say $.001 \leq Q^2 \leq 0.01$ in
order to improve the linear approximation mathematically, then the
numerical noise will have a stronger influence. So there is a trade off
between truncation error in the series expansion of the Bessel function
and numerical noise. We are satisfied with an overall numerical error of
the order of 1\%. In Table \ref{tab.1}, we summarize the numerical 
results of radii. For the $d$-variants, both $G_M$ and $G_Q$ diverges for 
$Q^2 \to 0$ as discussed above and an entry `div.' is given for those 
cases in this Table. The numerical estimates for the radii in these 
divergent cases, which do not include $Q^2 = 0$, give indeed values of 
the order of a hundred GeV${}^{-2}$.
\begin{table}[t]
\caption{Squared radii in GeV${}^{-2}$ for the different reference frames
 and variants. \label{tab.1}}
\begin{tabular}{|l|r|r|r|}
 Ref. frame, variant & $G_C$ & $G_M$ & $G_Q$ \\
\hline
 DYW, $G^b$ and $G^c$ &  7.63  & 9.73 & 12.6 \\
 DYW, $G^d$ &  9.56  & div. & div. \\
\hline
 & \multicolumn{3}{c}{$\theta = \pi/20$}\\
\hline
 BRT, tot          &  7.63  & 9.73 & 12.6 \\
 BRT, $G^b$, val   &  14.3  & 27.6 & 50.4 \\
 BRT, $G^c$, val   &  14.7  & 18.3 & 19.4 \\
 BRT, $G^d$, val   &  13.9  & div. & div. \\
\hline
 & \multicolumn{3}{c}{$\theta = 9\pi/20$}\\
\hline
 BRT, tot          &  7.63  & 9.73 & 12.6 \\
 BRT, $G^b$, val   &  8.42  & 12.0 & 17.7 \\
 BRT, $G^c$, val   &  8.48  & 10.7 & 13.6 \\
 BRT, $G^d$, val   &  10.0  & div. &  div. \\
\hline
 & \multicolumn{3}{c}{$\theta = \pi/20$}\\
\hline
 TRF, tot          &  7.63  & 9.73 & 12.6 \\
 TRF, $G^b$, val   &  14.3  & 27.7 & 50.6 \\
 TRF, $G^c$, val   &  14.8  & 18.3 & 19.5 \\
 TRF, $G^d$, val   &  14.1  & div. & div. \\
\hline
 & \multicolumn{3}{c}{$\theta = 9\pi/20$}\\
\hline
 TRF, tot          &  7.63  & 9.73 & 12.6 \\
 TRF, $G^b$, val   &  9.84  & 15.9 & 26.2 \\
 TRF, $G^c$, val   &  9.98  & 12.5 & 15.2 \\
 TRF, $G^d$, val   &  15.5  & div. &  div. \\
\end{tabular}
\end{table}

\section{Conclusion}
\label{sect.V}

In this work, we made a taxonomical analysis of spin-1 form factors
with respect to several different prescriptions (GK,CCKP,BH),
polarization vector choices (LF,IF), and reference frames
(DYW,BRT,TRF). We used the $J^+$ current for all of our analysis.

In the $q^+=0$ frame, we looked at both LF and
IF polarization vectors and made a comparative analysis
on the three prescriptions (GK,CCKP,BH) in relating the matrix elements
to the physical form factors. In the light-front gauge, $A^+ =0$, the
light-front helicity basis is the set of eigenvectors of the
light-front helicity operator. However, the instant-form polarization
vectors have been also used in the literature. We find that the
zero-mode contamination occurs minimally in the light-front gauge
because only the helicity zero to zero amplitude, i.e. $G^+_{00}$, gets
a zero-mode contribution, but all others
($G^+_{++},\,G^+_{+0},\,G^+_{+-}$) are immune from the zero-mode in the
light-front basis. Thus, one can find a prescription which doesn't
involve the zero mode contribution at all.  Indeed, the GK prescription
which doesn't use $G^+_{00}$ has precisely this property.
Consequently, the computation of only the valence contributions to the
form factors using the GK prescription yields results which coincide
exactly with the full results of the form factors as we have shown in
Figs.\ref{Fig_GC}-\ref{Fig_GQ}. These full results can be obtained by the 
other prescriptions,
CCKP and BH, only if the zero-mode contributions are added to the valence
contributions.  We have also computed the form factors using a
manifestly covariant Feynman method and explicitly shown that the full
results of the light-front calculations are fully in agreement
with the covariant one no matter what prescriptions we use. Although
the full results are also independent of the choice of the
polarization vectors, we find that the nice feature of the GK
prescription described above is lost in the instant-form basis.  As
shown in Fig.\ref{Fig_GM}, the valence contribution to the magnetic
form factor ($G_M (Q^2)$) computed in the instant-form basis
substantially differs from the full result even if the GK prescription
is used.  This is due to the fact that a larger number of matrix elements
are contaminated by the zero-mode in the instant-form basis
and in the particular case of $G_M (Q^2)$ the terms affected by the
zero-mode such as $-G^+_{zz}$ and $G^+_{zx}/\sqrt{\eta}$ in
Eq.~(\ref{G_GK_I}) do not cancel each other (See Eq.~(\ref{Sxx}).) but
rather add up.  We thus conclude that the GK prescription in the
light-front basis is certainly advantageous for model calculation
involving only the valence contributions to the spin-1 form factors.

We have also analyzed the frame dependence of the valence contribution
to the physical form factors and the angular conditions using the
light-front polarization vectors. Since the three presciptions
discussed above are defined only in the $q^+=0$ frame, we use the
prescriptions $b$, $c$, and $d$ defined in Section~\ref{sect.II} to work
in a general frame. In DYW frame, our $c$ prescription corresponds to
the GK prescription and the results on the form factors from our $b$
prescription coincide with those from the $c$ or GK prescription. Also,
some combinations of the $c$ and $d$ prescriptions correspond to the
CCKP and BH prescriptions depending on the coefficients of the
combinations. Again, only the $d$ prescription involves the zero-mode
and the valence result from the $d$ prescription differs from the
results of the $b$ and $c$ prescriptions that coincide each other 
exactly as shown in Figs.~\ref{fig.05} and  \ref{fig.06}. The results in 
the Breit frame
reproduce the DYW results if $\theta = \pi/2$, since they can be
transformed into each other by purely kinematic operators in LFD. If
$\theta \neq \pi/2$, however, the results are quite different from the
DYW results. As shown in Figs.~\ref{fig.07}-\ref{fig.10} more drastic
differences in the results among the $b$, $c$, and $d$ prescriptions
are found at $\theta = \pi/20$ than at $\theta = 9\pi/20$.  Similar
observations can be made also for the TRF.  However, the kinematic
equivalence to DWY obtains only at a special angle $\theta =\theta_0$
(See Ref.~\cite{BJ1}.) which depends on $Q^2$. Thus, it is rather
difficult to see the similarity of the results of TRF and DYW.
Although the angular condition $\Delta_{bc}$ is rather well satisfied,
the usual angular condition $\Delta_{bd}$ is severely broken in the
small $Q^2$ region. Singularities associated with those violations are
visible in our figures(See e.g. Figs.\ref{fig.18} and \ref{fig.19}.).

Nevertheless, both angular conditions $\Delta_{bc}$ and $\Delta_{bd}$
are well satisfied in the region $Q^2$ above a few GeV$^2$ 
except the tiny region near the singularity position discussed in
Section \ref{sect.IV.2.5}.
and thus the
results from all the prescriptions become consistent with each other.
Thus, one may
conclude that the zero-mode contributions are highly suppressed in the high
$Q^2$ region and the results are consistent with the perturbative QCD
predictions. This may justify the use of the BH prescription in the
analysis of high $Q^2$ form factors.  For the low and intermediate
$Q^2$ regions, however, the zero-mode contributions are very important
and the GK prescription with the light-front polarization vectors in
the DYW frame is certainly desirable for the form factor analyses. An
application of this observation to a more realistic model calculation
is under consideration.

\newpage

\acknowledgements
\noindent
This work was supported in part by a grant from the US Department of
Energy (DE-FG02-96ER 40947) and the National Science Foundation 
(INT-9906384).
This work was started when HMC and CRJ visited the Vrije Universiteit
and they want to thank the staff of the department of physics at
VU for their kind hospitality.
BLGB wants to thank the staff of the department of physics
at NCSU for their warm hospitality during a stay when this work was
completed. The North Carolina Supercomputing
Center and the National Energy Research Scientific Computer Center are also
acknowledged for the grant of Cray time.

\appendix

\section{Form factors in the instant-form basis}
\label{sect.A.1}

Using the instant-form linear polarization vectors ($h,h'=x,y,z$), 
the form factors
corresponding to Eqs.~(\ref{G_GK})-(\ref{G_BH}) in the light-front
polarization vectors are given by~\cite{MT}
\begin{eqnarray}\label{G_GK_I}
G_C^{\rm GK}&=& \frac{1}{2p^+}
\biggl[\frac{1}{3}G^+_{xx}
+ \frac{(2 - \eta)}{3} G^+_{yy}
+ \frac{\eta}{3} G^+_{zz} \biggr],
\nonumber\\
G_M^{\rm GK} &=& -\frac{1}{2p^+}\biggl[
G^+_{yy} - G^+_{zz} + \frac{G^+_{zx}}{\sqrt{\eta}}\biggr],
\nonumber\\
G_Q^{\rm GK}&=& \frac{1}{2p^+}
\biggl[\frac{1}{2\eta}G^+_{xx}
- \frac{(1+\eta)}{2\eta} G^+_{yy}
+ \frac{1}{2} G^+_{zz} \biggr],
\end{eqnarray}
\begin{eqnarray}\label{G_CCKP_I}
G_C^{\rm CCKP}&=& \frac{1}{2p^+}
\biggl[ \frac{1}{3} G^+_{xx}
+ \frac{1}{6}G^+_{yy} + \frac{1}{2} G^+_{zz} \biggr],
\nonumber\\
G_M^{\rm CCKP} &=&
-\frac{1}{2p^+}\frac{G^+_{zx}}{\sqrt{\eta}},\;\;
G_Q^{\rm CCKP}= \frac{1}{2p^+}
\frac{(G^+_{xx} - G^+_{yy})}{2\eta},
\end{eqnarray}
and
\begin{eqnarray}\label{G_BH_I}
G_C^{\rm BH}&=& \frac{1}{2p^+(1+2\eta)}
\biggl[ \frac{(1+2\eta)}{3} G^+_{xx}
+ \frac{(2\eta - 1)}{3} G^+_{yy}
+ \frac{(3 + 2\eta)}{3} G^+_{zz} \biggr],
\nonumber\\
G_M^{\rm BH} &=& \frac{-1}{2p^+(1+2\eta)}
\biggl[ \frac{(1+2\eta)}{\sqrt{\eta}} G^+_{zx}
- G^+_{yy} + G^+_{zz} \biggr],
\nonumber\\
G_Q^{\rm BH}&=& \frac{1}{2p^+(1+2\eta)}
\biggl[ \frac{(1+2\eta)}{2\eta} G^+_{xx}
- \frac{(1+\eta)}{2\eta} G^+_{yy}
- \frac{1}{2} G^+_{zz} \biggr],
\end{eqnarray}
where
we redefine the definition of the form factors $G_0,G_1$, and
$G_2$ in Ref.~\cite{MT} in terms of $G_C, G_M$, and $G_Q$ according
to footnote 1.
In order to calculate the form factors of a spin-1 particle in this
instant-form bases, the authors of Ref.~\cite{Melo1,MT}
used the reference frame specified in our Eq.~(\ref{eq.13}).
However, they use different (i.e. non LF gauge)
polarization vectors in the initial($\ep$) and final($\epp$)
states that are given by
\begin{eqnarray}\label{poli}
\ep^\mu_x&=&(-\sqrt{\eta},\sqrt{1+\eta},0,0),
\;\ep^\mu_y=(0,0,1,0),\; \ep^\mu_z=(0,0,0,1),\end{eqnarray}
and
\begin{eqnarray}\label{polf}
\epsilon'^\mu_x&=&(\sqrt{\eta},\sqrt{1+\eta},0,0),\;
\epsilon'^\mu_y=\ep^\mu_y,\;
\epsilon'^\mu_z=\ep^\mu_z.
\end{eqnarray}
(Here we use the component convention $p ^\mu =(p^0,p^1,p^2,p^3)$ etc.)
Even though these polarization vectors satisfy the correct orthonormality
and closure relations as well as the conditions $\ep\cdot p=\epp\cdot 
p'=0$, they cannot be obtained in the LF gauge.

Proceeding to calculate the trace terms
$S^+_{xx},S^+_{yy},S^+_{zz}$ and $S^+_{zx}$ using Eqs.~(\ref{poli}),
and~(\ref{polf}), we obtain
\begin{eqnarray}\label{Sxx}
S^+_{xx}&=& \underline{
-4k^-\eta\biggl[m^2_q+ \vec{k}^2_\perp - \eta m^2 \biggr] }
+ 4k^-p^+(p^+ - k^+)(1-x-\eta -\eta x)
\nonumber\\
&+&4p^+\biggl[ 4(x +\eta x -1)k^2_x -\eta(1+\eta)x^3 m^2
- (x+3\eta x-2)(m^2_q+ \vec{k}^2_\perp)\biggr],
\nonumber\\
S^+_{yy}&=& 4k^-(k^+ - p^+)^2
- 4p^+\biggl[4(1-x) k^2_y + \eta x m^2
- (2-x)(m^2_q+ \vec{k}^2_\perp)\biggr],
\nonumber\\
S^+_{zz}&=& \underline{
4k^-\biggl[m^2_q+ \vec{k}^2_\perp-\eta m^2 \biggr] }
+ 4(1-x)p^+\biggl[x(1-x)(1+\eta)m^2 + 2(m^2_q+ \vec{k}^2_\perp)\biggr],
\nonumber\\
S^+_{zx}&=& \underline{ -4k^-\sqrt{\eta}
\biggl[m^2_q+ \vec{k}^2_\perp - \eta m^2 \biggr] }
+ 4\sqrt{\eta}k^-p^+(k^+ -p^+)
\nonumber\\
&+& 4\sqrt{\eta}p^+\biggl[ 2k^2_x - x^2(1-x)(1+\eta)m^2
- 2(m^2_q +  \vec{k}^2_\perp)\biggr],
\end{eqnarray}
where $\surd 2 p^+=p^0=m\sqrt{1+\eta}$ and we omit the terms of odd
power
in $\vec{k}_\perp$ since they do not contribute to the integral.
Note also that we do not separate the on-shell propagating part from
the instantaneous one in this instant-form calculation.  As we
discussed, only the underlined terms in $S^+_{xx},S^+_{zz}$ and
$S^+_{zx}$ contribute to the zero-mode part and the $S^+_{yy}$
component is immune to the zero-mode contributions.

\section{Derivation of the zero-mode}
\label{sect.A.2}
In this appendix, we derive Eq.~(\ref{jjinv}) in detail.
Performing the $k^-$-integration in Eq.~(\ref{jjinv}),
the residue at the pole $k^-= p'^+ +
[(\vec{k}_\perp-\vec{p'}_\perp)^2+\Lambda^2]/2(k^+-p'^+)$
gives
\begin{eqnarray}\label{B1}
G^{+{\rm z.m.}}_{h'h}
&=&i \int\frac{d^4k}{(2\pi)^4}\frac{k^-}
{D_k D_\Lambda D'_\Lambda}\nonumber\\
&=&\frac{i} {(2\pi)^4}
\int\frac{ dk^+ dk^- d^2\vec{k}_\perp}
{2k^+ 2(k^+-p^+) 2(k^+-p'^+)}
\nonumber\\
&&\times
\frac{k^-}{\biggl[k^- -\frac{ \vec{k}^2_\perp + m^2_q-i\vep}{2k^+}\biggr]
\biggl[k^- - p^- -
\frac{ (\vec{k}_\perp-\vec{p}_\perp)^2 + \Lambda^2-i\vep}
{2(k^+-p^+)}\biggr]
\biggl[k^- - p'^- -
\frac{(\vec{k}_\perp-\vec{p'}_\perp)^2 + \Lambda^2-i\vep}
{2(k^+-p'^+)}\biggr]},
\nonumber\\
&=&-\frac{1} {(2\pi)^3}
\int^{p'^+}_{p^+}\frac{ dk^+ d^2 \vec{k}_\perp}
{2k^+ 2(k^+-p^+) 2(k^+-p'^+)}
\nonumber\\
&&\times
\frac{k^-}{\biggl[p'^-
+ \frac{( \vec{k}_\perp-\vec{p'}_\perp)^2 + \Lambda^2}
{2(k^+-p'^+)}
-\frac{ \vec{k}^2_\perp + m^2_q}{2k^+}\biggr]
\biggl[p'^- - p^-
+ \frac{(\vec{k}_\perp-\vec{p'}_\perp)^2 + \Lambda^2}
{2(k^+-p'^+)}
-\frac{ (\vec{k}_\perp-\vec{p}_\perp)^2 + \Lambda^2}
{2(k^+-p^+)}\biggr]},
\nonumber\\
&=&-\frac{1} {2(2\pi)^3}
\int^{p'^+}_{p^+}dk^+ d^2 \vec{k}_\perp
p^+(p'^+)^2(k^+-p'^+)k^-
\nonumber\\
&&\times
\frac{1} {\biggl\{k^+(k^+-p'^+) (m^2 + \vec{p'}^2_\perp)
+ p'^+k^+[(\vec{k}_\perp-\vec{p'}_\perp)^2+\Lambda^2]
-p'^+(k^+-p'^+)(\vec{k}^2_\perp+m^2_q)\biggr\}}
\nonumber\\
&&\times \frac{1}
{\biggl\{ p^+(k^+-p^+)(k^+-p'^+)(m^2 +\vec{p'}^2_\perp)
- p'^+(k^+-p^+)(k^+-p'^+)(m^2 +\vec{p}^2_\perp)}\nonumber\\
&&\hspace{0.5cm}+
p^+ p'^+(k^+ - p^+)[ (\vec{k}_\perp-\vec{p'}_\perp)^2 + \Lambda^2]
- p^+ p'^+(k^+ - p'^+)[ (\vec{k}_\perp-\vec{p}_\perp)^2 + \Lambda^2]
\biggr\}
\nonumber\\
&=&-\frac{1} {2(2\pi)^3}
\int^{p'^+}_{p^+} dk^+ d^2\vec{k}_\perp\; p^+(p'^+)^2(k^+ - p'^+)
k^-
\nonumber\\
&&\times
\frac{1}{\biggl\{k^+(k^+-p'^+) (m^2+\vec{p}^2_\perp)
+ p'^+k^+[(\vec{k}_\perp-\vec{p'}_\perp)^2+\Lambda^2]
-p'^+(k^+-p'^+)(\vec{k}^2_\perp+m^2_q)\biggr\}}
\nonumber\\
&&\times\frac{1}
{\biggl\{ (-q^+)(k^+-p^+)(k^+-p'^+)(m^2+\vec{p}^2_\perp)
+ p^+ p'^+(k^+ - p^+)[ (\vec{k}_\perp-\vec{p'}_\perp)^2 + \Lambda^2]}
\nonumber\\
&&\hspace{0.5cm}
- p^+ p'^+(k^+ - p'^+)[ (\vec{k}_\perp-\vec{p}_\perp)^2 + \Lambda^2]
\biggr\},
\end{eqnarray}
where we use $\vec{p}^2_\perp=\vec{p'}^2_\perp$ and $q^+=p'^+ - p^+$
in the derivation of the last term in Eq.~(\ref{B1}).
Writing $q^+  = \delta p^+$ and $p'^+ = (1+\delta) p^+$, $\delta \to 0$
at the end, we obtain
\begin{eqnarray}\label{B2}
G^{+{\rm z.m.}}_{h'h}
&=&-\frac{1} {2(2\pi)^3} {\rm lim}_{\delta\to0}
\int^{1+\delta}_{1}dx\; d^2\vec{k}_\perp
(x-1-\delta)(1+\delta)^2
\biggl[\frac{m^2 +\vec{p'}^2_\perp}{(1+\delta)p^+} +
\frac{( \vec{k}_\perp-\vec{p'}_\perp)^2+\Lambda^2}{(x-1-\delta)p^+}
\biggr]
\nonumber\\
&&\times\frac{1}
{\biggl\{ x(x-1-\delta)(m^2 +\vec{p}^2_\perp)
+ (1+\delta)x[(\vec{k}_\perp -\vec{p'}_\perp)^2 + \Lambda^2]
- (1+\delta)(x-1-\delta)(\vec{k}^2_\perp + m^2_q)\biggr\}}
\nonumber\\
&&\times\frac{1}
{\biggl\{ -\delta(x-1)(x-1-\delta)(m^2+\vec{p}^2_\perp)
+ (1+\delta)(x-1)[(\vec{k}_\perp -\vec{p'}_\perp)^2 + \Lambda^2]}
\nonumber\\
&&\hspace{0.5cm} - (1+\delta)(x-1-\delta)
  [(\vec{k}_\perp-\vec{p}_\perp)^2 + \Lambda^2]\biggr\}.
\end{eqnarray}

If we write $x= 1+\delta\, y$ and $dx=\delta\,dy$, then the integral over
$y$ runs from 0 to 1 as $x$ runs from 1 to $1+\delta$. Therefore, we get
\begin{eqnarray}\label{B3}
G^{+{\rm z.m.}}_{h'h}
&=& -\frac{1} {2p^+(2\pi)^3}
\int^{1}_{0}dy\; d^2\vec{k}_\perp
\frac{1} { \biggl\{ y[(\vec{k}_\perp -\vec{p'}_\perp)^2 + \Lambda^2]
+(1-y)[(\vec{k}_\perp-\vec{p}_\perp)^2 + \Lambda^2]\biggr\} }
\nonumber\\
&=&\frac{1} {2p^+(2\pi)^3}
\int\; d^2\vec{k}_\perp
\frac{ {\rm ln}\biggl[
\frac{(\vec{k}_\perp-\vec{p'}_\perp)^2+\Lambda^2}
{(\vec{k}_\perp -\vec{p}_\perp)^2 + \Lambda^2}\biggr] }
{ [(\vec{k}_\perp -\vec{p}_\perp)^2 + \Lambda^2]
-[(\vec{k}_\perp-\vec{p'}_\perp)^2 + \Lambda^2] }.
\end{eqnarray}

\newpage
\begin{figure}[t]
\vspace{-2cm}
\psfig{figure=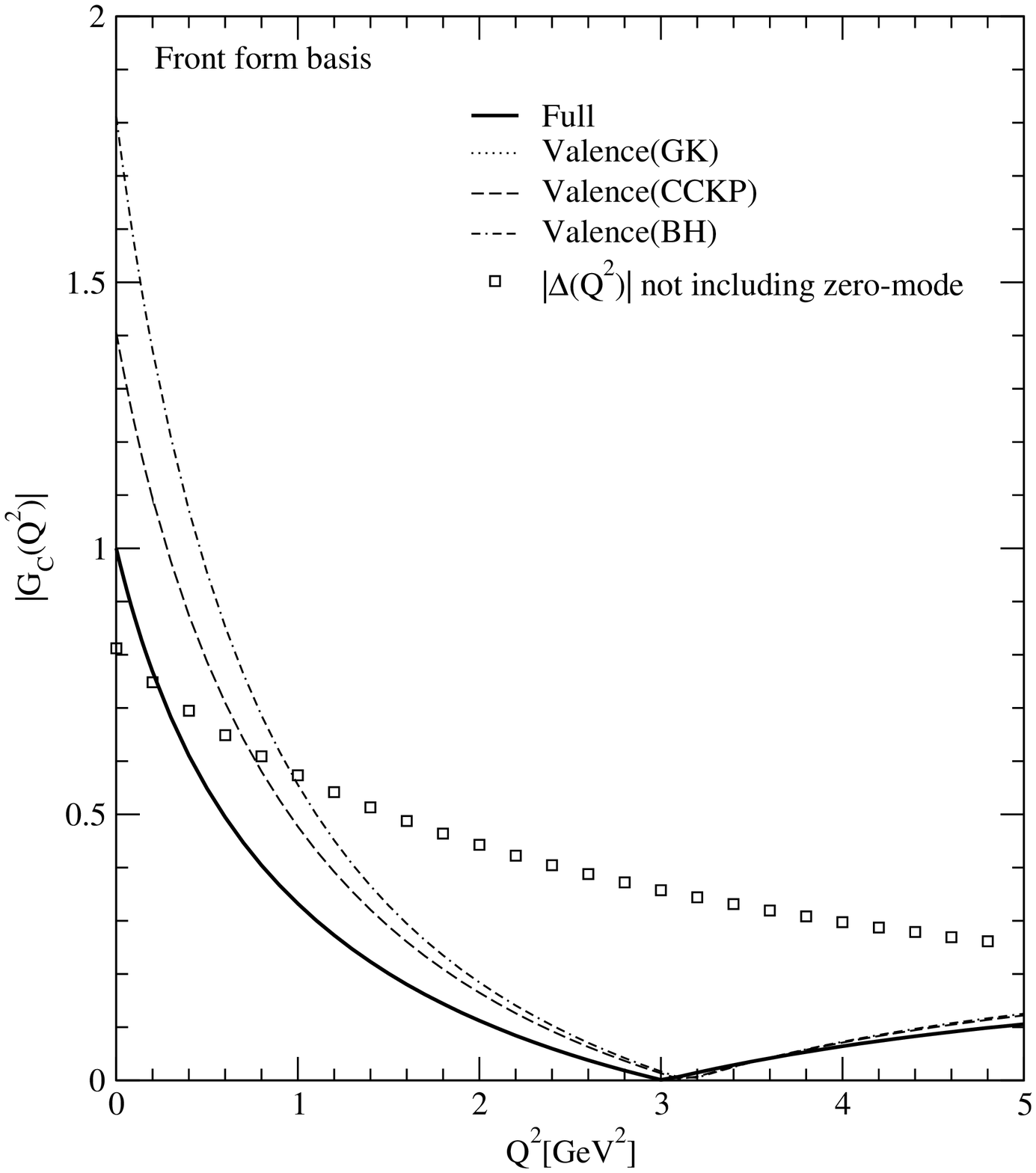,height=10cm,width=8cm}
\psfig{figure=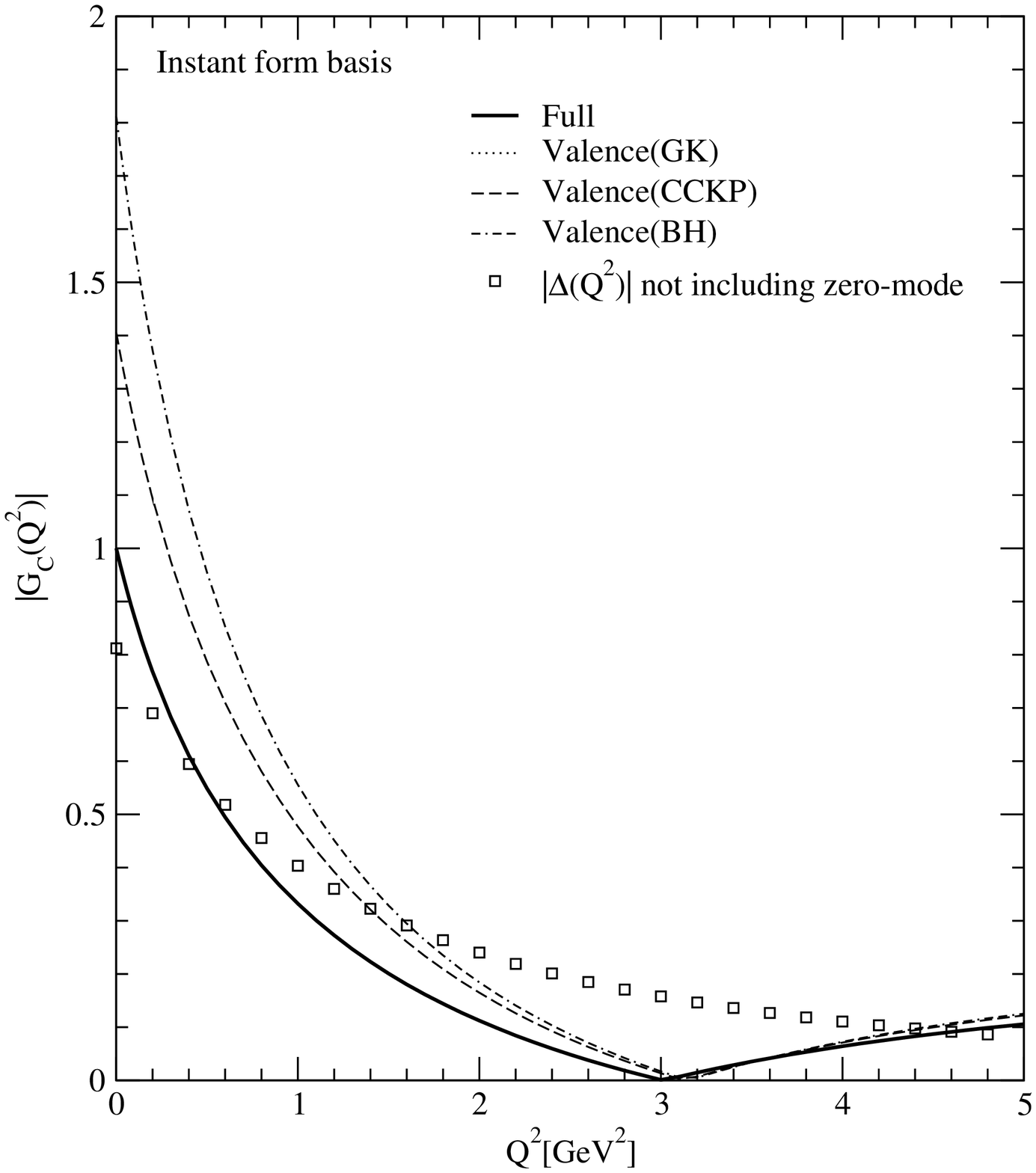,height=10cm,width=8cm}
\caption{The charge form factor $|G_C(Q^2)|$ obtained from the light-front
(left) and the instant-form (right) spin bases: The thick solid line
represents the full (i.e. valence+zero-mode in LF = covariant)
solution. The dotted, long-dashed, and dot-dashed
lines represent the valence contributions only, where we use the same
normalization as for the full solution $G_C(0)=1$.
The small squares represent the angular condition in
Eq.~(\protect\ref{eq.16}) without including
the zero-mode contribution.\label{Fig_GC}}
\end{figure}
\begin{figure}[t]
\psfig{figure=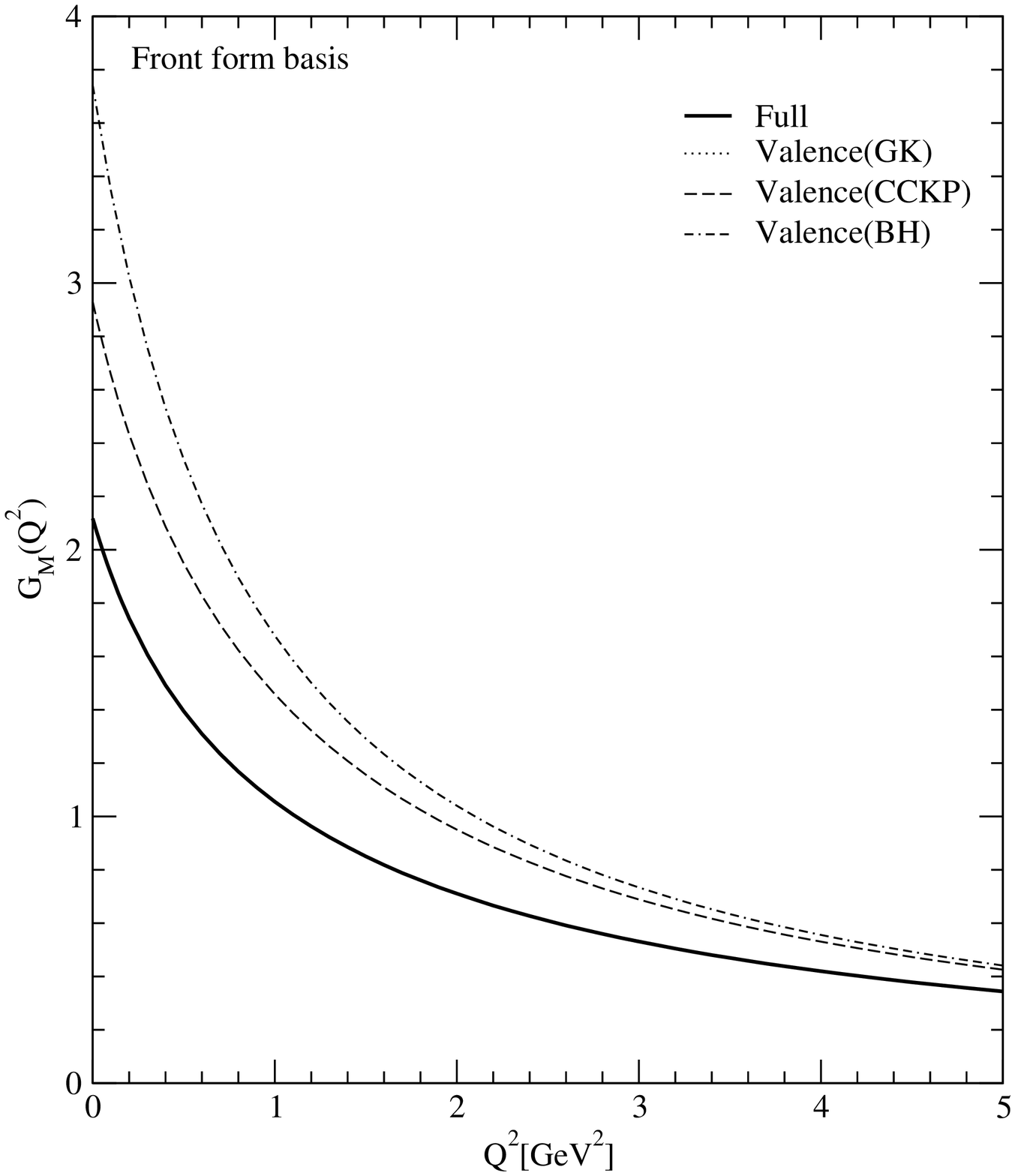,height=10cm,width=8cm}
\psfig{figure=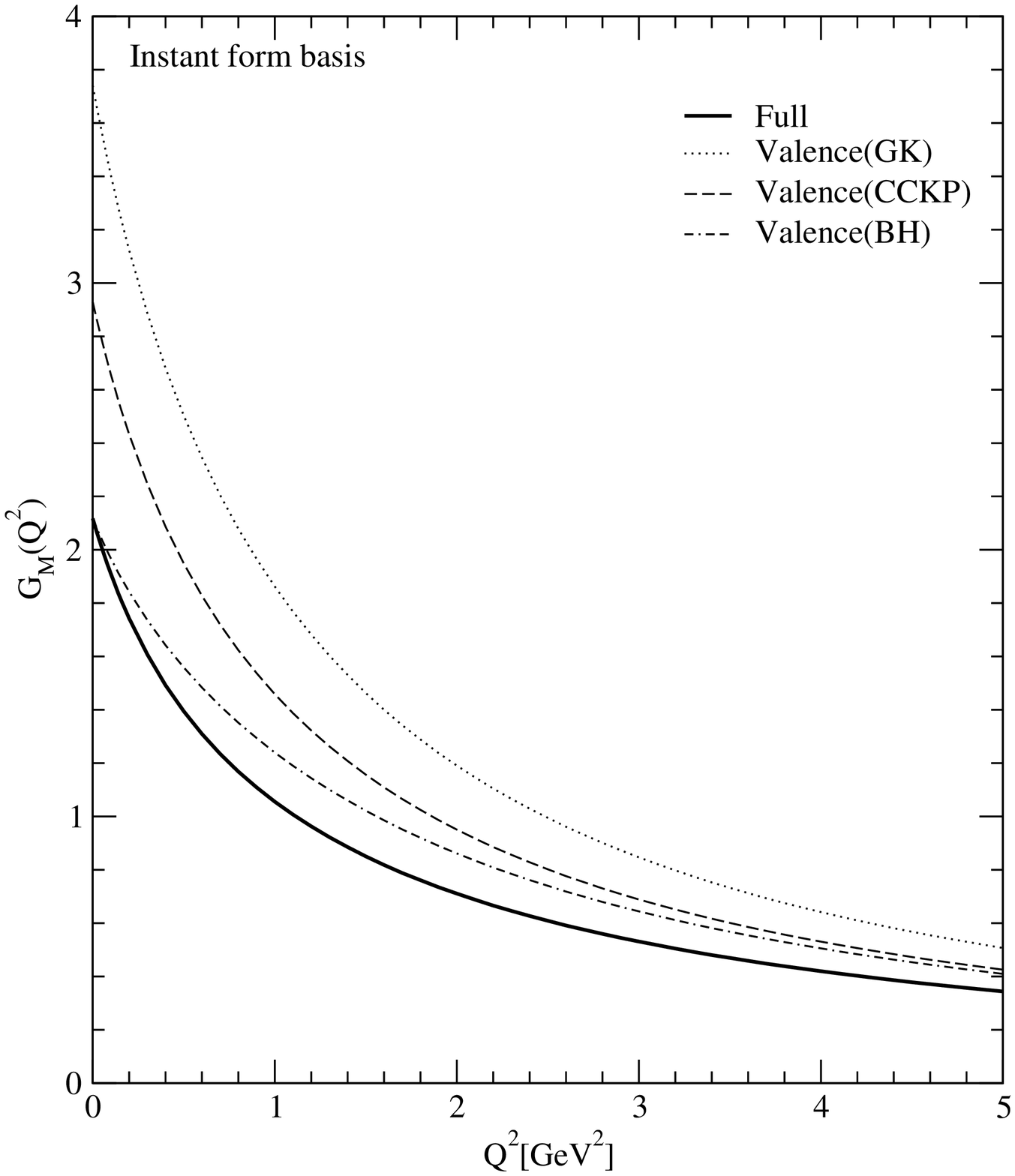,height=10cm,width=8cm}
\caption{The magnetic form factor $G_M(Q^2)$ obtained from the light-front
(left) and the instant form (right) spin bases. The same lines are used as
in Fig~\protect\ref{Fig_GC}.
 \label{Fig_GM}}
\end{figure}
\begin{figure}[t]
\psfig{figure=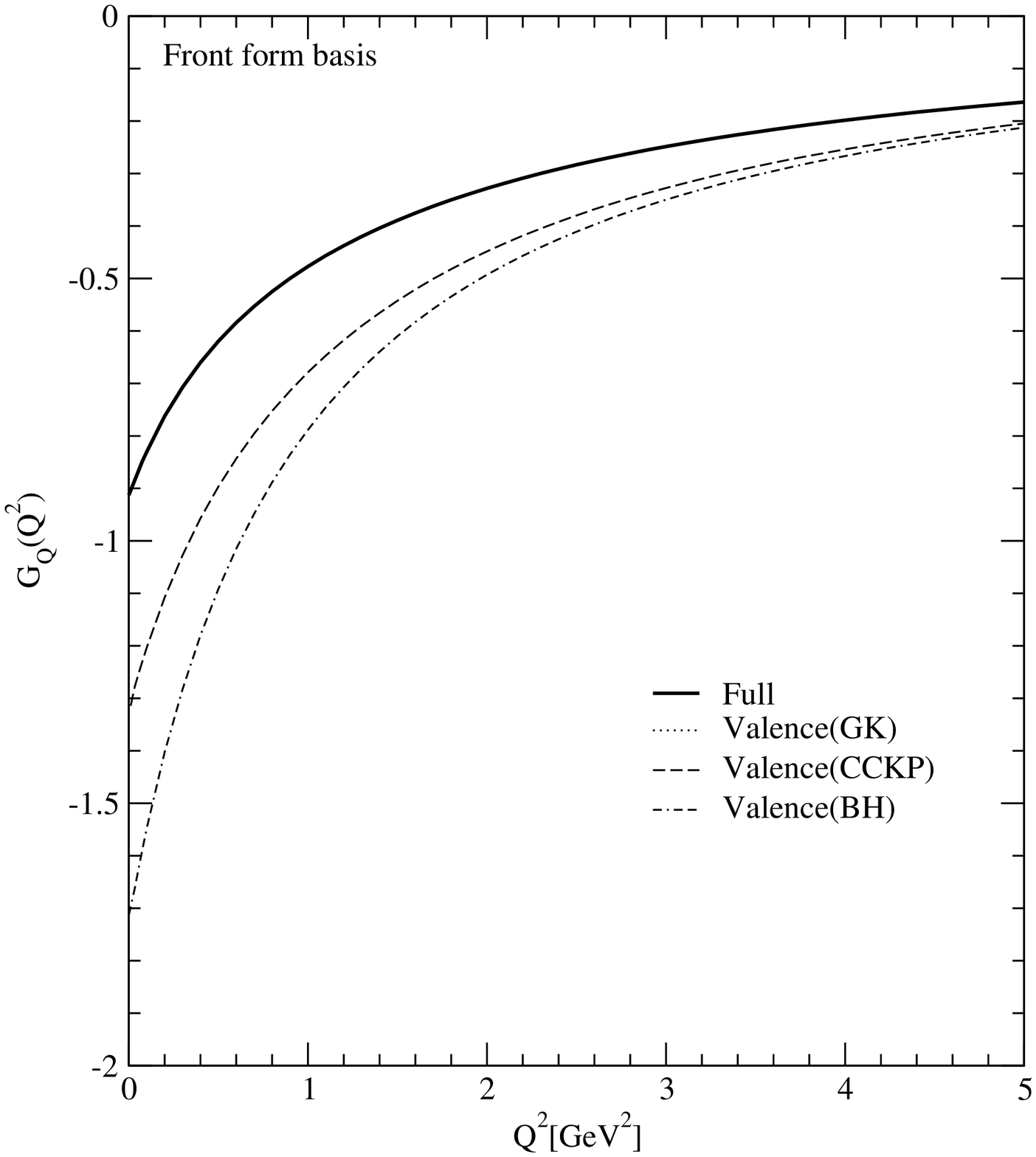,height=10cm,width=8cm}
\psfig{figure=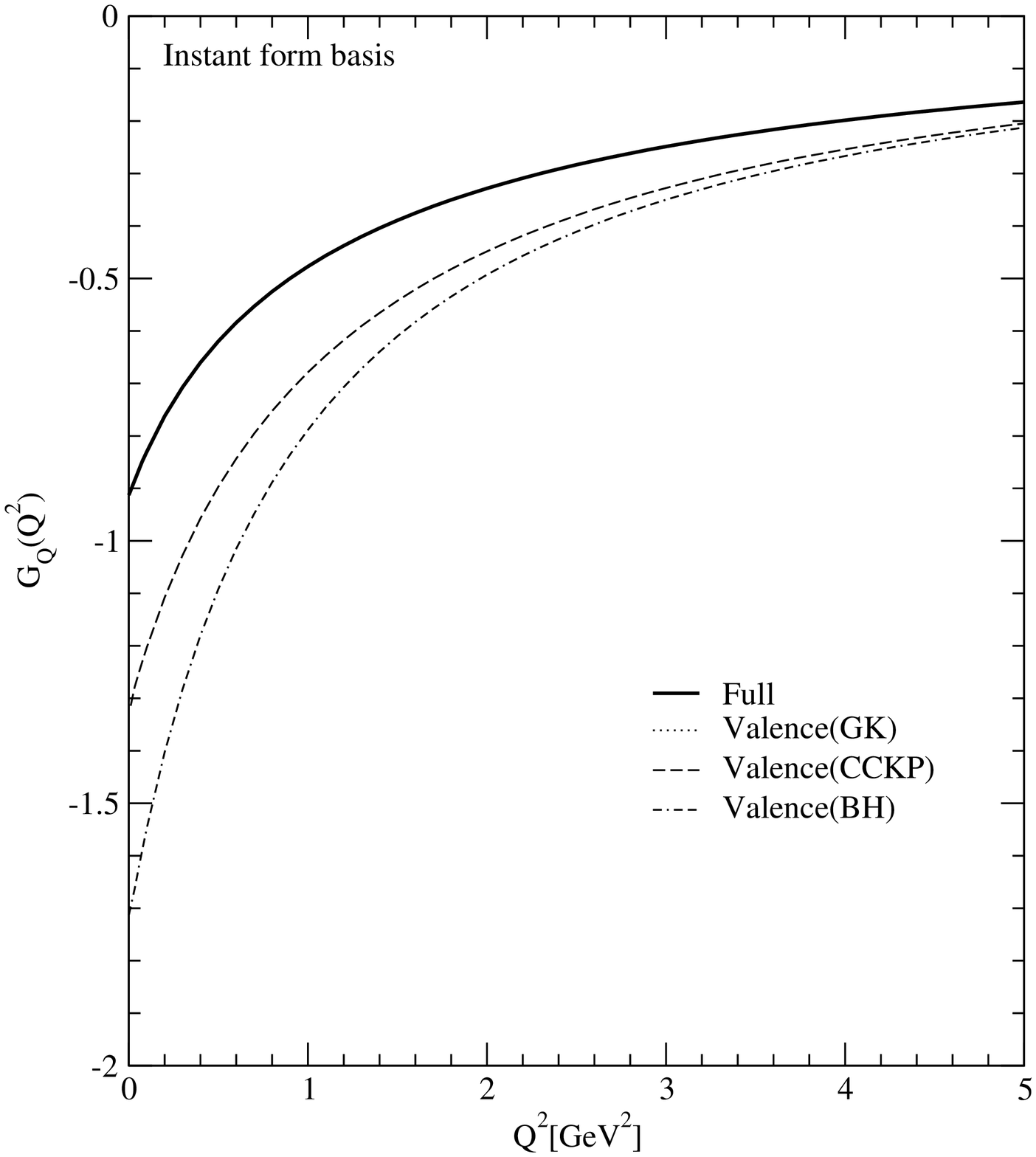,height=10cm,width=8cm}
\caption{The quadrupole form factor $G_Q(Q^2)$ obtained from the light-front
(left) and the instant form (right) spin bases. The same lines are used as
in Fig~\protect\ref{Fig_GC}.
 \label{Fig_GQ}}
\end{figure}
\begin{figure}[t]
\psfig{figure=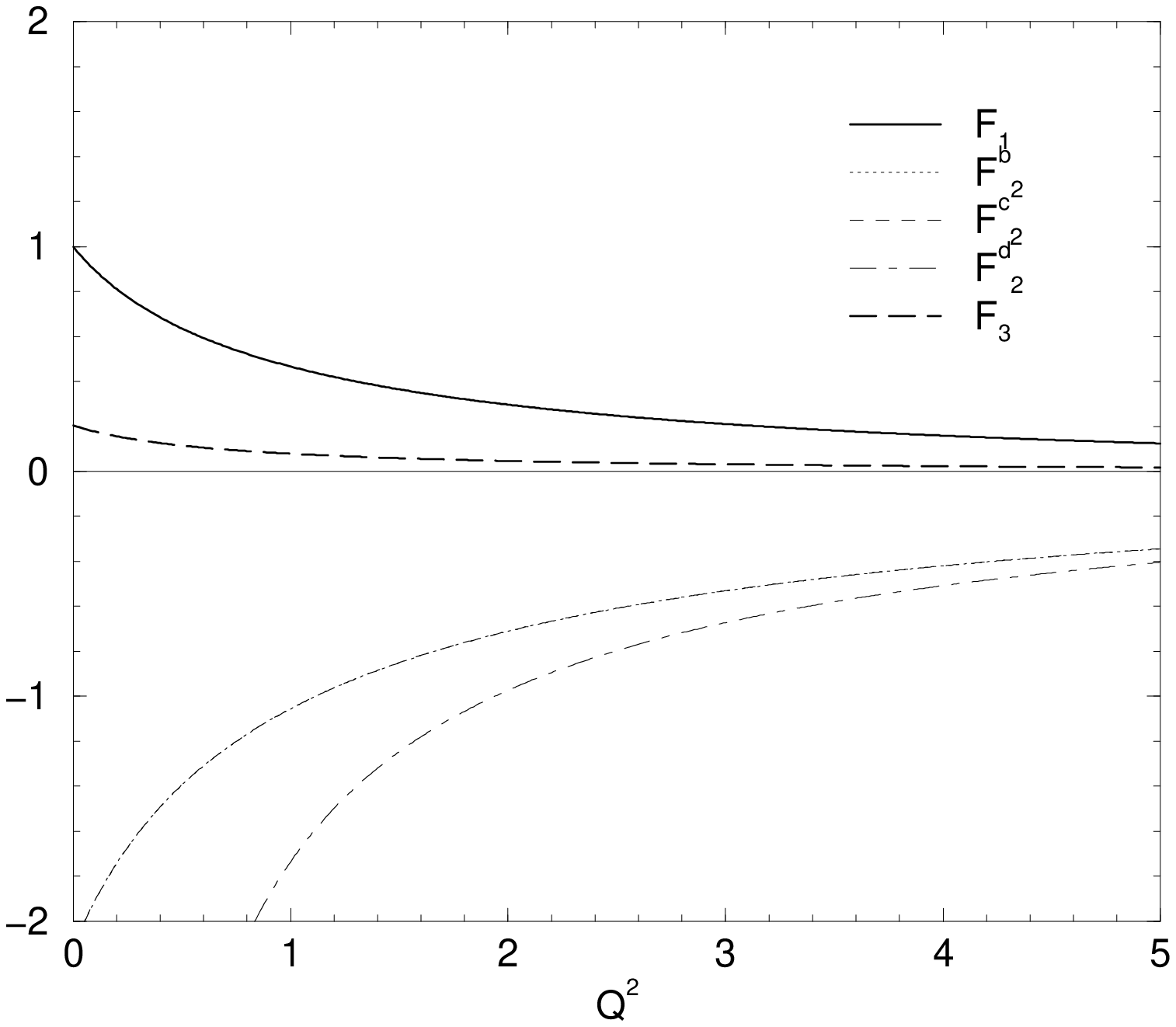,height=10cm,width=8cm}
\psfig{figure=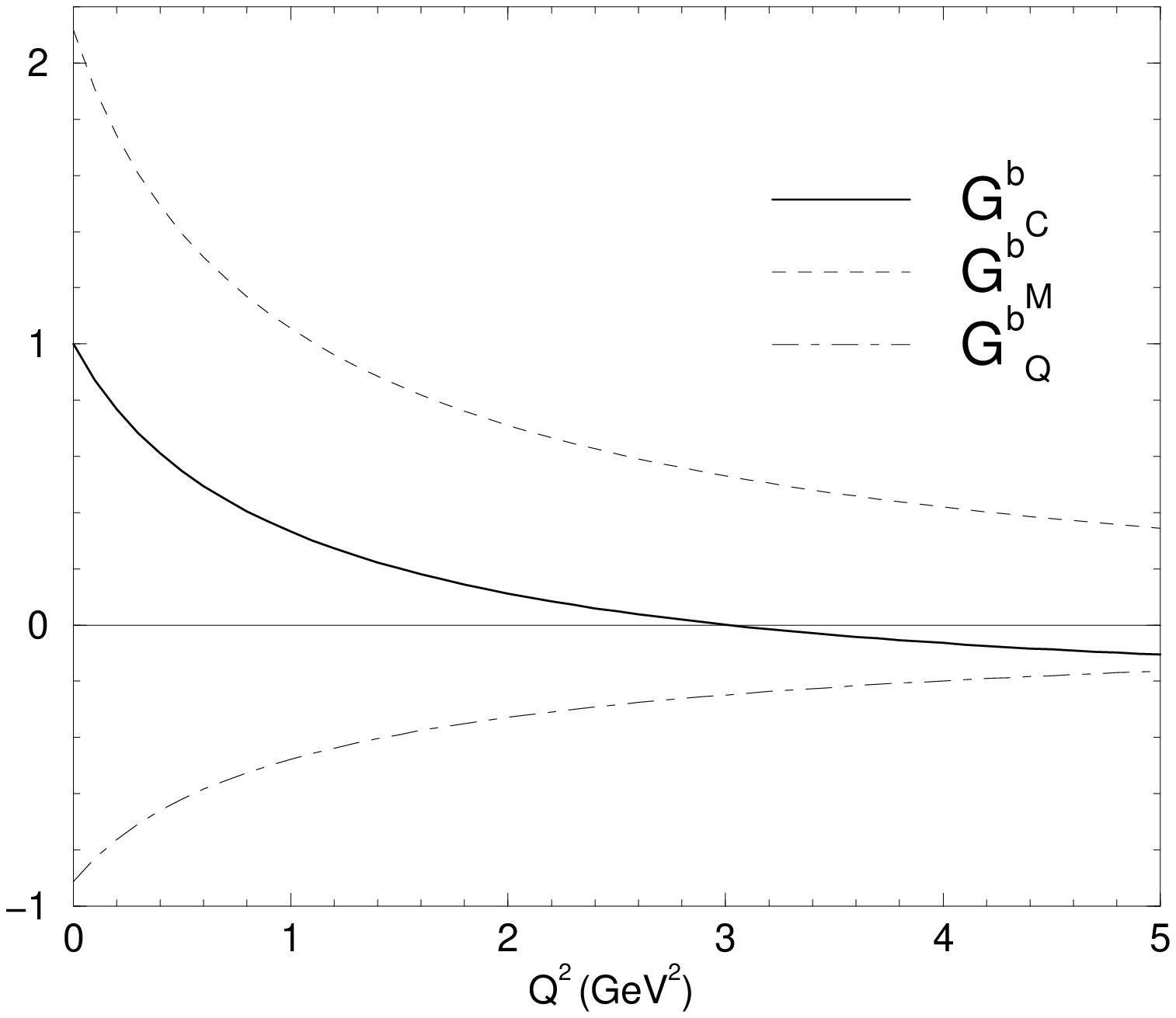,height=10cm,width=8cm}
\caption{Invariant form factors $F_1$, $F^b_2$, $F^c_2$, $F^d_2$, and
$F_3$ (left) and physical form factors $G_C$, $G_M$, and  $G_Q$ from
$F^b_2$ (right) calculated in the Drell-Yan-West frame. Valence parts only.
 \label{fig.05}}
\end{figure}
\begin{figure}[t]
\psfig{figure=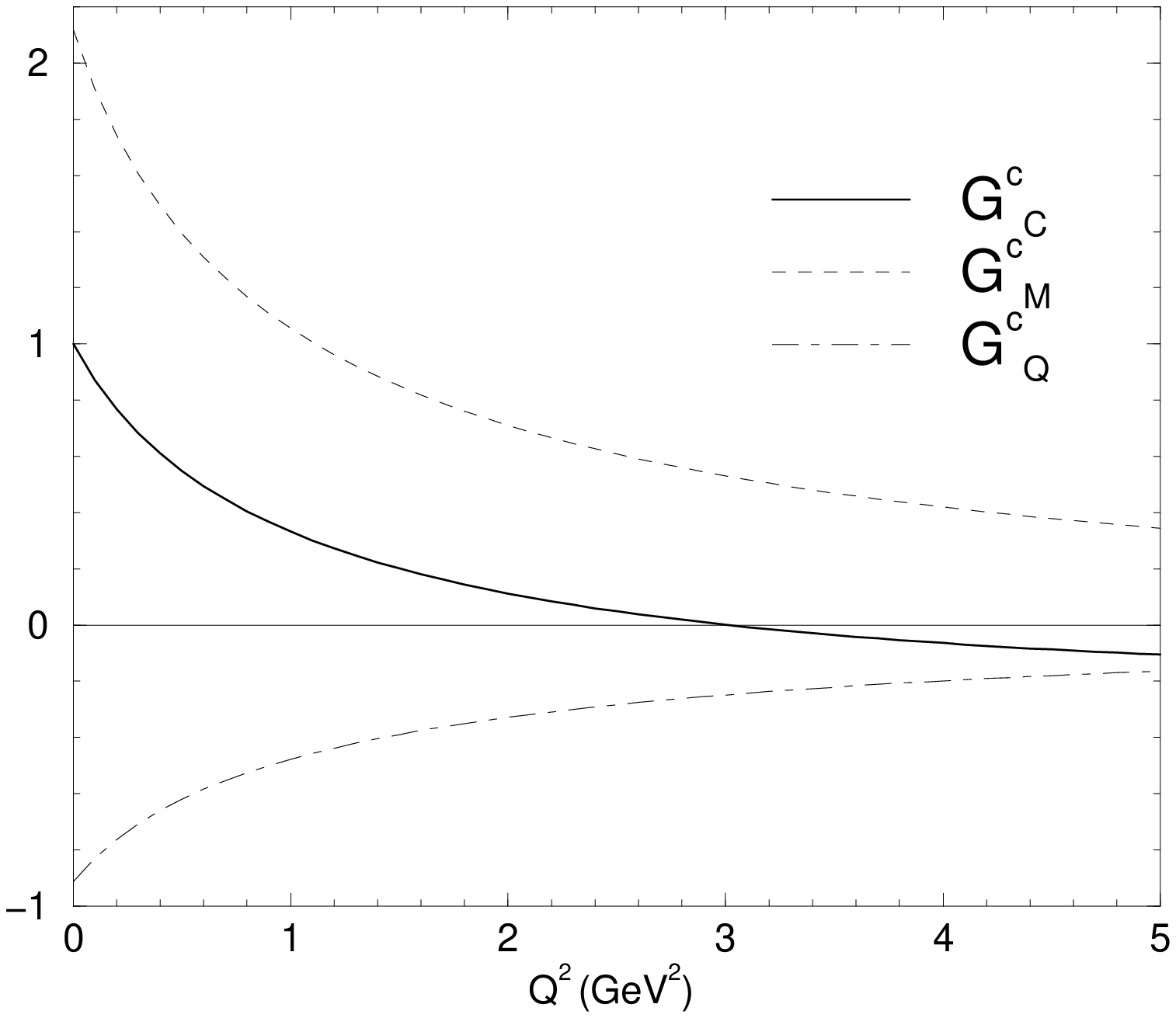,height=10cm,width=8cm}
\psfig{figure=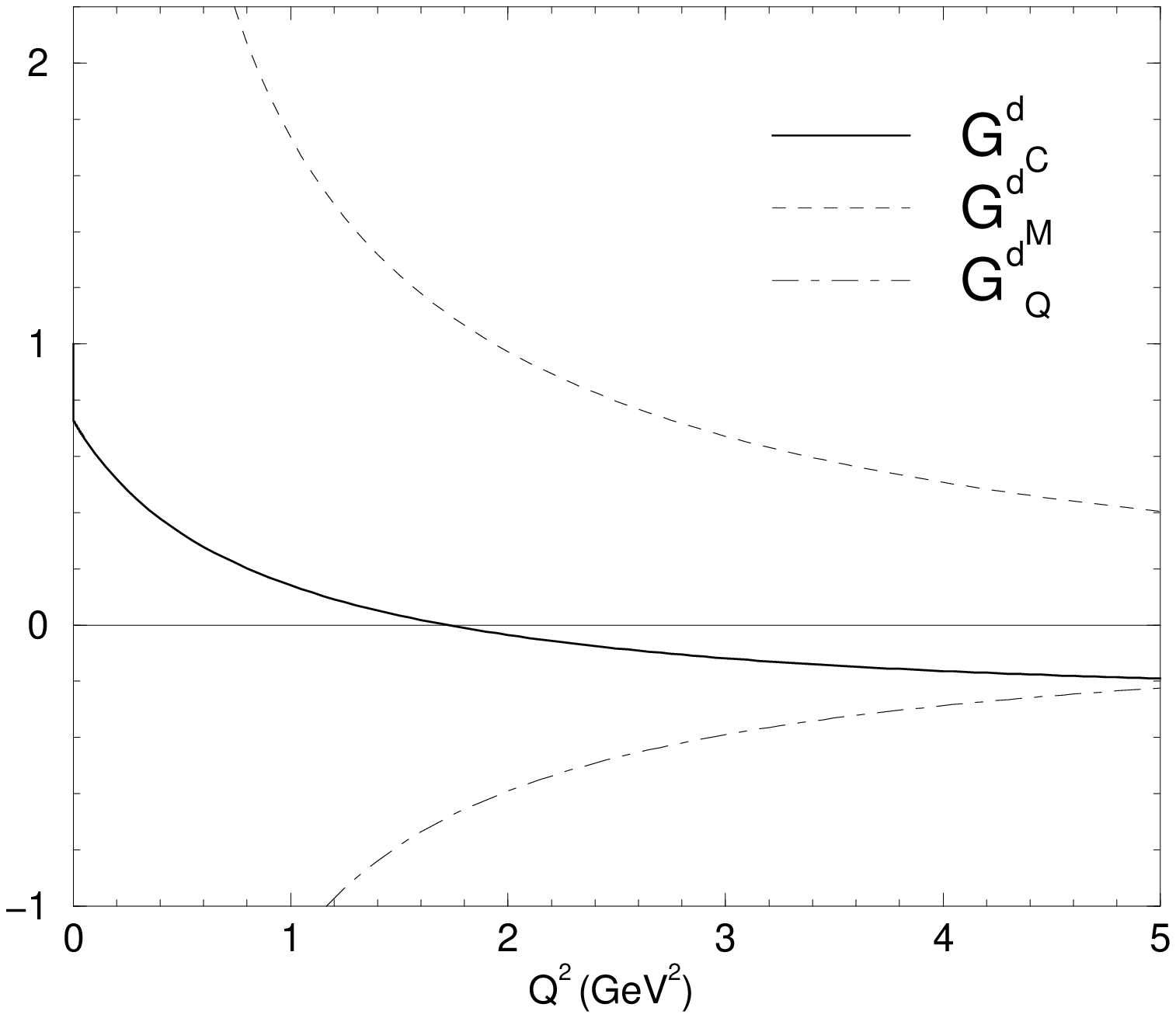,height=10cm,width=8cm}
\caption{Physical form factors $G_C$, $G_M$, and  $G_Q$ from $F^c_2$
(left) and $F^d_2$ (right) calculated in the Drell-Yan-West frame.
Valence parts only.
\label{fig.06}}
\end{figure}
\begin{figure}[t]
 \psfig{figure=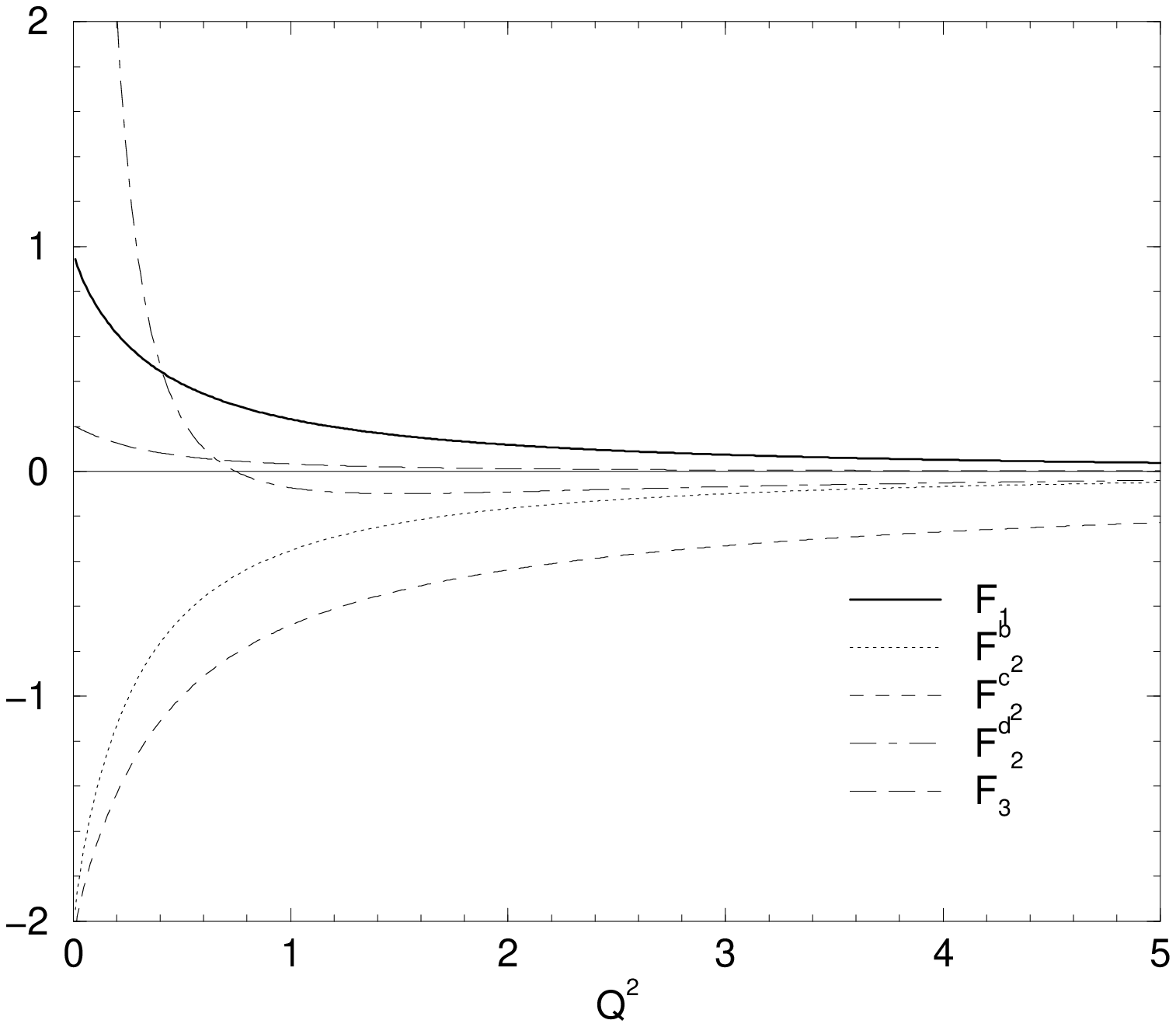,height=10cm,width=8cm}
 \psfig{figure=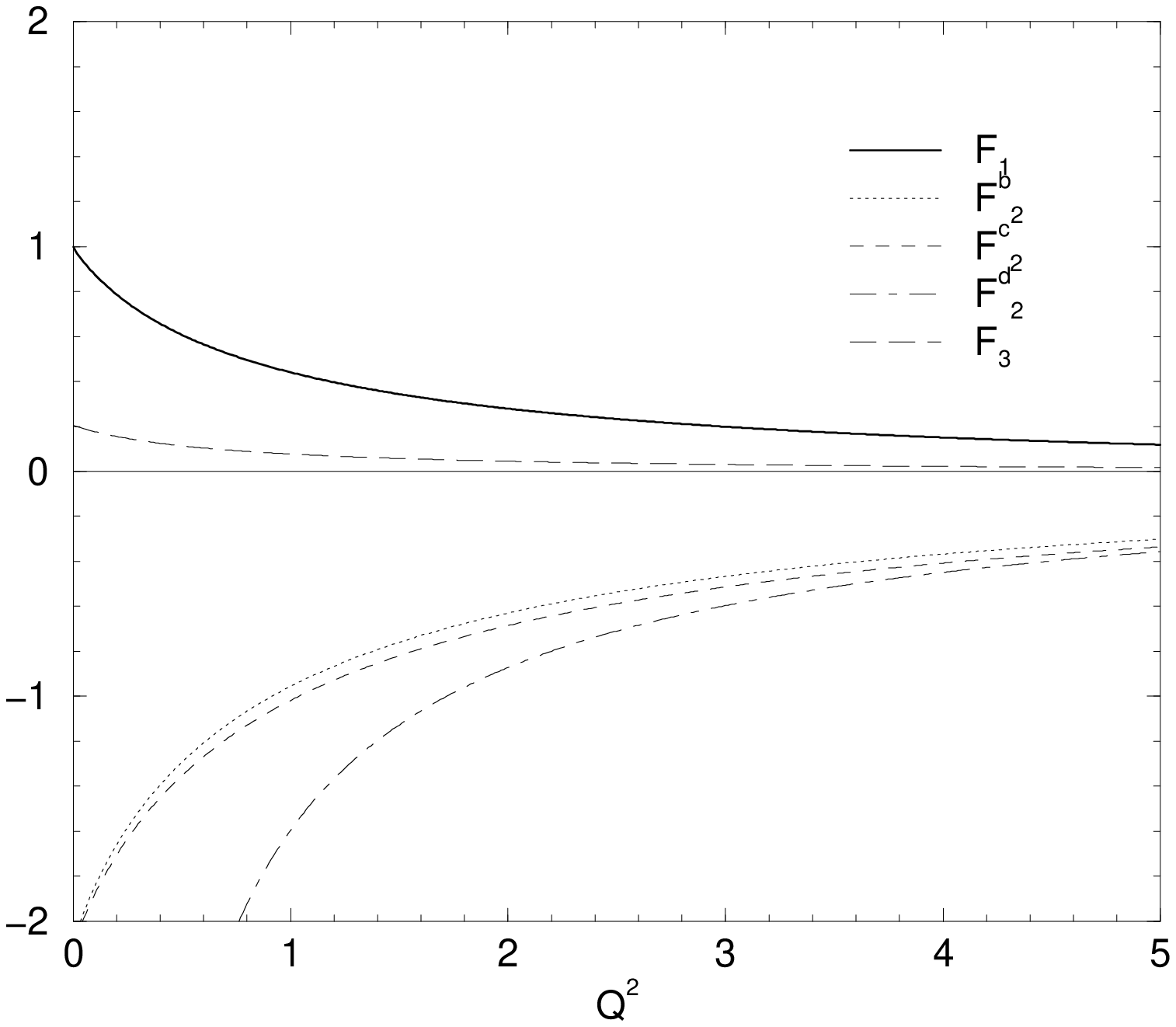,height=10cm,width=8cm}
 \caption{Invariant form factors $F_1$, $F^b_2$, $F^c_2$, $F^d_2$, and
 $F_3$ for $\theta = \pi/20$ (left) and $\theta = 9\pi/20$ (right)
 calculated in the Breit frame. Valence parts only.
 \label{fig.07}}
\end{figure}
\begin{figure}[t]
 \psfig{figure=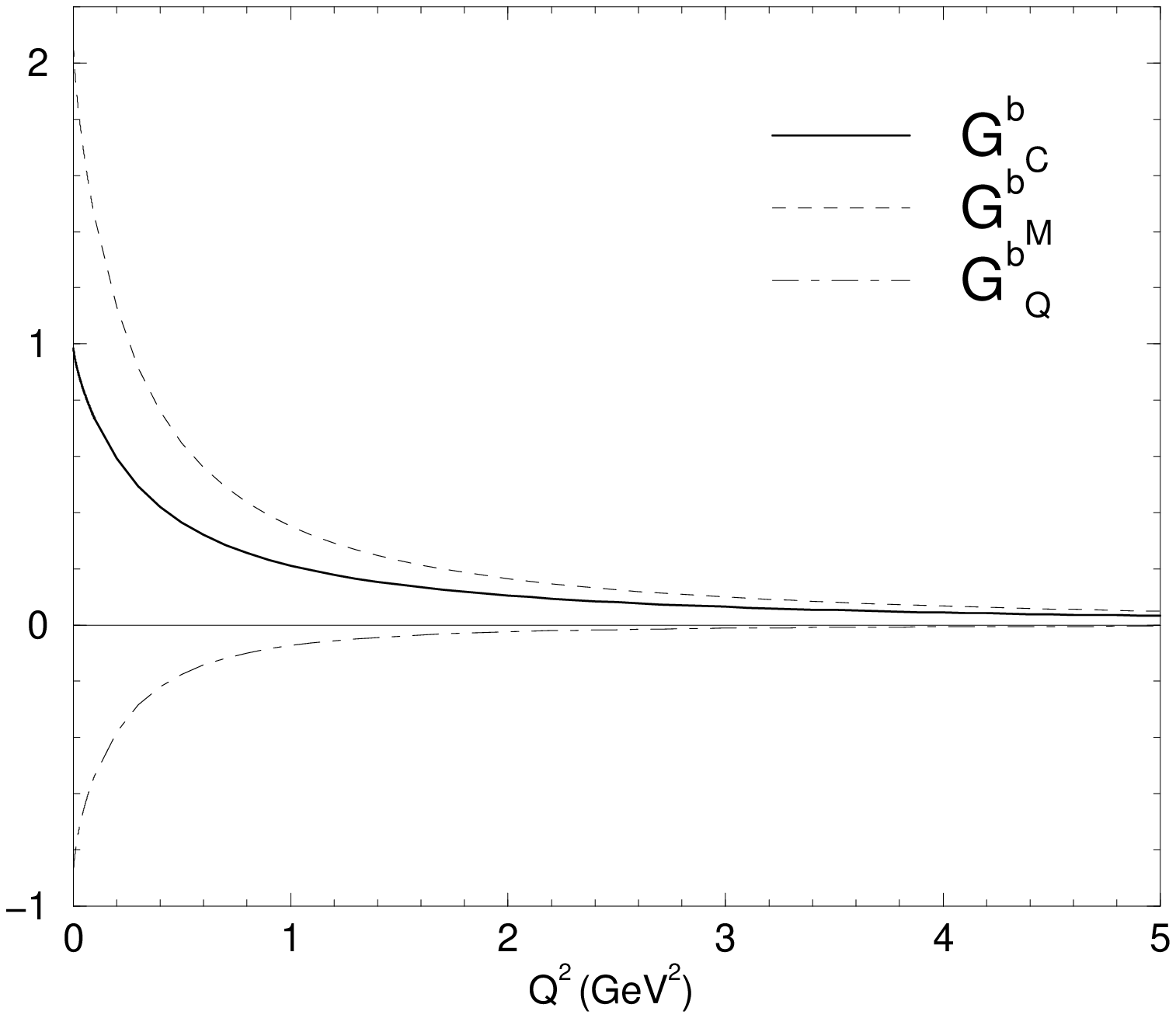,height=10cm,width=8cm}
 \psfig{figure=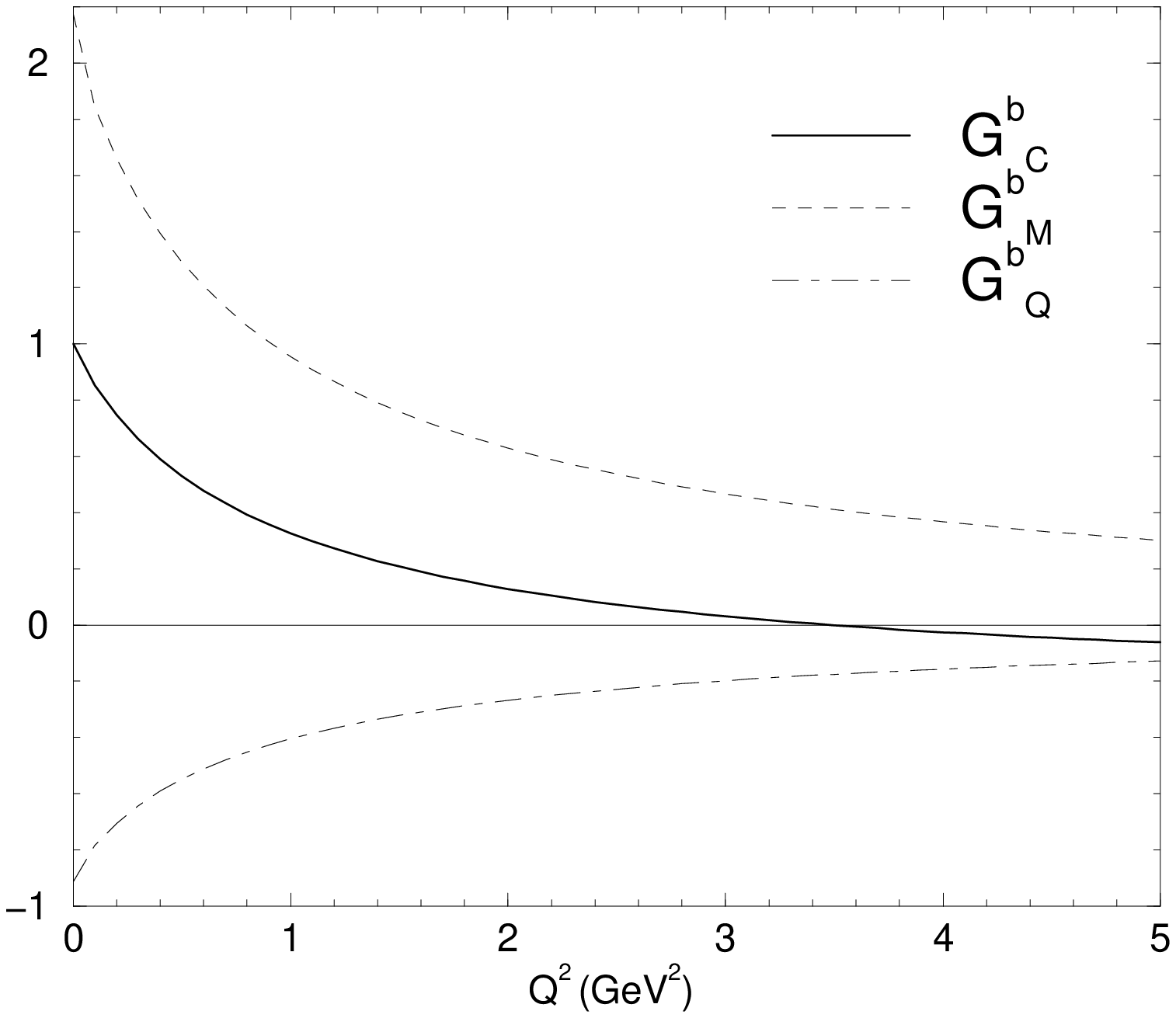,height=10cm,width=8cm}
 \caption{Physical form factors $G_C$, $G_M$, and  $G_Q$ from $F^b_2$
 calculated in the Breit frame.
 Left $\theta = \pi/20$, right $\theta = 9\pi/20$. Valence parts only.
 \label{fig.08}}
\end{figure}
\begin{figure}[t]
 \psfig{figure=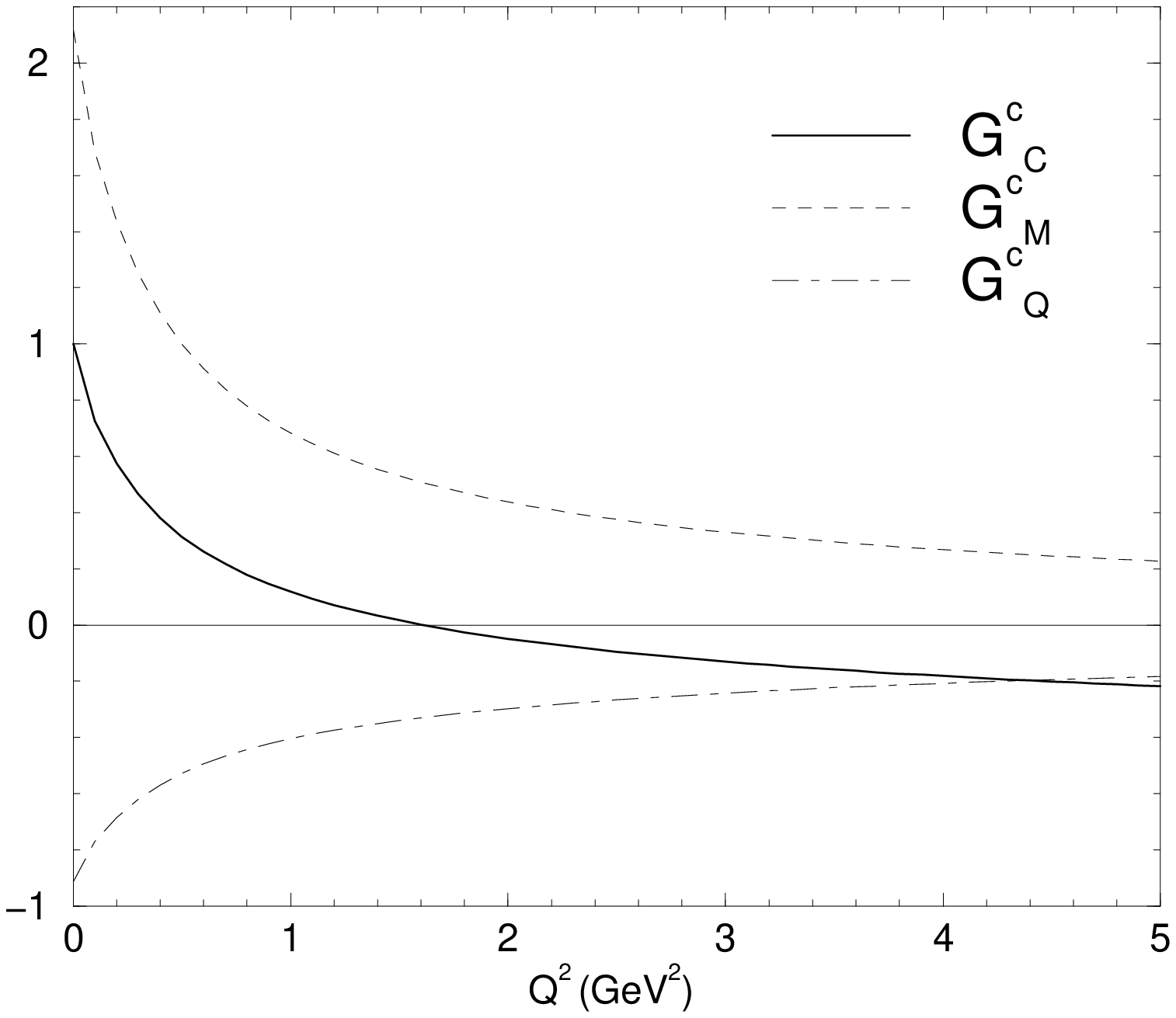,height=10cm,width=8cm}
 \psfig{figure=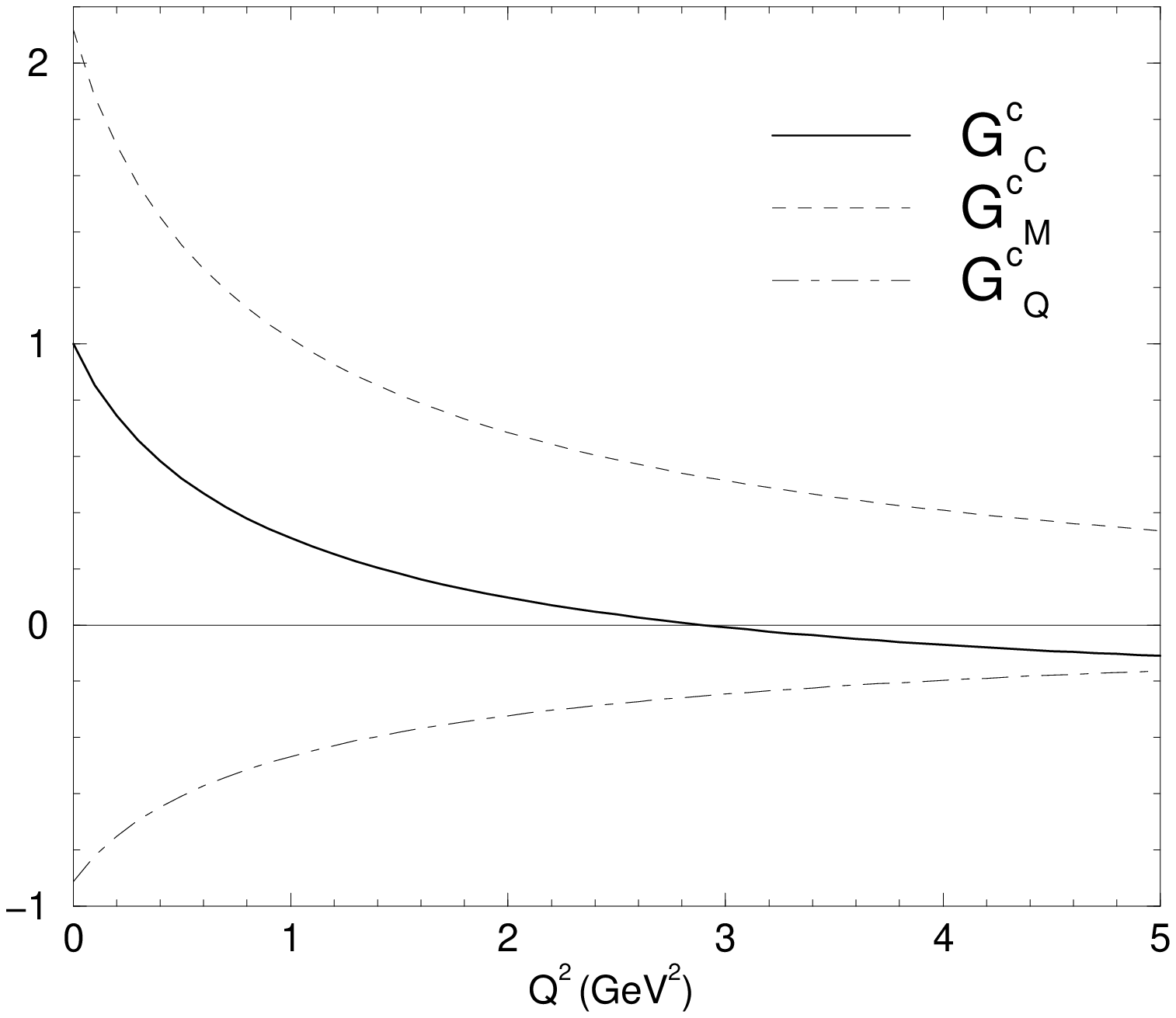,height=10cm,width=8cm}
 \caption{Physical form factors $G_C$, $G_M$, and  $G_Q$ from $F^c_2$
 calculated in the Breit frame.
 Left $\theta = \pi/20$, right $\theta = 9\pi/20$. Valence parts only.
 \label{fig.09}}
\end{figure}
\begin{figure}[t]
 \psfig{figure=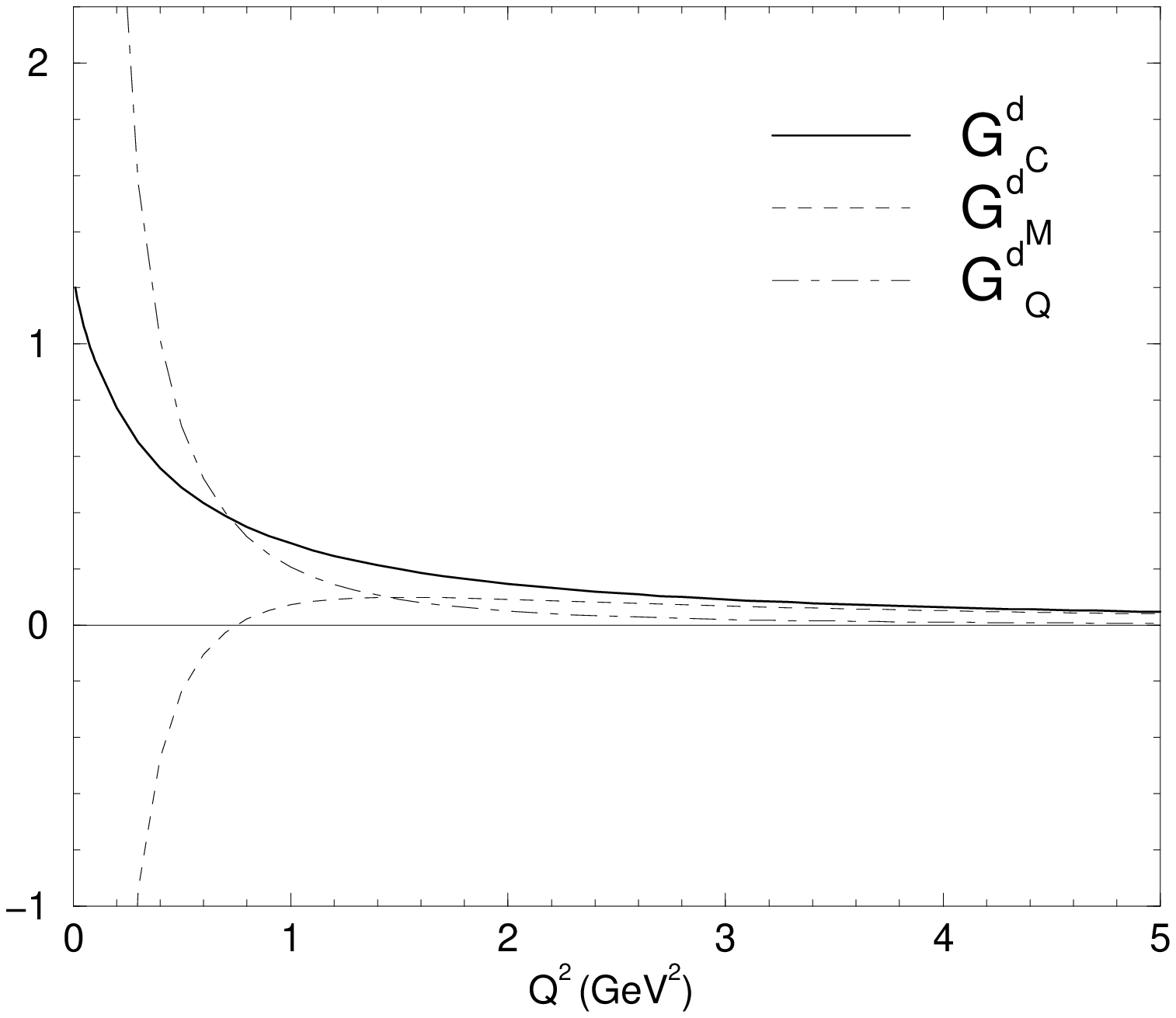,height=10cm,width=8cm}
 \psfig{figure=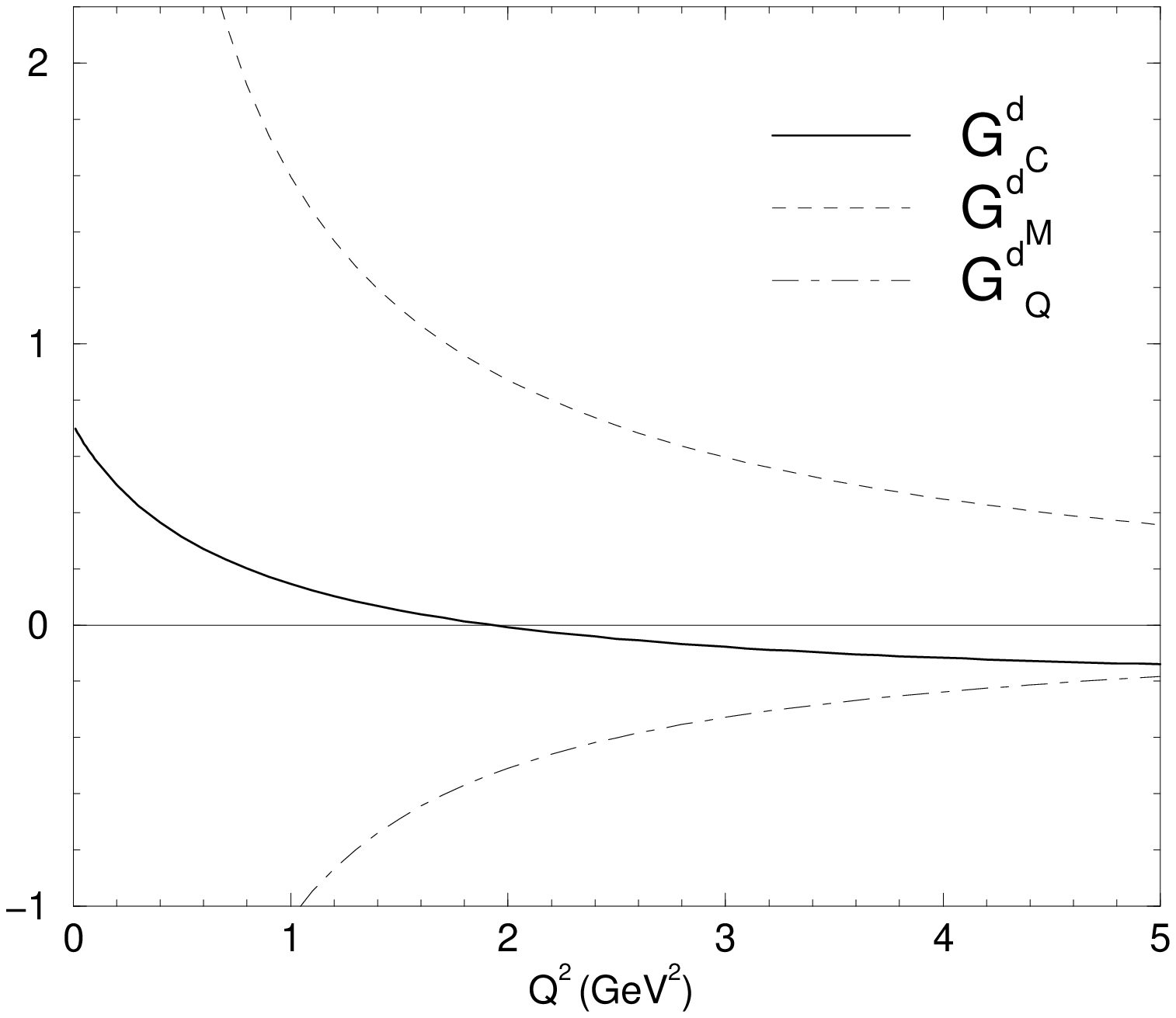,height=10cm,width=8cm}
 \caption{Physical form factors $G_C$, $G_M$, and  $G_Q$ from $F^d_2$
 calculated in the Breit frame.
 Left $\theta = \pi/20$, right $\theta = 9\pi/20$. Valence parts only.
 \label{fig.10}}
\end{figure}
\begin{figure}[t]
 \psfig{figure=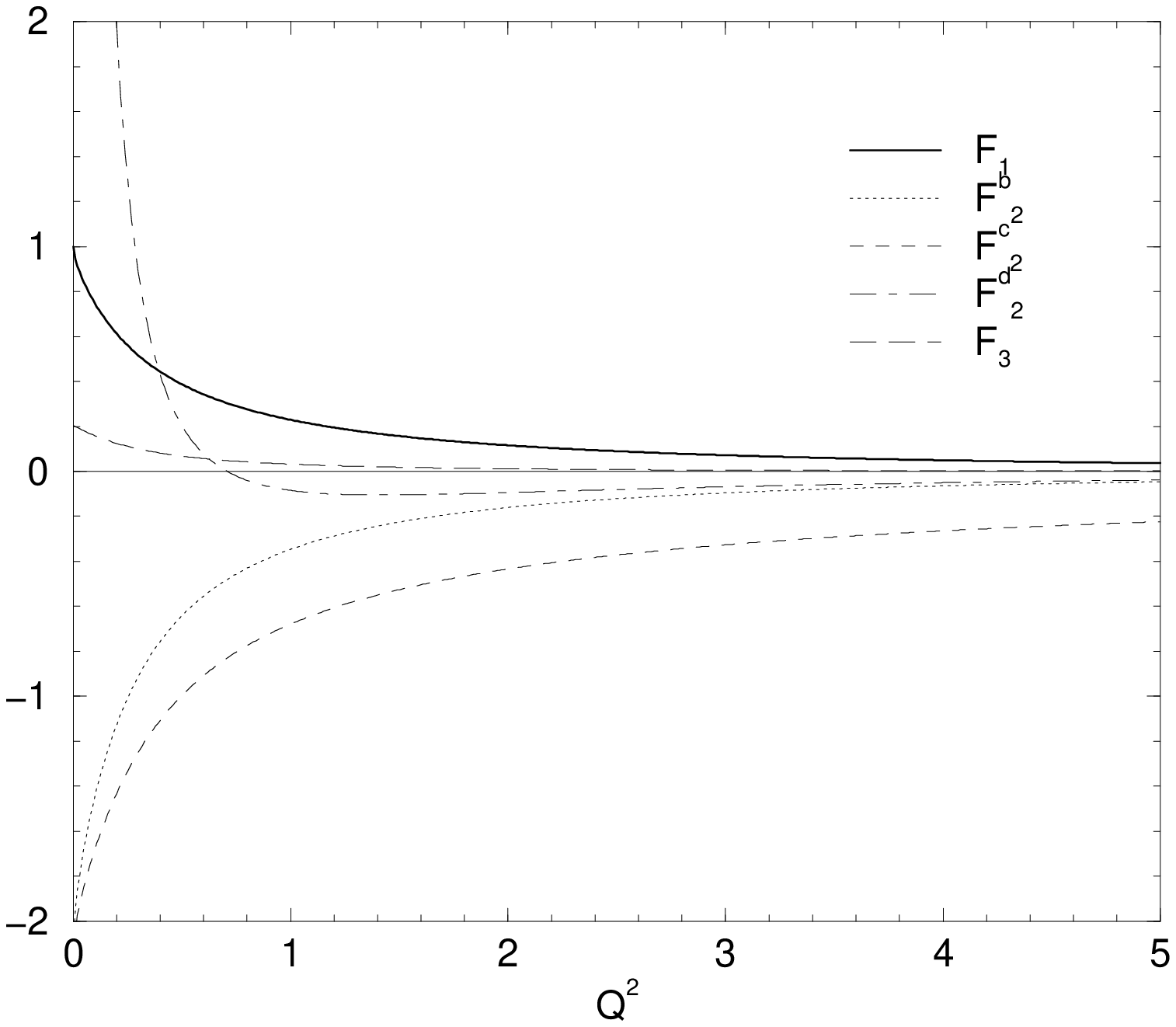,height=10cm,width=8cm}
 \psfig{figure=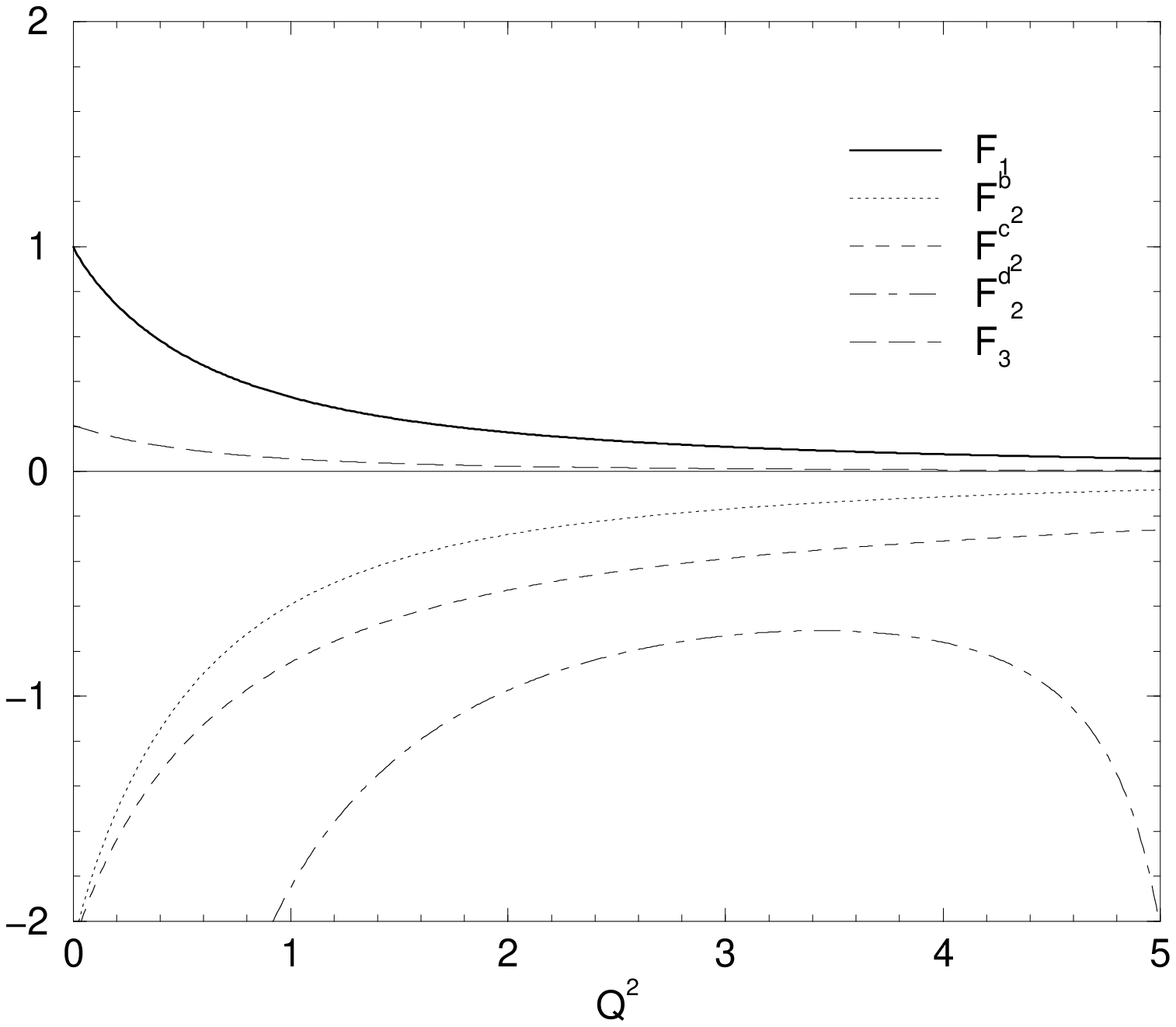,height=10cm,width=8cm}
 \caption{Invariant form factors $F_1$, $F^b_2$, $F^c_2$, $F^d_2$, and
 $F_3$ for $\theta = \pi/20$ (left) and $\theta = 9\pi/20$ (right)
 calculated in the target-rest-frame. Valence parts only.
 \label{fig.11}}
\end{figure}

\begin{figure}[t]
 \psfig{figure=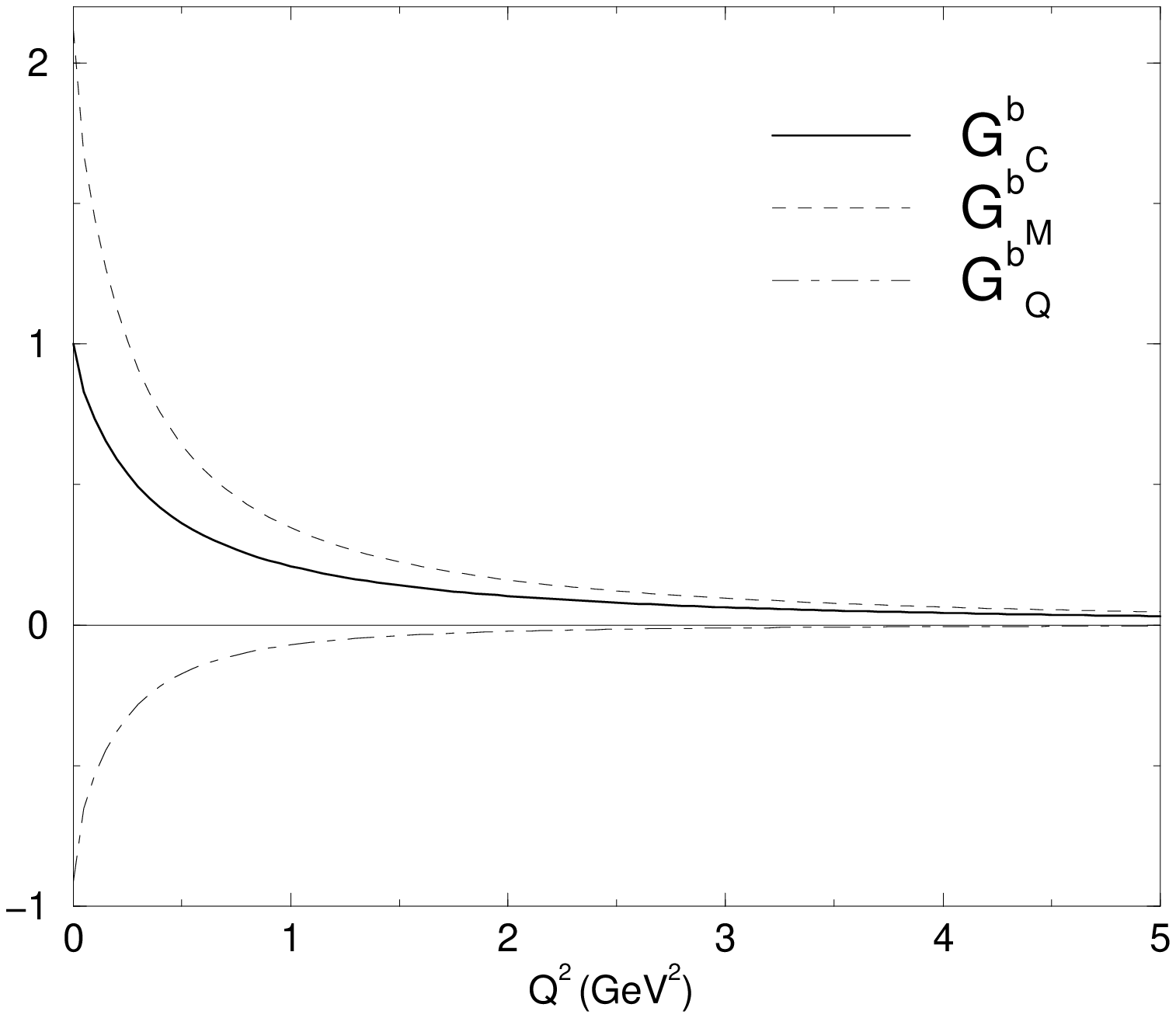,height=10cm,width=8cm}
 \psfig{figure=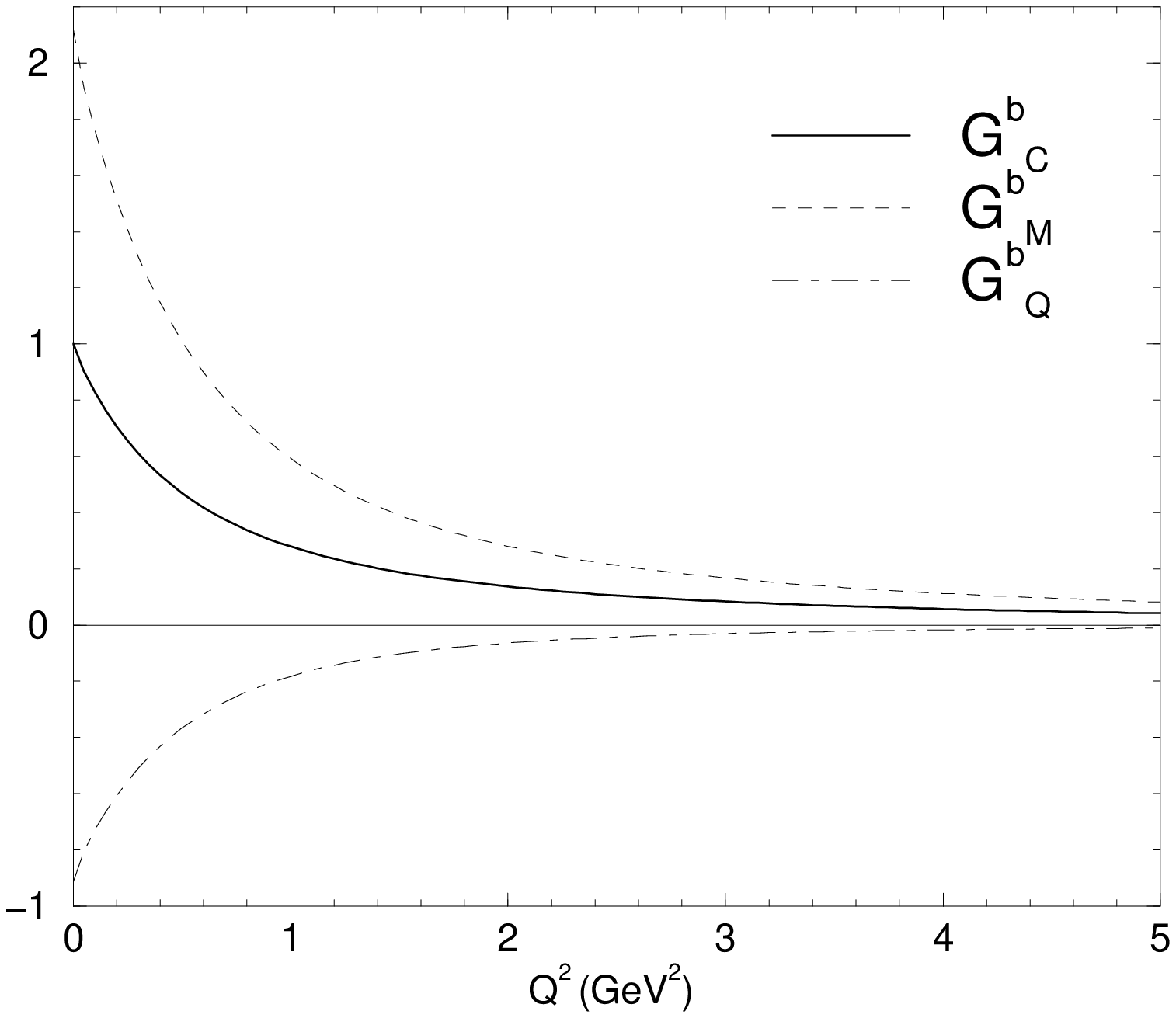,height=10cm,width=8cm}
 \caption{Physical form factors $G_C$, $G_M$, and  $G_Q$ from $F^b_2$
 calculated in the target-rest-frame. Version $b$.
 Left $\theta = \pi/20$, right $\theta = 9\pi/20$. Valence parts only.
 \label{fig.12}}
\end{figure}

\begin{figure}[t]
 \psfig{figure=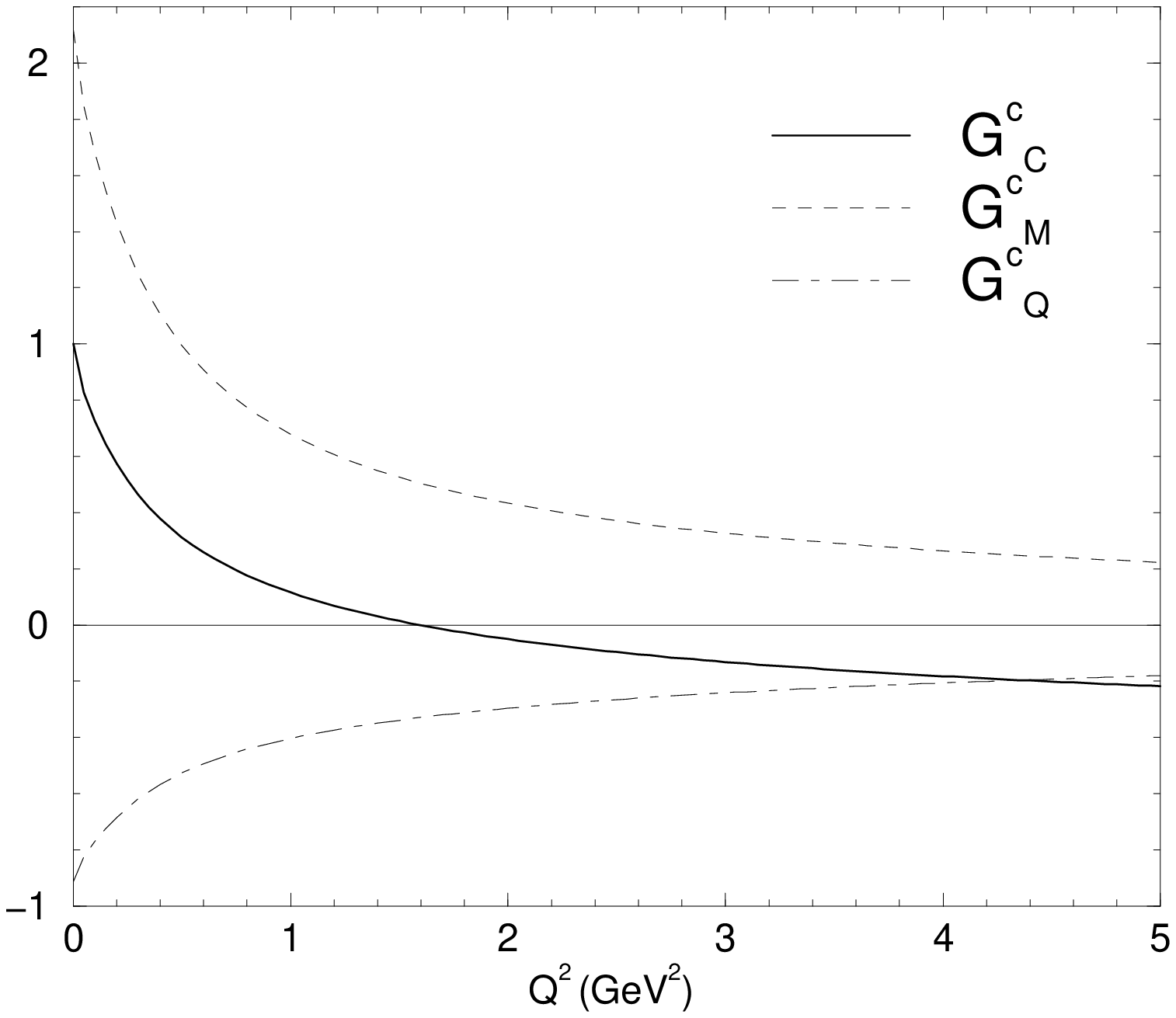,height=10cm,width=8cm}
 \psfig{figure=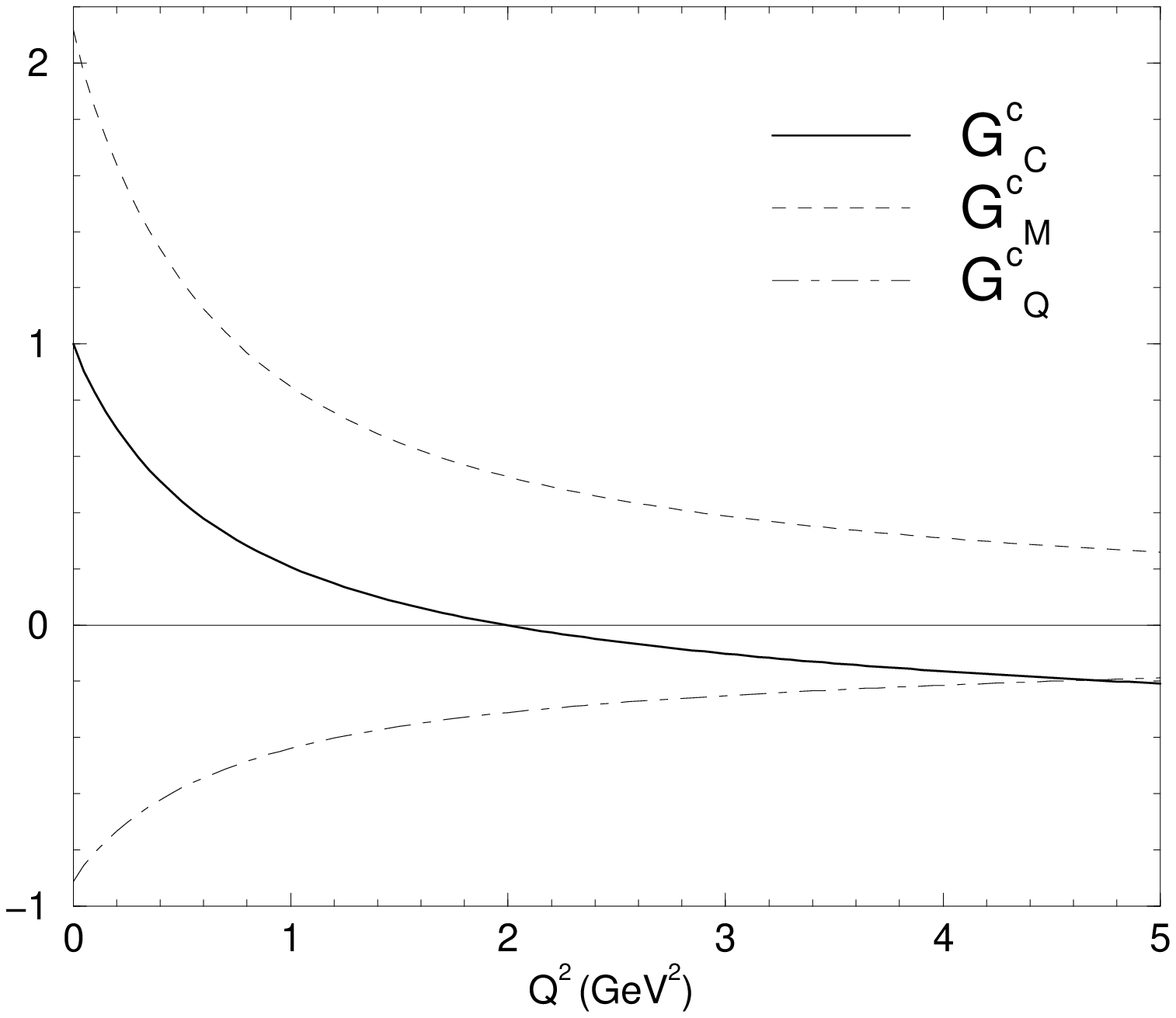,height=10cm,width=8cm}
 \caption{Physical form factors $G_C$, $G_M$, and  $G_Q$ from $F^c_2$
 calculated in the target-rest-frame.
 Left $\theta = \pi/20$, right $\theta = 9\pi/20$. Valence parts only.
 \label{fig.13}}
\end{figure}

\begin{figure}[t]
\psfig{figure=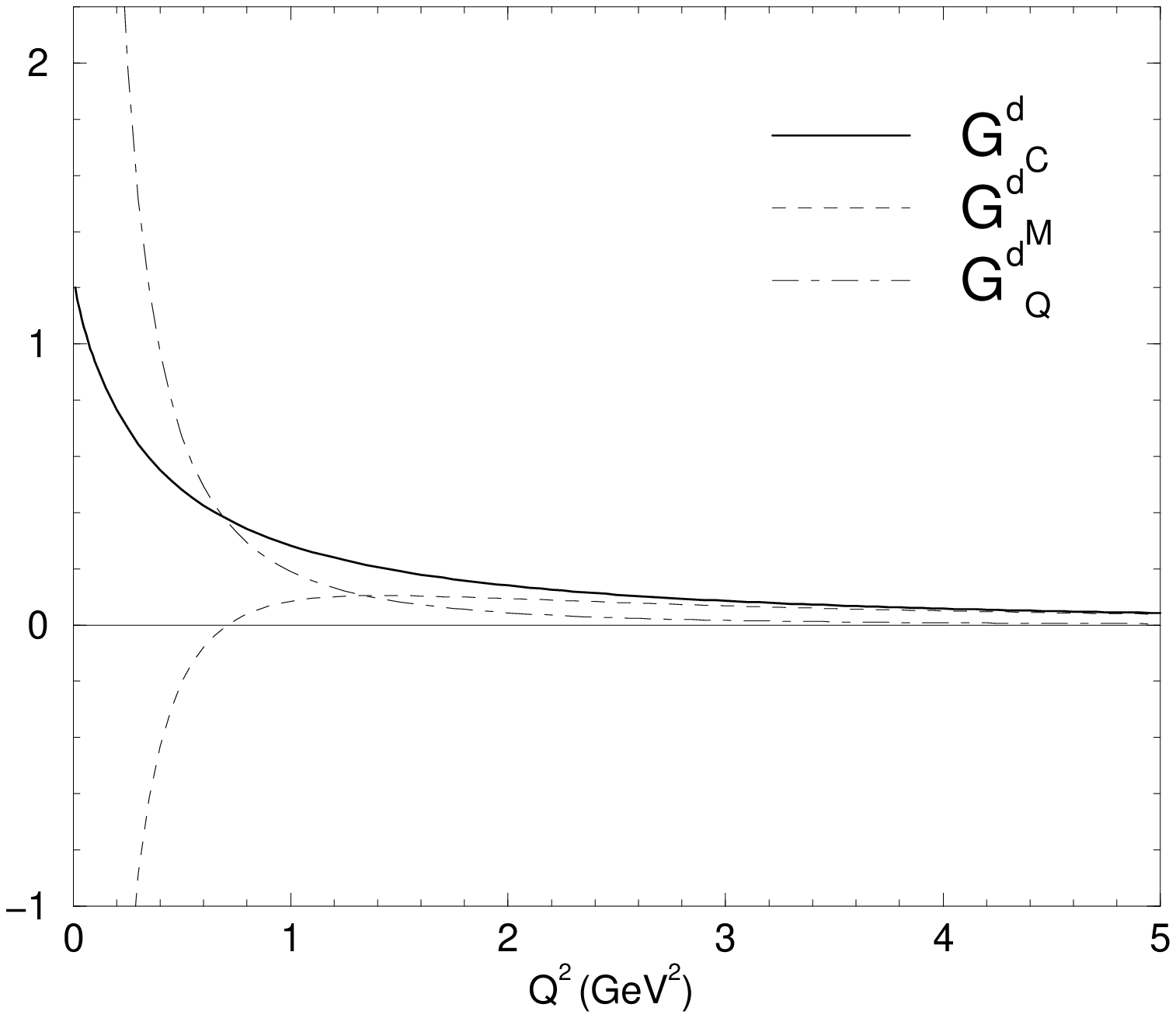,height=10cm,width=8cm}
\psfig{figure=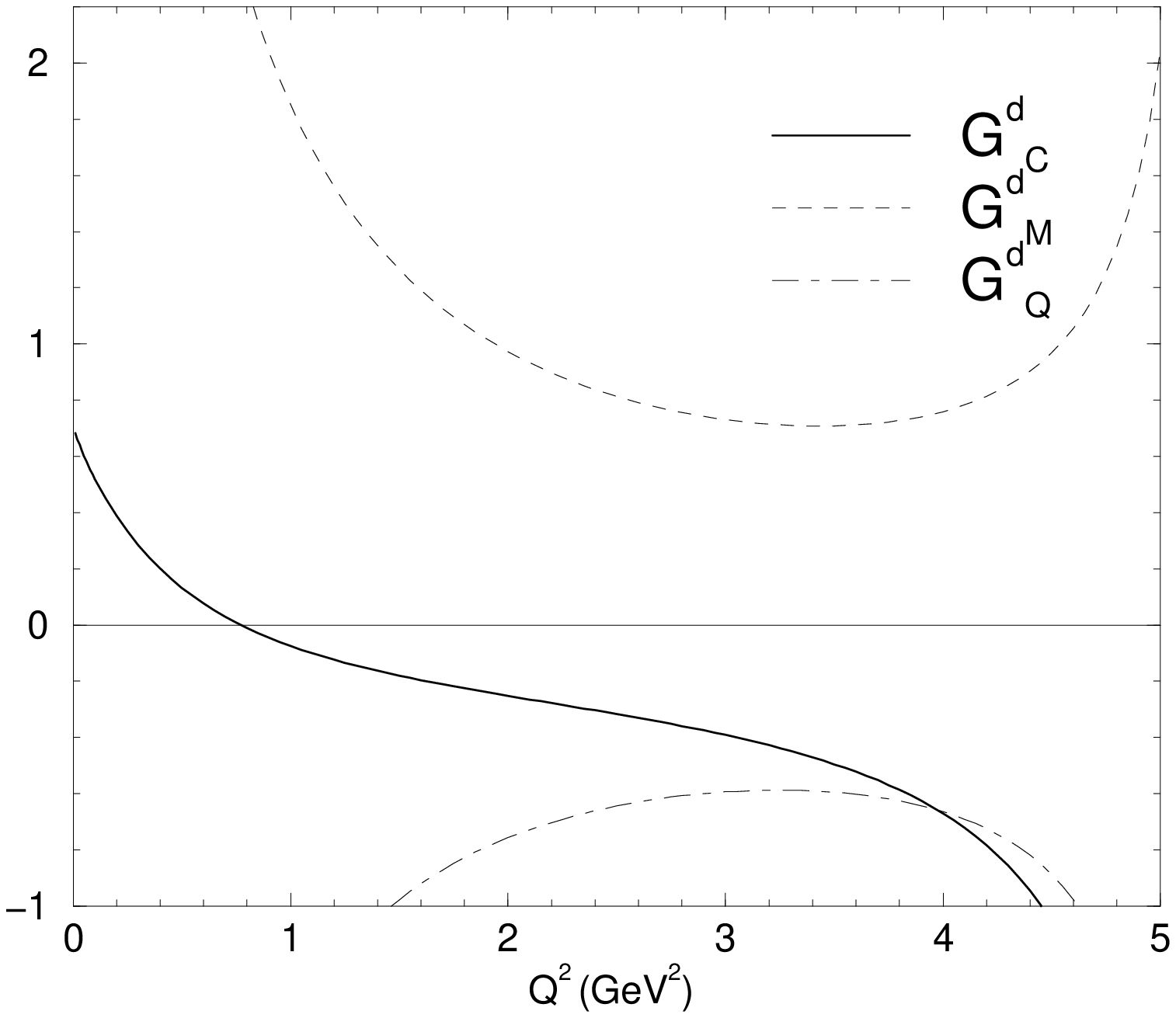,height=10cm,width=8cm}
 \caption{Physical form factors $G_C$, $G_M$, and  $G_Q$ from $F^d_2$
 calculated in the target-rest-frame.
 Left $\theta = \pi/20$, right $\theta = 9\pi/20$. Valence parts only.
 \label{fig.14}}
\end{figure}
\begin{figure}[t]
\begin{center}
\psfig{figure=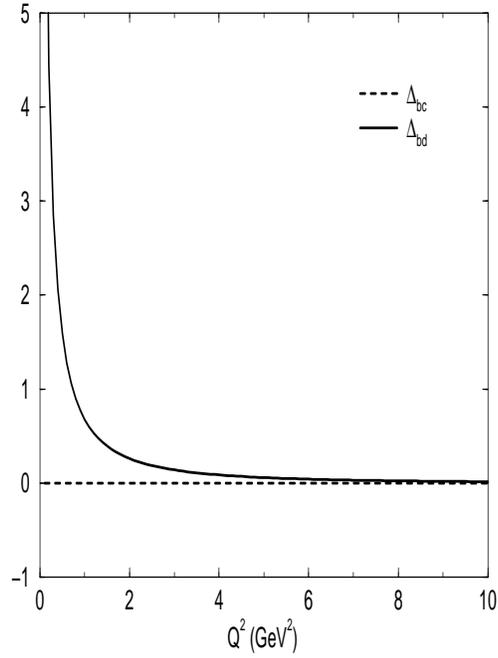,height=10cm,width=8cm}
 \caption{Angular conditions $\Delta_{\rm bc}$ and $\Delta_{\rm bd}$ in the
 Drell-Yan-West frame.
 \label{fig.15}}
\end{center}
\end{figure}
\begin{figure}[t]
\psfig{figure=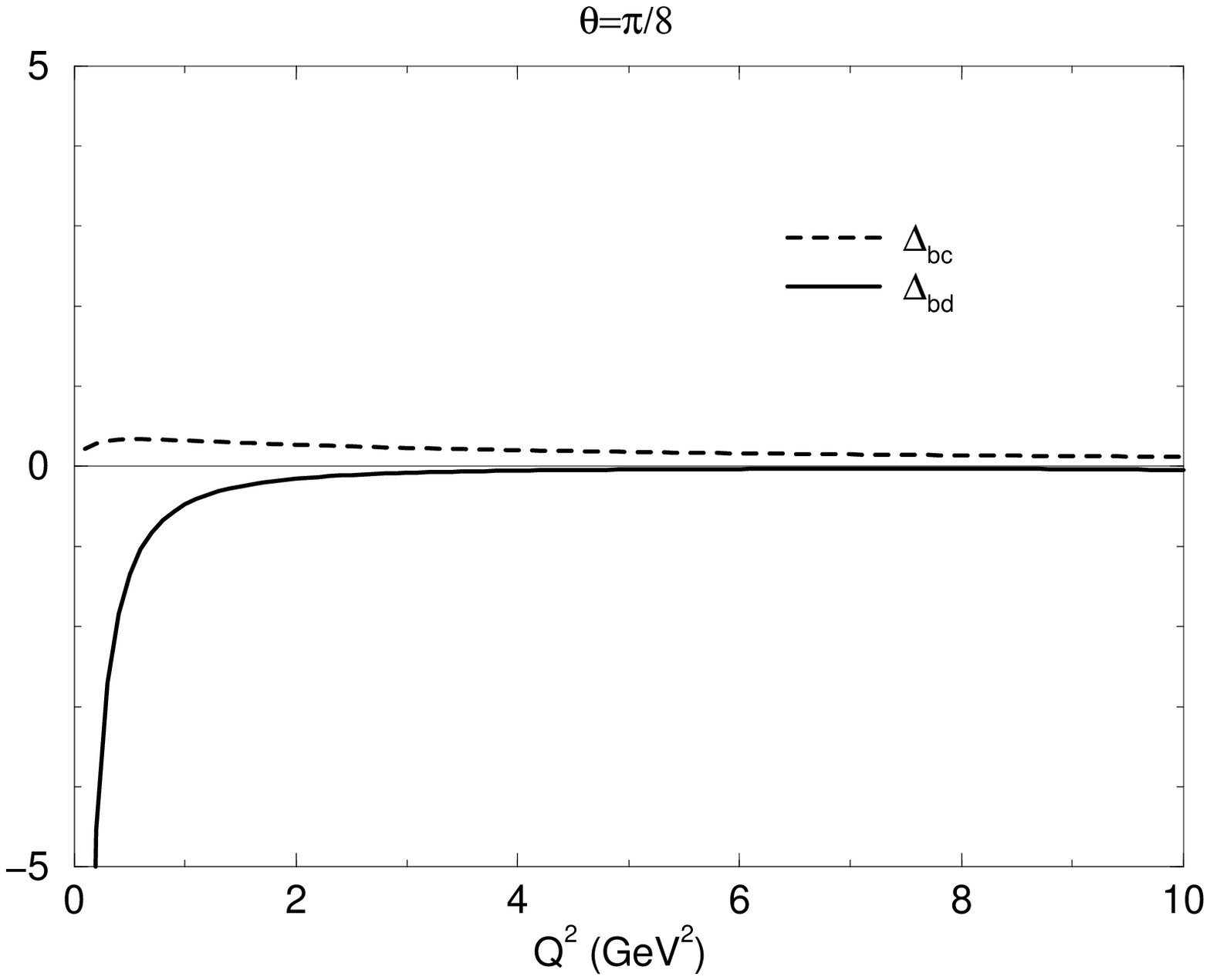,height=10cm,width=8cm}
\psfig{figure=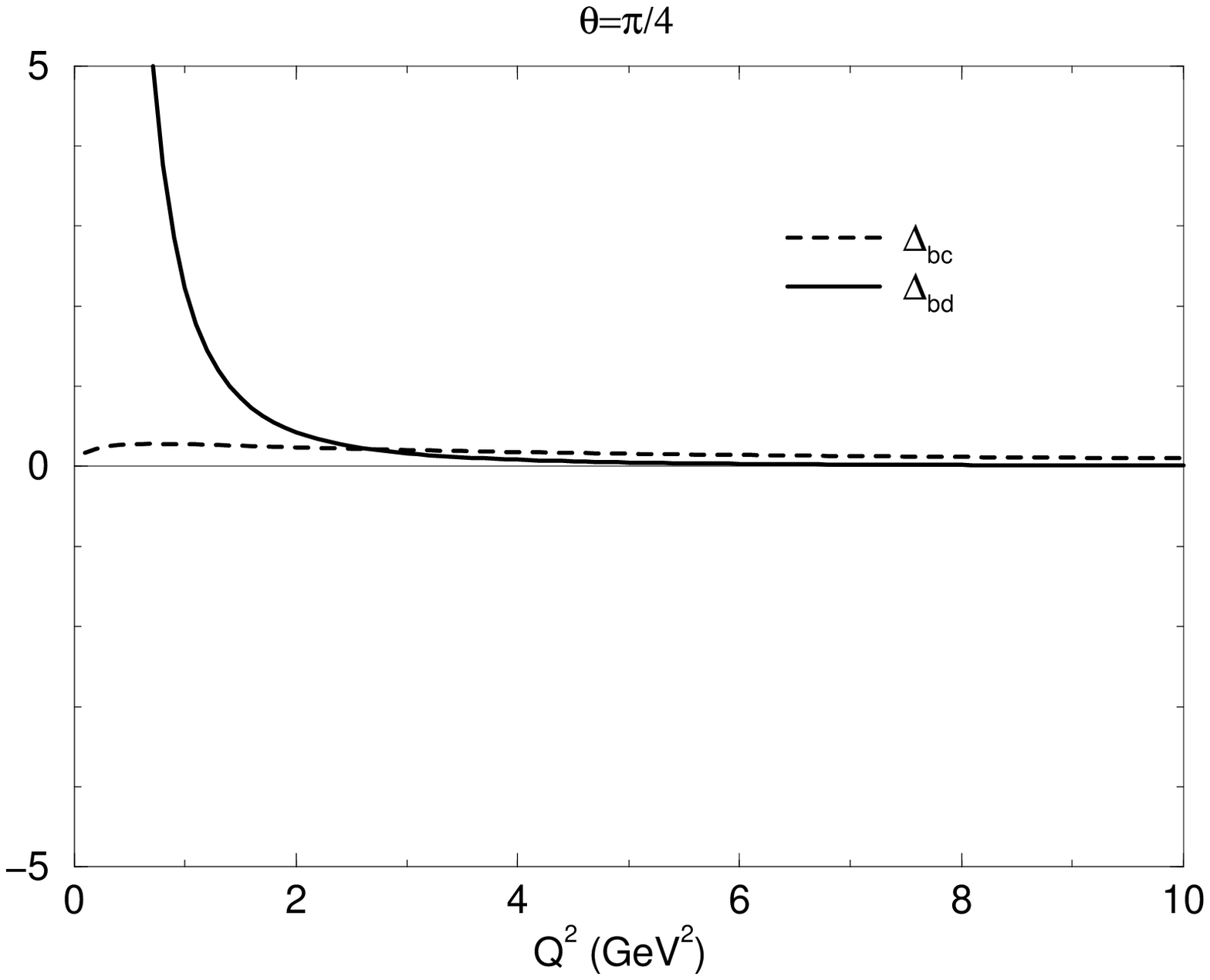,height=10cm,width=8cm}
\psfig{figure=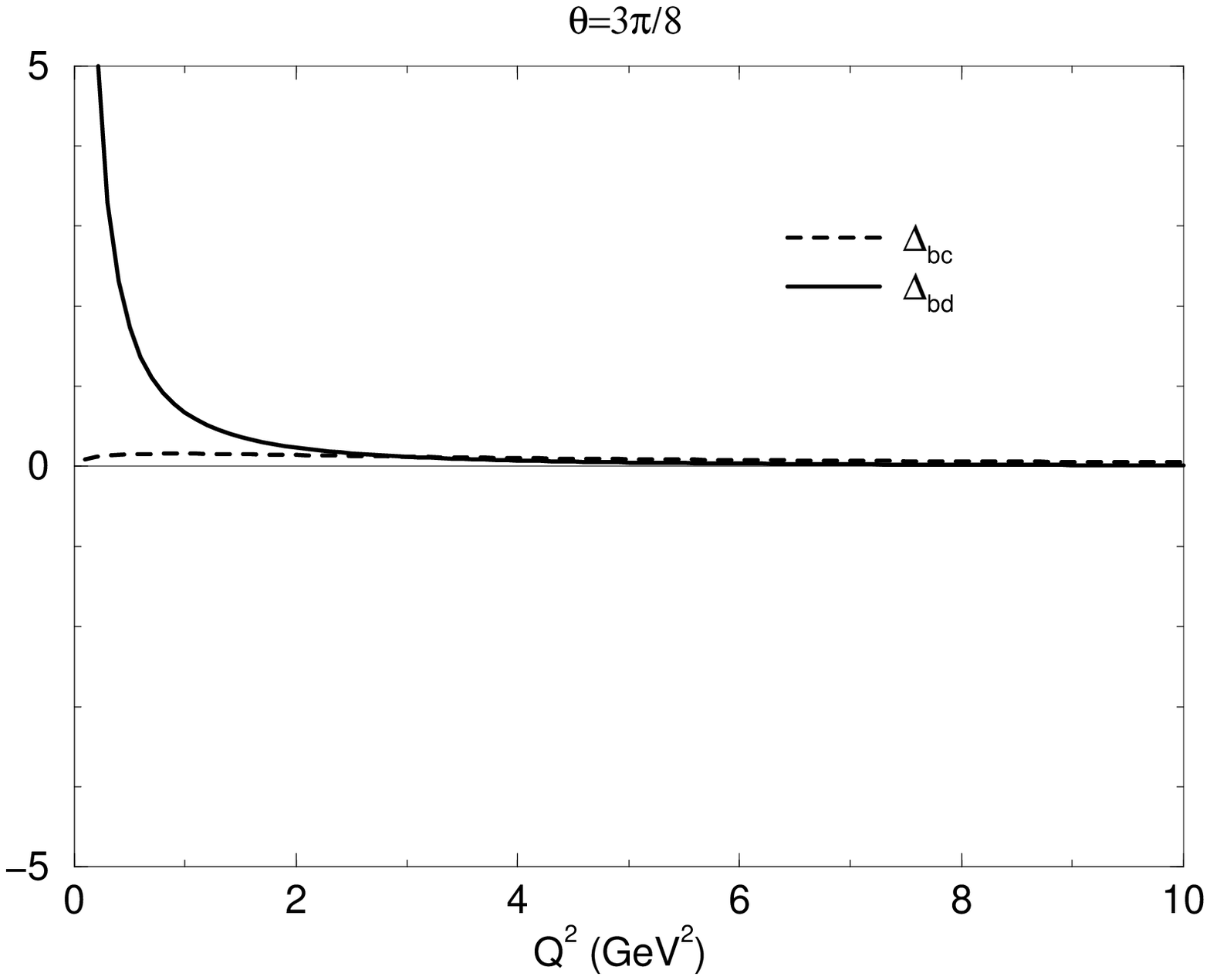,height=10cm,width=8cm}
\psfig{figure=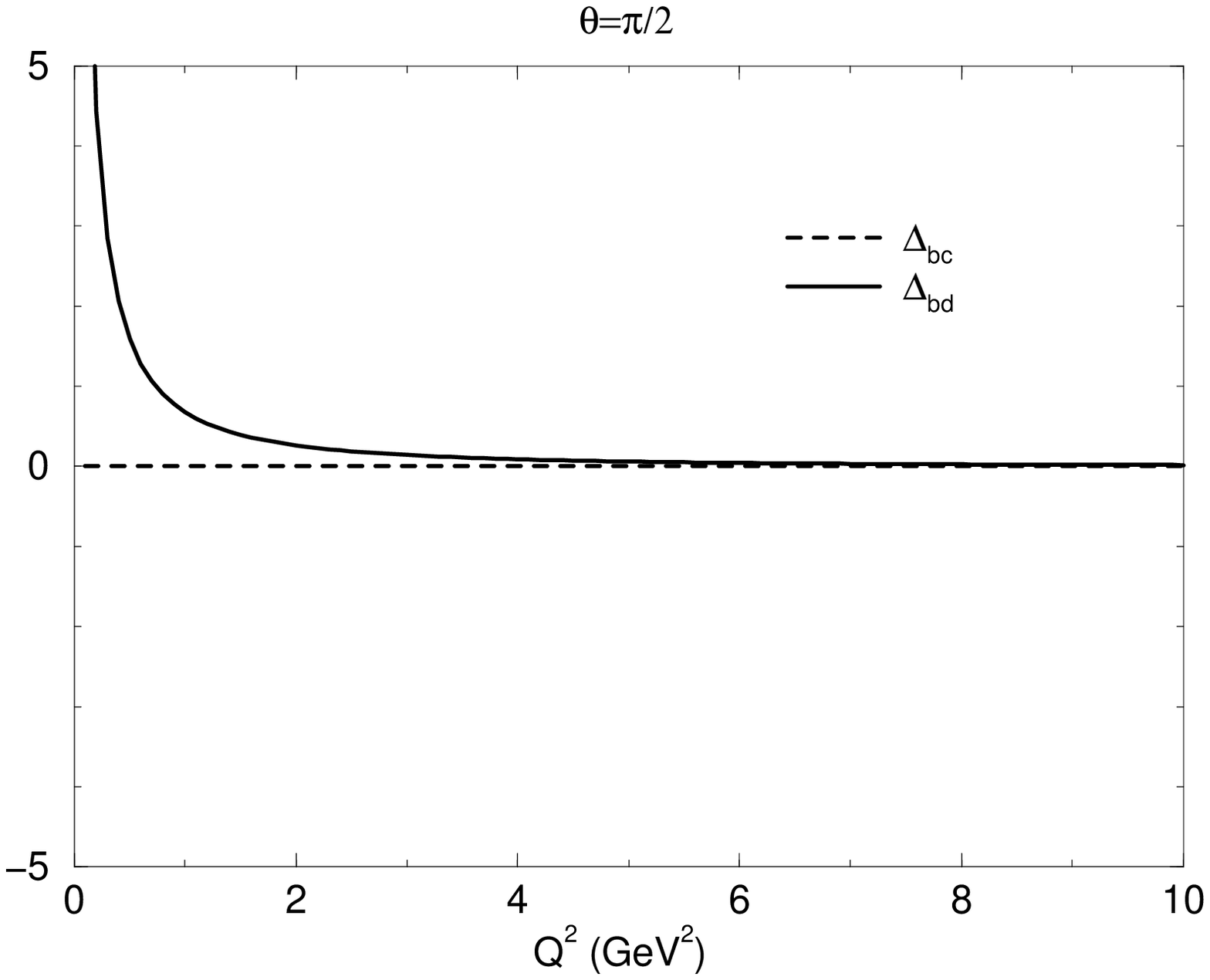,height=10cm,width=8cm}
 \caption{Dependence of the angular conditions $\Delta_{\rm bc}$ and
 $\Delta_{\rm bd}$ in the Breit frame on $Q^2$ for four different angles
 $\theta$.
 \label{fig.16}}
\end{figure}
\begin{figure}[t]
\psfig{figure=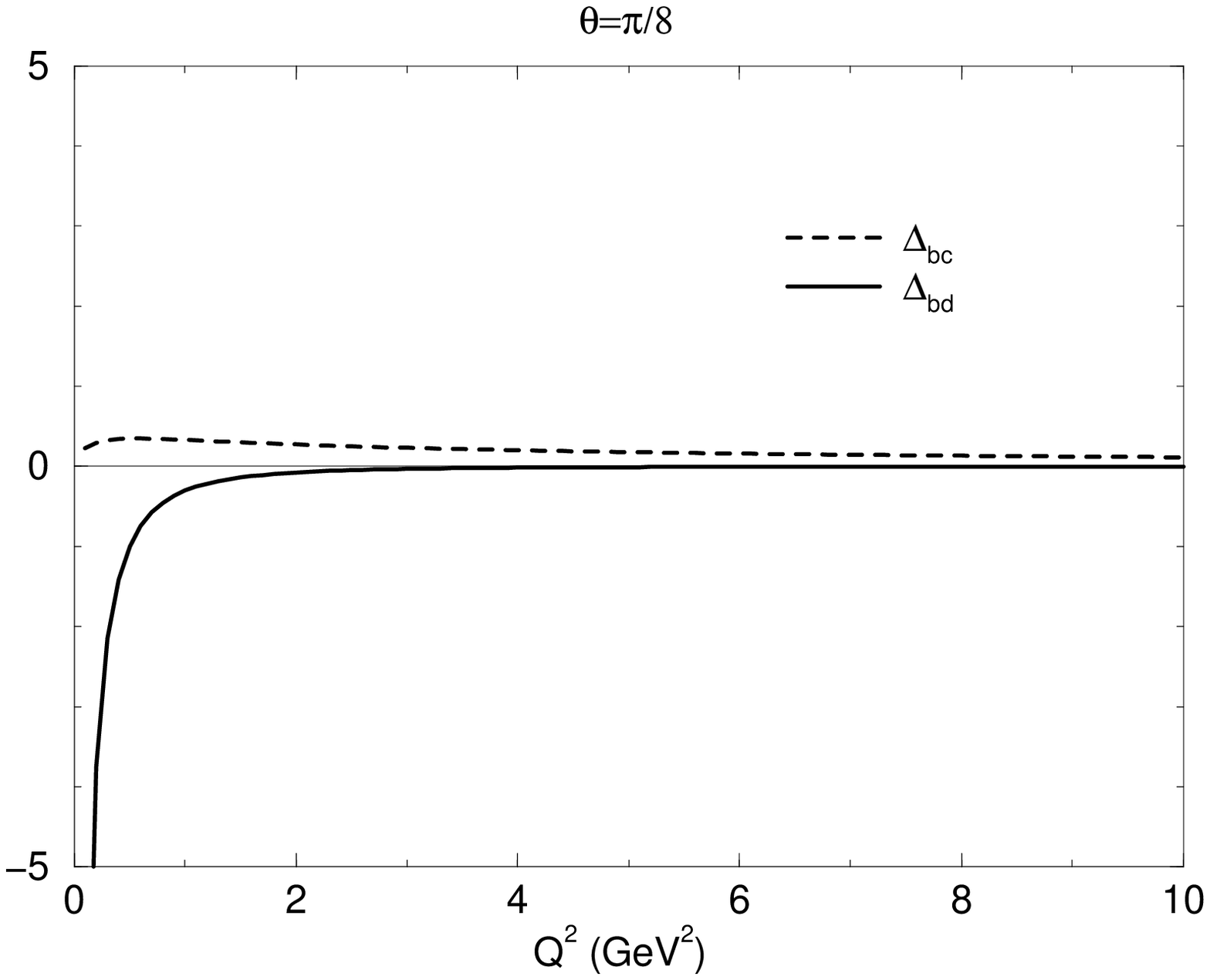,height=10cm,width=8cm}
\psfig{figure=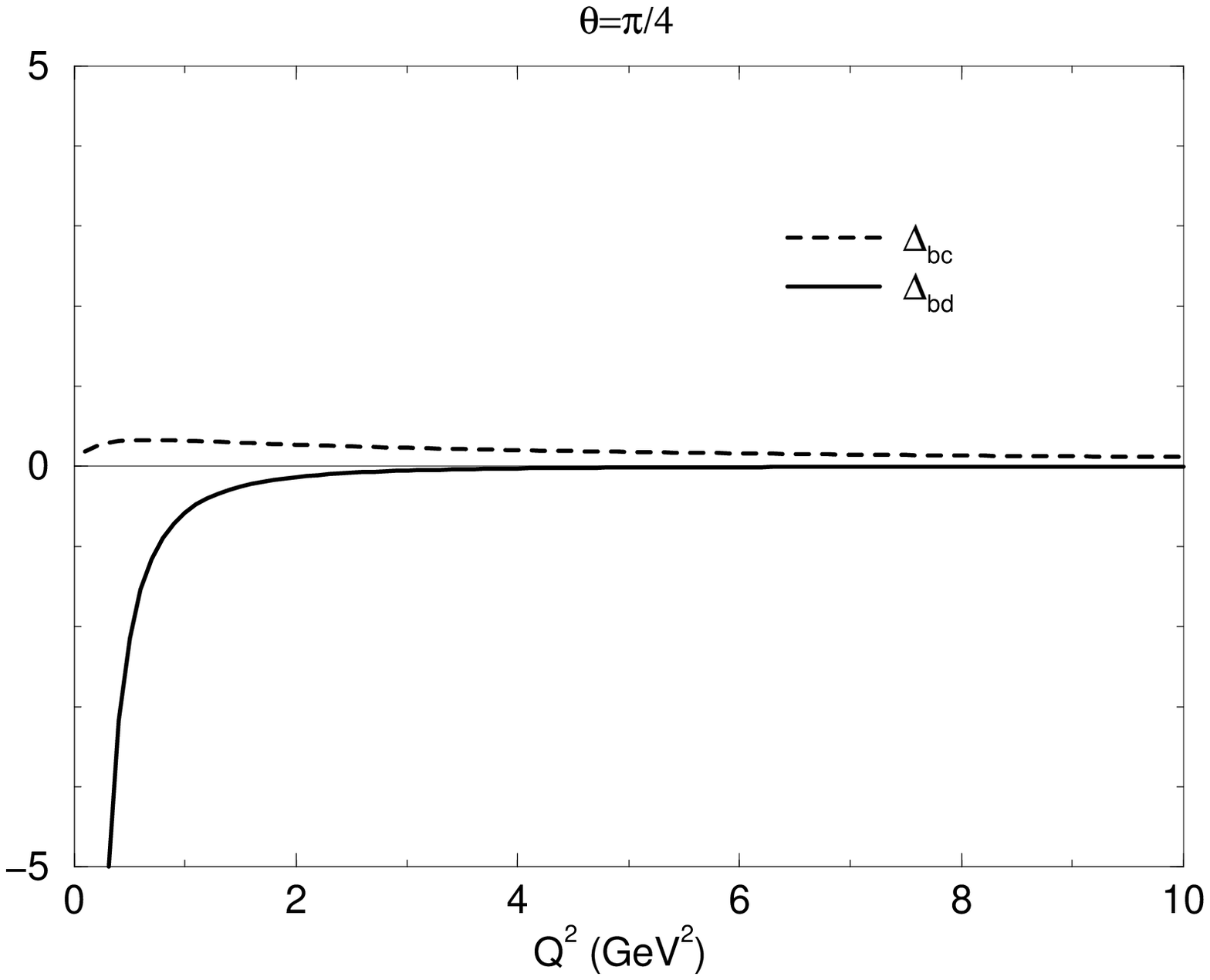,height=10cm,width=8cm}
\psfig{figure=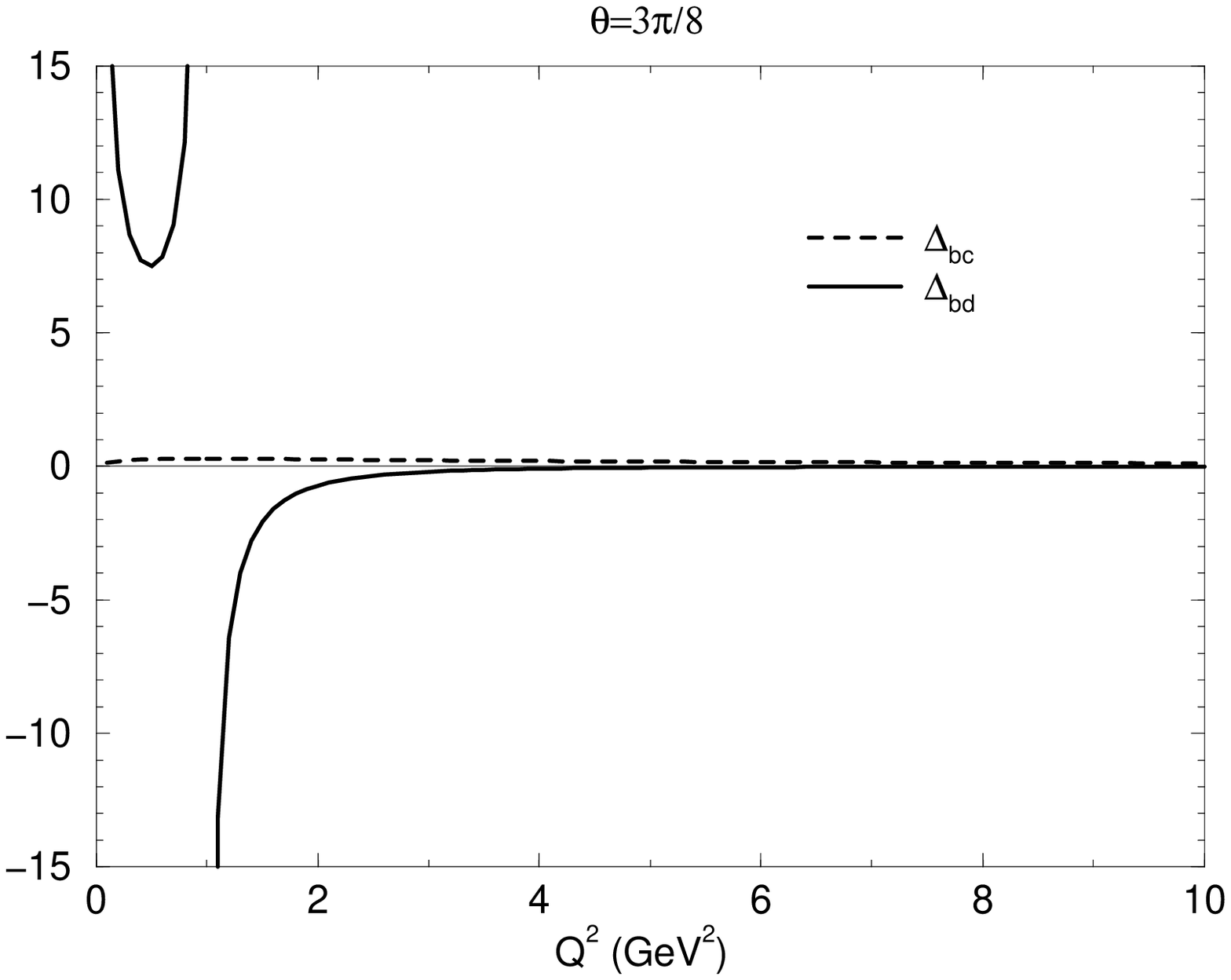,height=10cm,width=8cm}
\psfig{figure=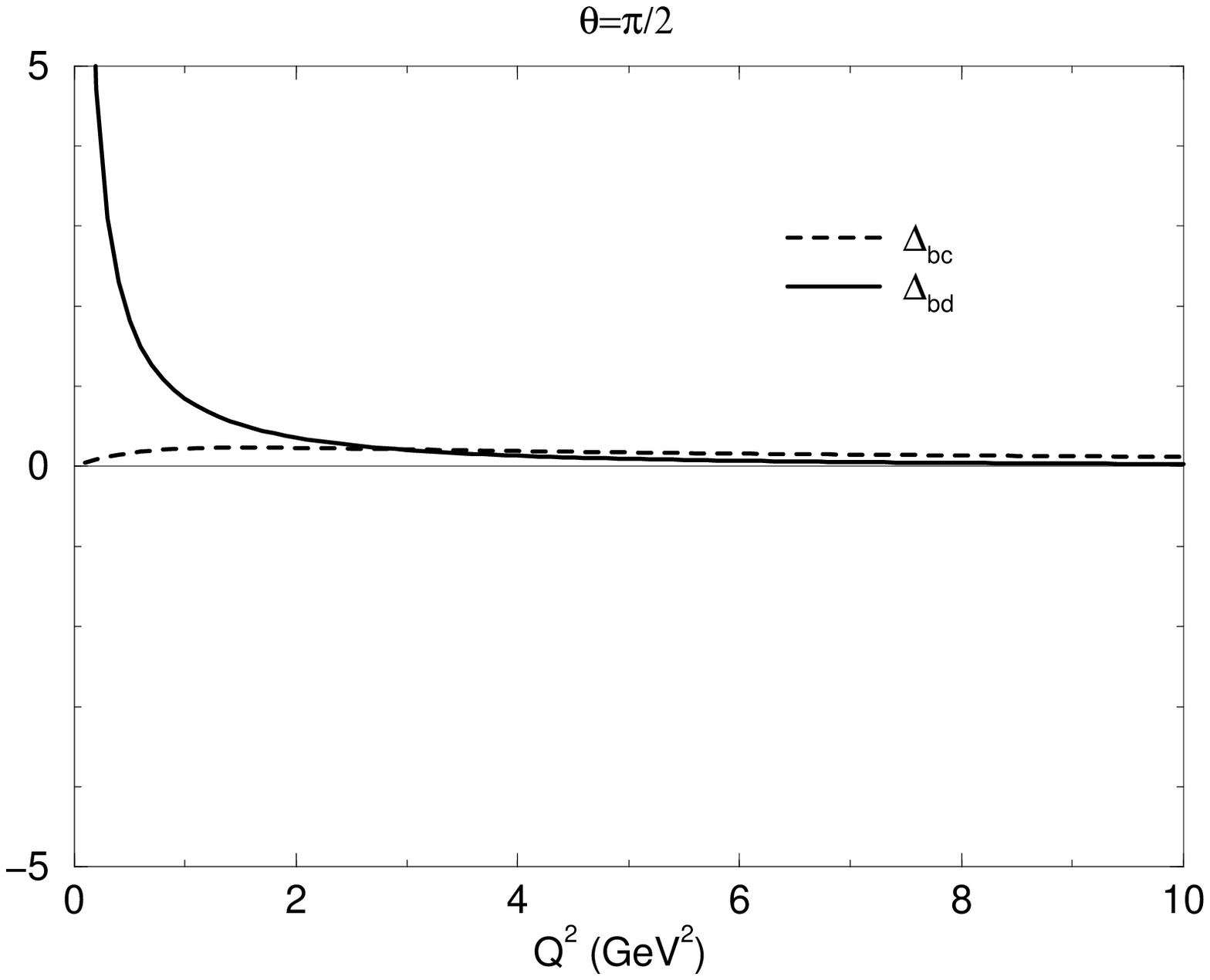,height=10cm,width=8cm}
 \caption{Dependence of the angular conditions $\Delta_{\rm bc}$ and
 $\Delta_{\rm bd}$ in the target-rest frame on $Q^2$ for four different angles
 $\theta$. Note the changed scale in the case $\theta = 3 \pi/8$. The
 singlularity is clearly visible there.
 \label{fig.17}}
\end{figure}
\begin{figure}[t]
\psfig{figure=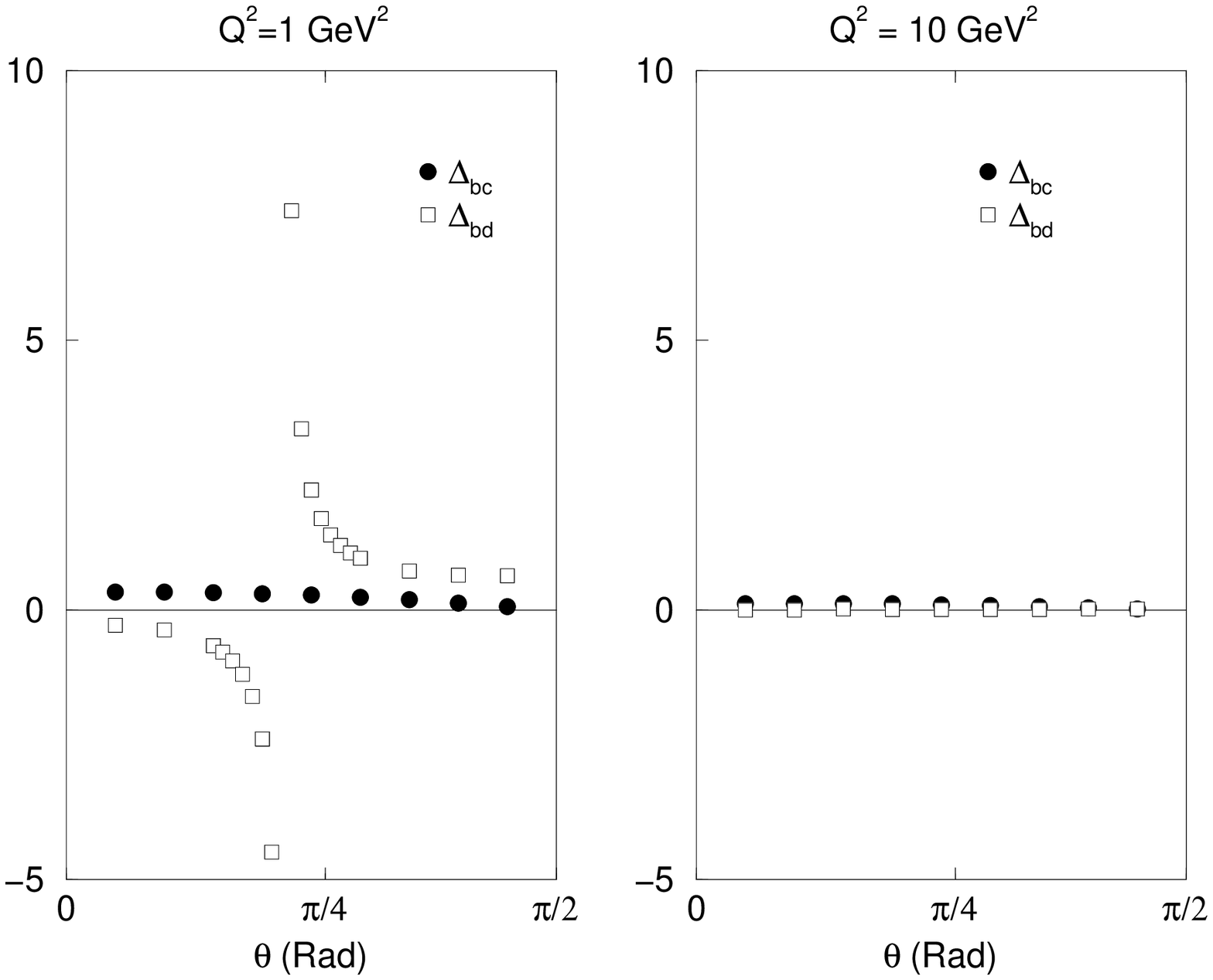,height=10cm,width=16cm}
 \caption{Angular conditions $\Delta_{\rm bc}$ and $\Delta_{\rm bd}$ in the
 Breit frame for $Q^2 = 1.0$ GeV${}^2$ and $Q^2 = 10.0$ GeV${}^2$
 for different angles $\theta$.
 \label{fig.18}}
\end{figure}

\begin{figure}[t]
\psfig{figure=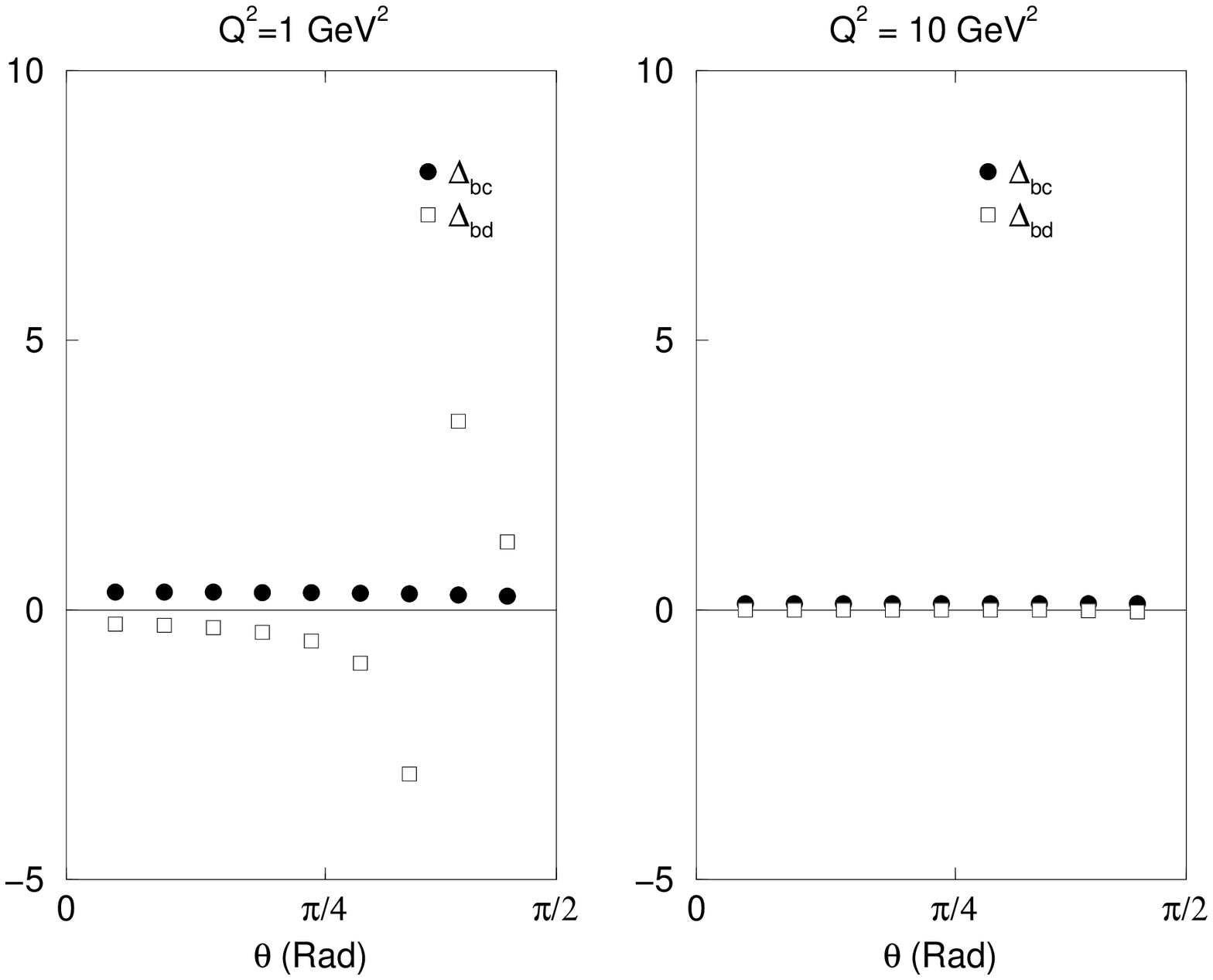,height=10cm,width=16cm}
 \caption{Angular conditions $\Delta_{\rm bc}$ and $\Delta_{\rm bd}$ in the
 target-rest frame for $Q^2 = 1.0$ GeV${}^2$ and $Q^2 = 10.0$ GeV${}^2$
 for different angles $\theta$.
 \label{fig.19}}
\end{figure}

\begin{figure}[t]
\begin{center}
\epsfig{figure=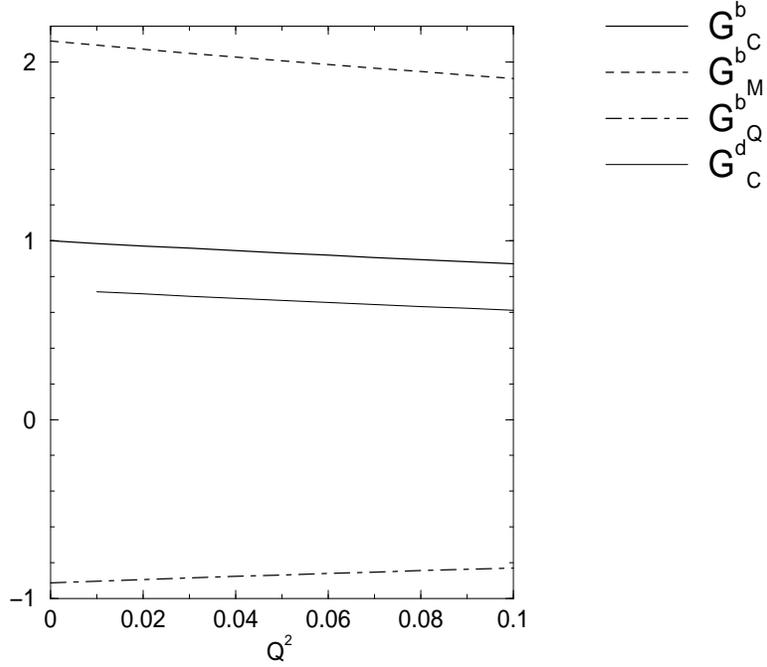,height=10cm,width=10cm}
 \caption{Valence contributions to physical form factors for small values of 
$Q^2$ in DYW frame. While $b$ and $d$ variants are shown here, the  
$d$-variant magnetic($G^d_M$) and quadrupole($G^d_Q$) form factors are 
out of scale because they diverge as $Q^2 \to 0$.
 \label{fig.1}}
\end{center}
\end{figure}

\begin{figure}[t]
\begin{center}
\epsfig{figure=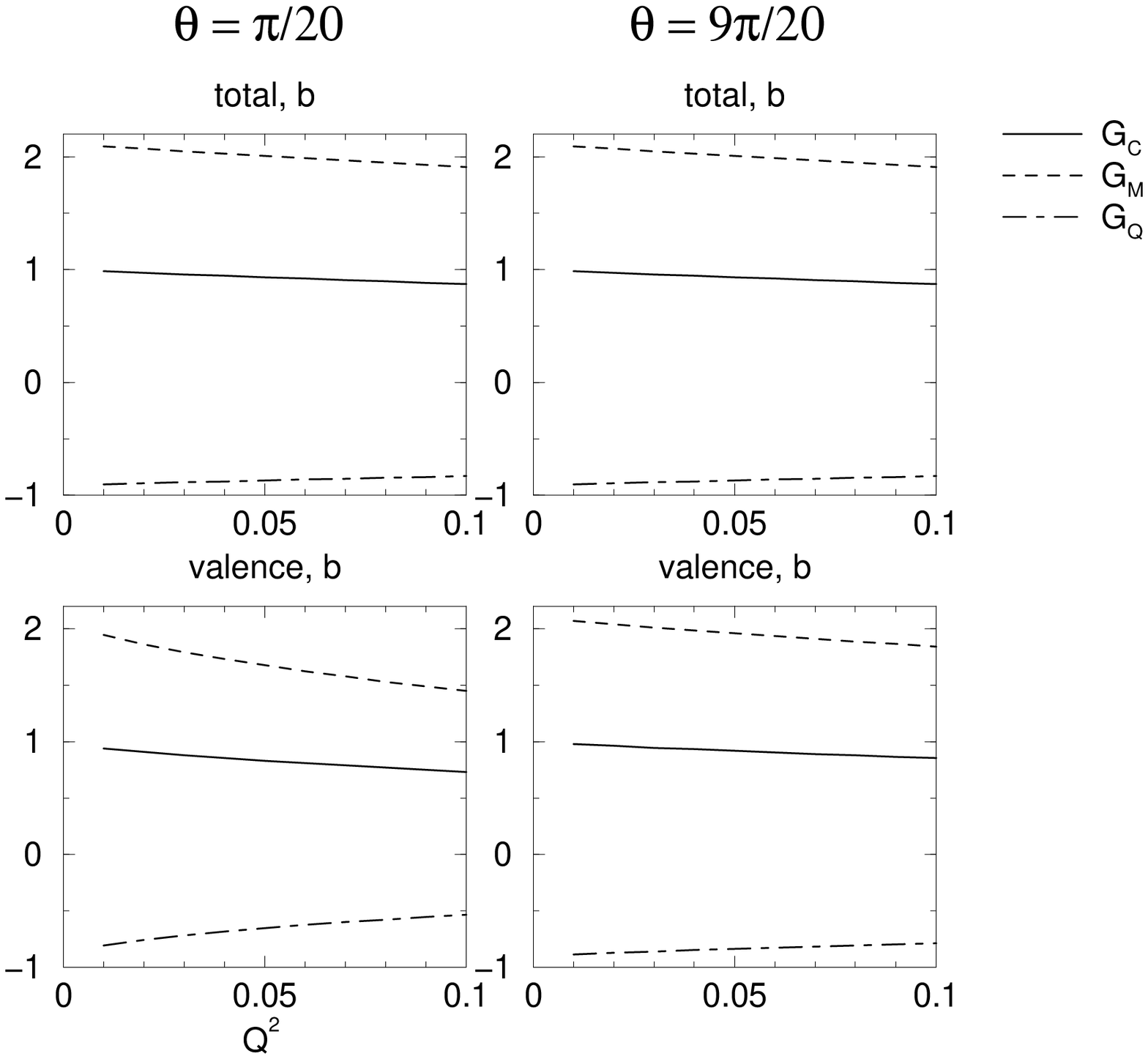,height=10cm,width=10cm}
 \caption{Physical form factors for small values of $Q^2$.
 BRT frame, variant $b$.
 \label{fig.2}}
\end{center}
\end{figure}
\begin{figure}[t]
\begin{center}
\epsfig{figure=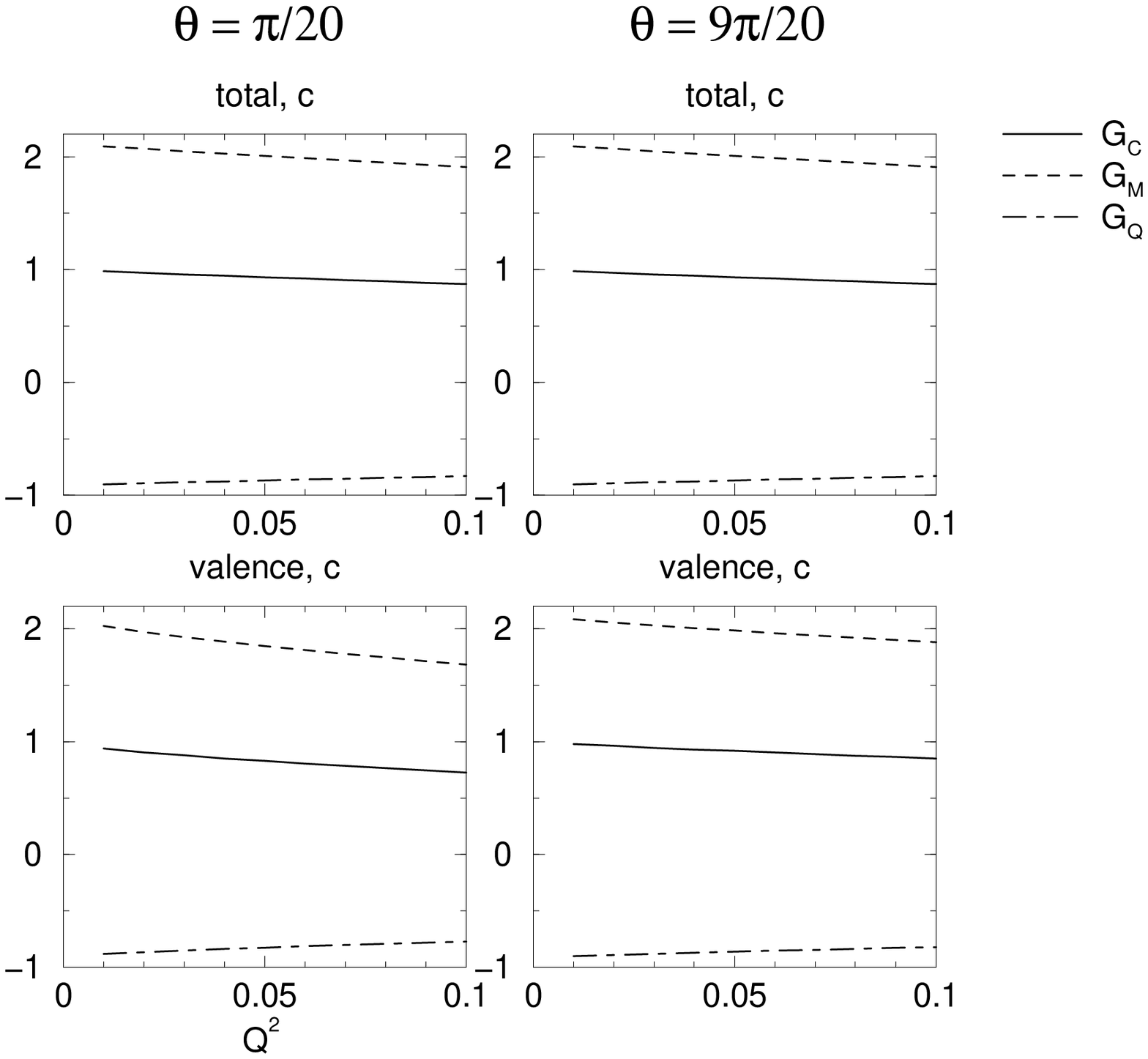,height=10cm,width=10cm}
 \caption{Physical form factors for small values of $Q^2$.
 BRT frame, variant $c$.
 \label{fig.3}}
\end{center}
\end{figure}
\begin{figure}[t]
\begin{center}
\epsfig{figure=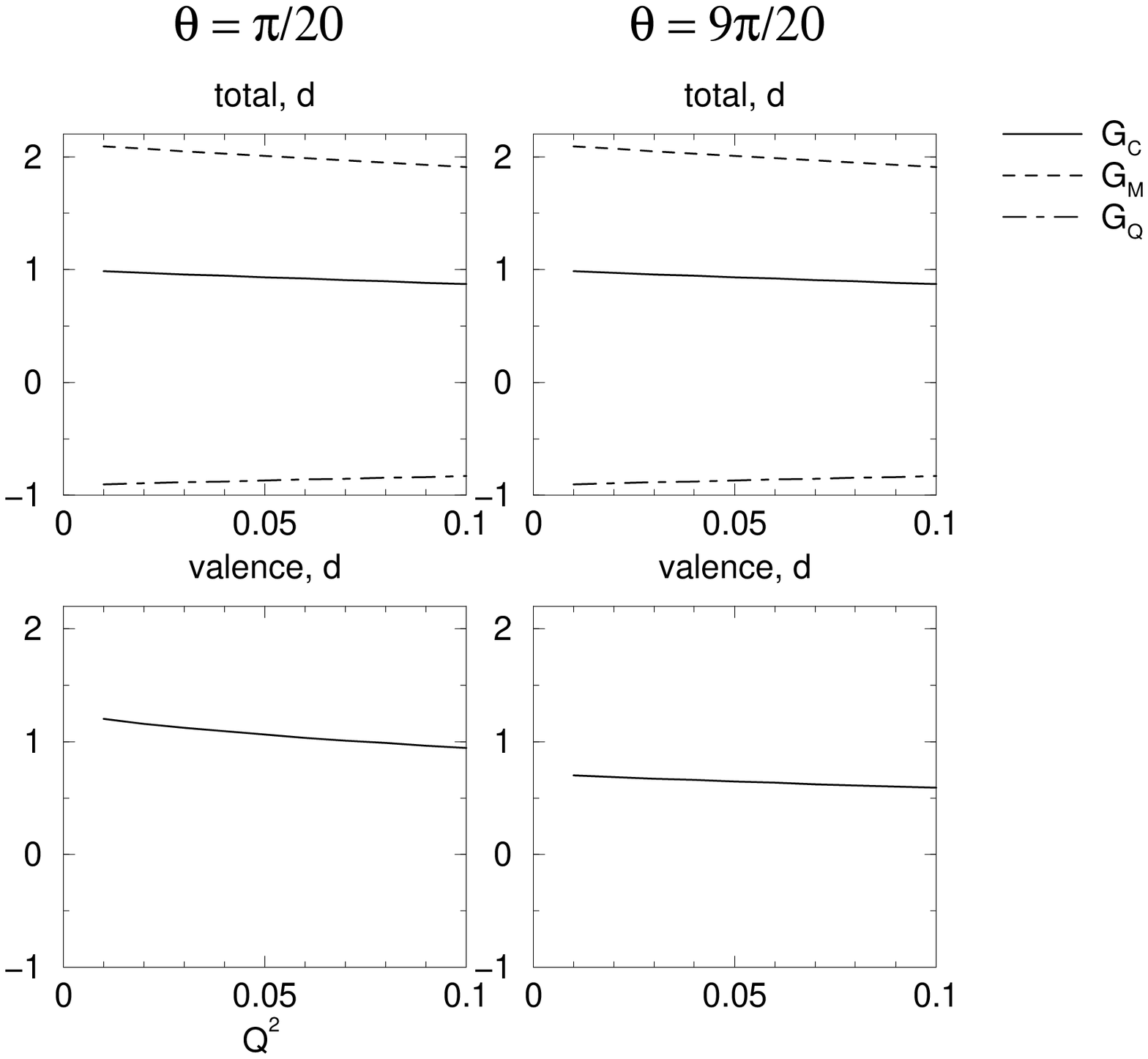,height=10cm,width=10cm}
 \caption{Physical form factors for small values of $Q^2$.
 BRT frame, variant $d$.
 \label{fig.4}}
\end{center}
\end{figure}
\begin{figure}[t]
\begin{center}
\epsfig{figure=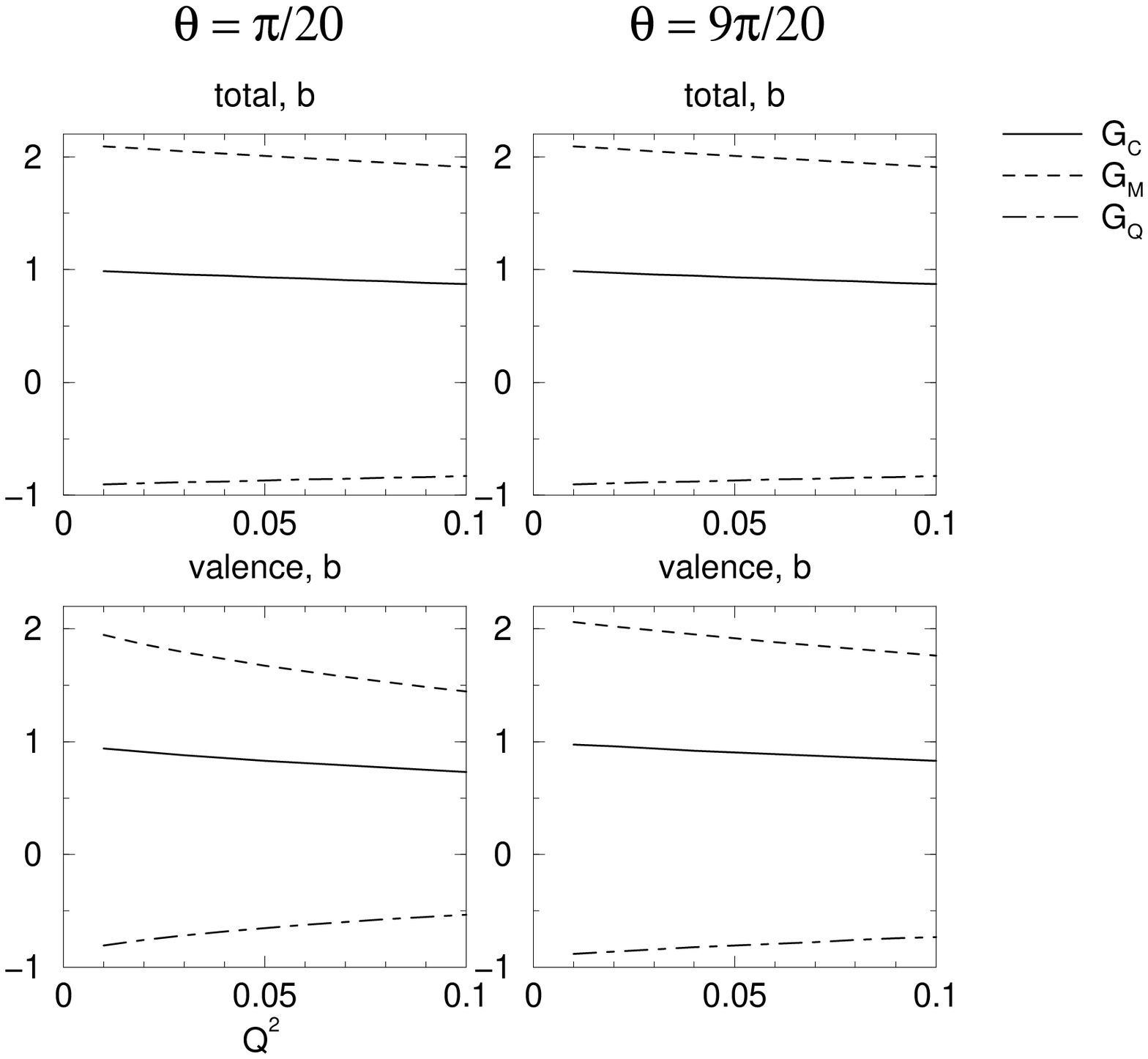,height=10cm,width=10cm}
 \caption{Physical form factors for small values of $Q^2$.
 TRF, variant $b$.
 \label{fig.5}}
\end{center}
\end{figure}
\begin{figure}[t]
\begin{center}
\epsfig{figure=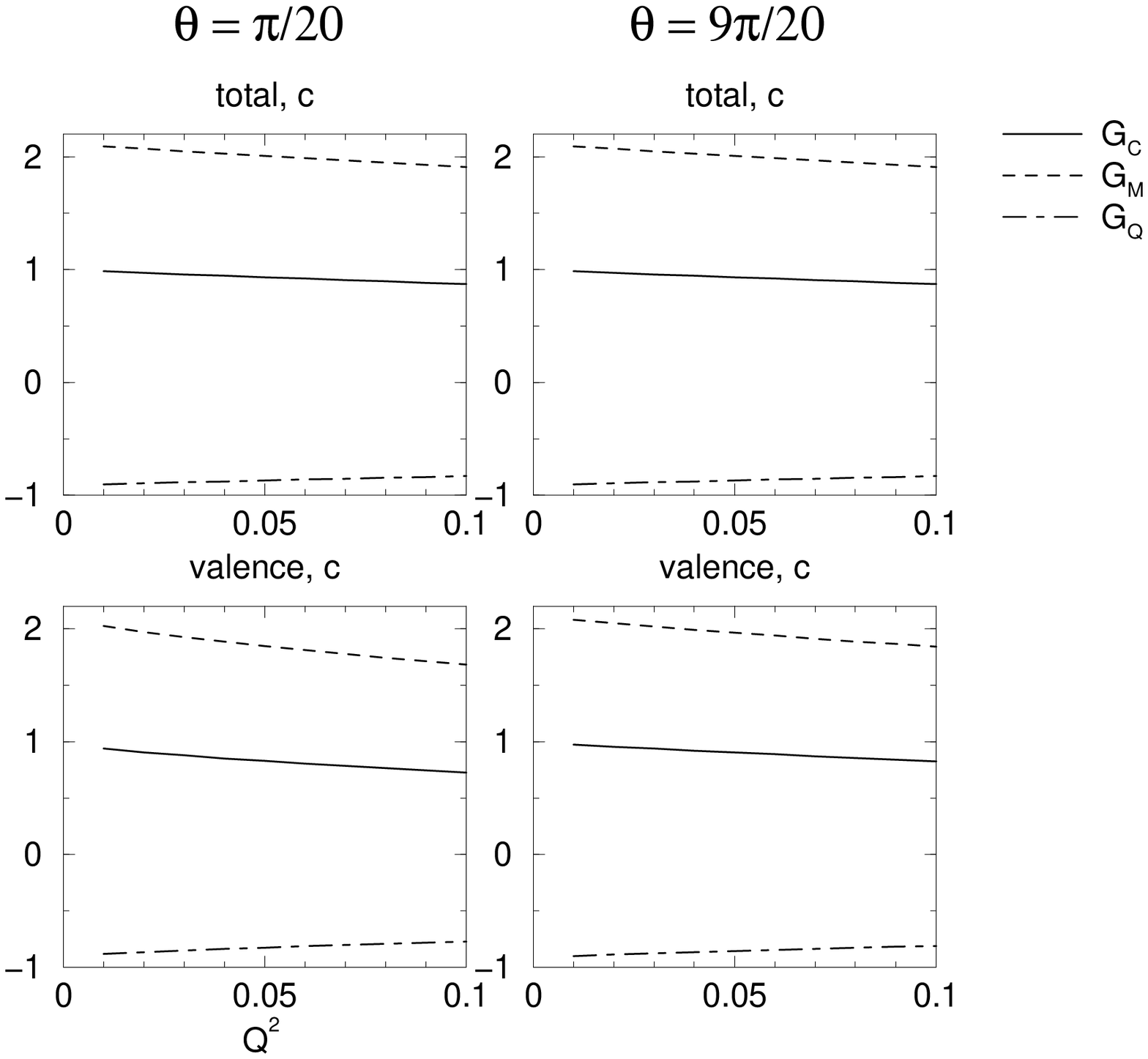,height=10cm,width=10cm}
 \caption{Physical form factors for small values of $Q^2$.
 TRF, variant $c$.
 \label{fig.6}}
\end{center}
\end{figure}
\begin{figure}[t]
\begin{center}
\epsfig{figure=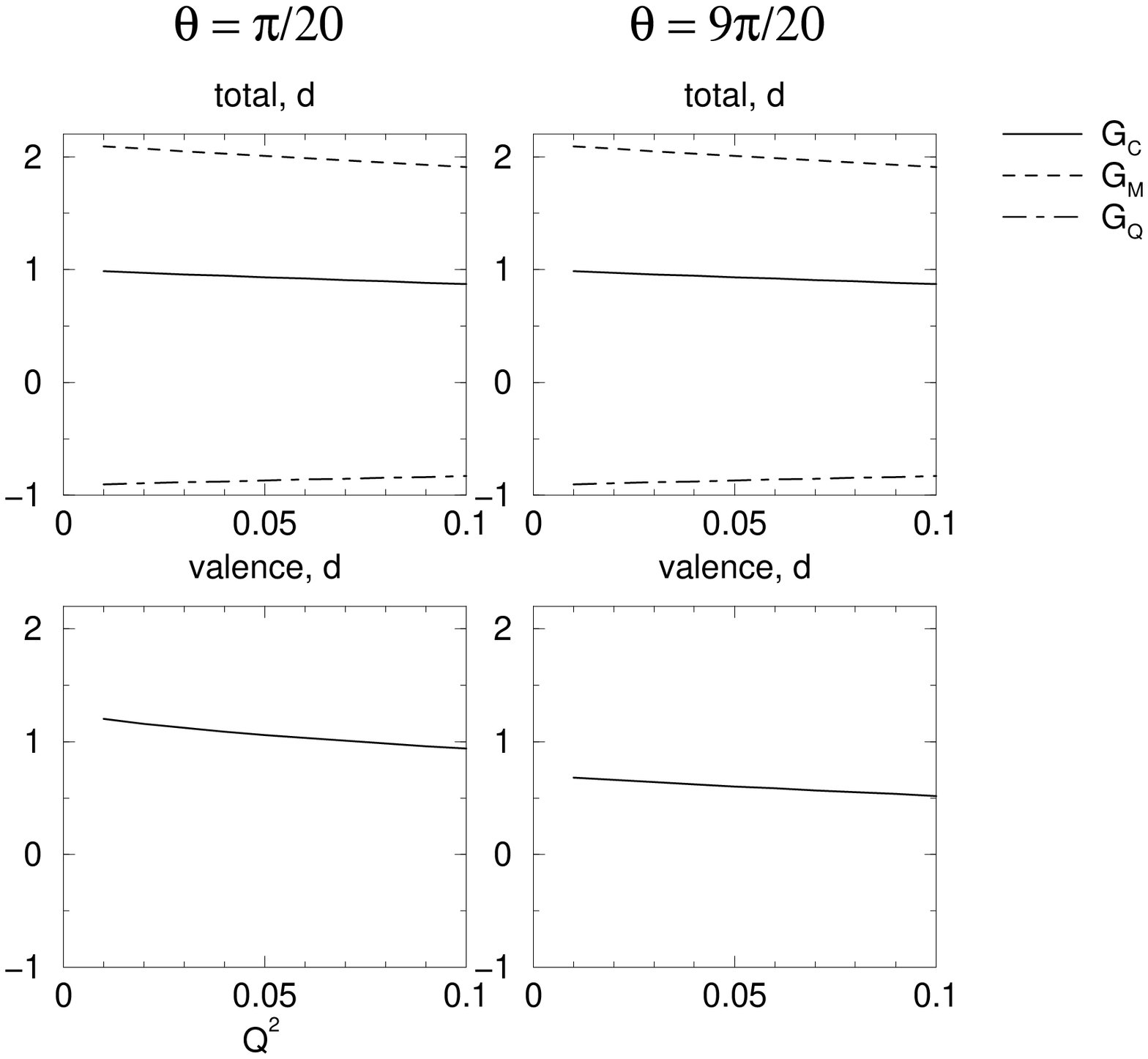,height=10cm,width=10cm}
 \caption{Physical form factors for small values of $Q^2$.
 TRF, variant $d$.
 \label{fig.7}}
\end{center}
\end{figure}


\begin{thebibliography}{99}

\bibitem{BCJ1} B. L. G. Bakker, H.-M. Choi, and C.-R. Ji,
\Journal{\PRD}{63}{074014}{2001}.

\bibitem{GK} I. L. Grach and L. A. Kondratyuk,
Sov. J. Nucl. Phys. {\bf 39}, 198 (1984).

\bibitem{CCKP} P. L. Chung, F. Coester, B. D. Keister
and W. N. Polyzou, \Journal{\PRC}{37}{2000}{1988}.

\bibitem{BH} S. J. Brodsky and J. R. Hiller,
\Journal{\PRD}{46}{2141}{1992}.

\bibitem{Card} F. Cardarelli, I. L. Grach, I. M. Narodetskii,
G. Salme, and S. Simula, \Journal{\PLB}{349}{393}{1995}.

\bibitem{BJ2} B. L. G. Bakker and C.-R. Ji,
\Journal{\PRD}{65}{03xxxx}{2002} [hep-ph/0109005].

\bibitem{Melo1} J.P.B.C. de Melo et al.,
\Journal{\NPA}{631}{574c}{1998}; \Journal{\NPA}{660}{219}{1999}.

\bibitem{MT} J.P.B.C. de Melo and T. Frederico,
\Journal{\PRC}{55}{2043}{1997}.


\bibitem{MS} D. Melikhov and S.Simula, hep-ph/0112044.

\bibitem{ACG} R. G. Arnold, C. E. Carlson and
F. Gross, \Journal{\PRC}{21}{1426}{1980}.

\bibitem{Kei} B. D. Keister, \Journal{\PRD}{49}{1500}{1994}.

\bibitem{CJNPA} H.-M. Choi and C.-R. Ji, \Journal{\NPA}{618}{291}{1997}.

\bibitem{FFS} L. L. Frankfurt, T. Frederico, and M. Strikman,
\Journal{\PRC}{48}{2182}{1993}.

\bibitem{CM} S.-J. Chang and T.-M. Yan,
\Journal{\PRD}{7}{1147}{1973}; \Journal{\PRD}{7}{1780}{1973}.

\bibitem{Bur} M. Burkardt, \Journal{\NPA}{504}{762}{1989}.

\bibitem{BHw} S. J. Brodsky and D. S. Hwang, \Journal{\NPB}{543}{239}{1998}.

\bibitem{SB} N. C. J. Schoonderwoerd and B. L. G. Bakker,
\Journal{\PRD}{57}{4965}{1998}; \Journal{\PRD}{58}{025013}{1998}.

\bibitem{CJZ} H.-M. Choi and C.-R. Ji,
\Journal{\PRD}{58}{071901}{1998}.

\bibitem{BJ1} B. L. G. Bakker and C.-R. Ji,
\Journal{\PRD}{62}{074014}{2000}.

\bibitem{Ja} W. Jaus, \Journal{\PRD}{60}{054026}{1999}.

\bibitem{pdg} Particle Data Group, C. Caso et al., Eur. Phys. J. C3,
1 (1998).

\end{thebibliography}
\end{document}